\documentstyle [12pt,epsfig]{article} 
\makeatletter
\@addtoreset{equation}{section}

\makeatother

\newcommand{\ba}{\begin{eqnarray}}
\newcommand{\ea}{\end{eqnarray}}

\begin{document}
\newcommand{\BS}{\bigskip}
\newcommand{\SECTION}[1]{\BS{\large\section{\bf #1}}}
\newcommand{\SUBSECTION}[1]{\BS{\large\subsection{\bf #1}}}
\newcommand{\SUBSUBSECTION}[1]{\BS{\large\subsubsection{\bf #1}}}
\newcommand {\rb}  {\overline{r}}

\newcommand {\sbx}  {\overline{s}}
\newcommand {\vb}  {\overline{v}}
\newcommand {\ab}  {\overline{a}}
\newcommand {\sbf}  {\overline{s}_f}
\newcommand {\sbn}  {\overline{s}_{\nu}}
\newcommand {\rbf}  {\overline{r}_f}
\newcommand {\vbf}  {\overline{v}_f}
\newcommand {\vbn}  {\overline{v}_{\nu}}
\newcommand {\abf}  {\overline{a}_f}
\newcommand {\abn}  {\overline{a}_{\nu}}
\newcommand {\sbq}  {\overline{s}_q}
\newcommand {\vbq}  {\overline{v}_q}
\newcommand {\abq}  {\overline{a}_q}
\newcommand {\rbQ}  {\overline{r}_Q}
\newcommand {\vbQ}  {\overline{v}_Q}
\newcommand {\abQ}  {\overline{a}_Q}
\newcommand {\rbl}  {\overline{r}_l}

\newcommand {\vbl}  {\overline{v}_l}
\newcommand {\abl}  {\overline{a}_l}
\newcommand {\sbl}  {\overline{s}_l}

\newcommand {\rbc}  {\overline{r}_{\rm{c}}}
\newcommand {\vbc}  {\overline{v}_{\rm{c}}}
\newcommand {\abc}  {\overline{a}_{\rm{c}}}
\newcommand {\sbc}  {\overline{s}_{\rm{c}}}
\newcommand {\rbb}  {\overline{r}_{\rm{b}}}
\newcommand {\vbb}  {\overline{v}_{\rm{b}}}
\newcommand {\abb}  {\overline{a}_{\rm{b}}}
\newcommand {\sbb}  {\overline{s}_{\rm{b}}}
\newcommand {\sbQ}  {\overline{s}_Q}
\newcommand {\sbnbp}  {\overline{s}_{\rm{nb}}'}
\newcommand {\gbl}  {\overline{g}_{\rm{b}}^L}
\newcommand {\gbr}  {\overline{g}_{\rm{b}}^R}
\newcommand {\gcl}  {\overline{g}_{\rm{c}}^L}
\newcommand {\gcr}  {\overline{g}_{\rm{c}}^R}
\newcommand {\gll}  {\overline{g}_{\rm{l}}^L}
\newcommand {\glr}  {\overline{g}_{\rm{l}}^R}
\newcommand {\gnl}  {\overline{g}_{\rm{\nu}}^L}
\newcommand {\Afbf}  {A_{FB}^{0,\rm{f}}}
\newcommand {\Afbl}  {A_{FB}^{0,l}}
\newcommand {\Afbc}  {A_{FB}^{0,\rm{c}}}
\newcommand {\Afbb}  {A_{FB}^{0,\rm{b}}}
\newcommand {\AfbQ}  {A_{FB}^{0,\rm{Q}}}
\newcommand {\tpol}  {$\tau$-polarisation}
\newcommand {\alps}  {\alpha_s(m_{\rm{Z}})}
\newcommand {\alp}  {\alpha(m_{\rm{Z}})}
\begin{titlepage}
%{\bf L3 Note 2445   16/7/1999}
%
%{\bf Revised Version 14/10/99}
%\hspace*{2cm} {UGVA-DPNC 1999/7-183 July 1999}
%\newline
%\hspace*{10cm} {hep-ex/9907018}
%\hspace*{0cm} {\bf{L3 Note 2209~(19/01/98)~~~~~~~~~Corrected Version 7/10/98}}
%\newline
\begin{center}
\vspace*{2cm}
{\large \bf
The Experimental Status of the Standard 
  Electroweak Model at the End of the LEP-SLC Era}

\vspace*{1.5cm}
\end{center}
\begin{center}
{\bf J.H.Field}
\end{center}
\begin{center}
{ 
D\'{e}partement de Physique Nucl\'{e}aire et Corpusculaire
 Universit\'{e} de Gen\`{e}ve .
 24, quai Ernest-Ansermet CH-1211 Gen\`{e}ve 4. CH-1211}
 \newline
  E-mail: john.field@cern.ch
\end{center}
\vspace*{2cm}
\begin{abstract}
 A method is proposed to calculate the confidence level for agreement of data with the Standard
 Model (SM) by combining information from direct and indirect Higgs Boson searches. Good agreement
 with the SM is found for $m_H \simeq 120$ GeV using the observables most sensitive
 to $m_H$: $A_l$ and $m_W$. In particular, quantum corrections, as predicted by the SM,
 are observed with a statistical significance of forty-four standard deviations.
   However, apparent deviations from the SM of
 3.7$\sigma$ and 2.8$\sigma$ are found for the Z$\nu \overline{\nu}$ and right-handed 
 Zb$\overline{{\rm b}}$ couplings respectively. The maximum confidence level for agreement with the 
 SM of the entire data set considered is $\simeq 0.006$ for $m_H \simeq 180$ GeV.
  The reason why confidence levels about an order of magnitude higher than this have been claimed
  for global fits to similar data sets is explained.        
\end{abstract}
\vspace*{1cm}

\end{titlepage}
\SECTION{\bf{Introduction}}
 
       \par It is now almost three decades since the the first accelerator~\cite{LEPPHYS1,LEPACC}
        and physicss~\cite{LEPPHYS1,LEPPHYS2} studies, that eventually lead to the construction and operation
     of the LEP $e^+e^-$ collider at CERN were performed. Now, some roughly twenty
      man-millenia of work by physicists and engineers later, the almost final
      results of the LEP and concurrent SLC (Stanford Linear Collider) experimental
      programs are available~\cite{EWWG,EWPDG}. By general consensus, the most 
       scientifically important of these results concern high precision tests of the
       Standard Electroweak Model (SM)~\cite{Glashow,Weinberg,Salam}. During the same period,
      important contributions to this subject (discovery of, and measurement of the mass, $m_t$,
      of the top quark, measurement of the mass, $m_W$, of the W boson and the NuTeV neutrino-quark
       scattering results) were also made at the FERMILAB laboratory. 
      \par The LEP program consisted of two stages. In the first, `Z-pole', running a total
     of $\simeq 1.7 \times 10^{7}$ Z decays into pairs of all the fundamental fermions
     (the matter fields of the SM), except the top quark, were collected by the four LEP 
     detectors, ALEPH, DELPHI, L3 and OPAL. At SLC, the same processes were studied with lower statistics but 
     with precisely controlled electron beam polarisation, allowing experiments of
     very high sesitivity to be performed, so producing results of comparable
      statistical accuracy to those of LEP. The final outcome of all this 
     work may be summarised, in what concerns the SM, in only seven numbers, which,
     for definiteness can be taken to be the right- and left-handed couplings to the Z of the 
     charged lepton, c-quark and b-quark pairs and the (left-handed) coupling of the Z
     to neutrino pairs. The busy reader, who would like to go straight to the 
      final conclusions, is invited to look directly at Tables 19 and 20 below
     where the measured values of these coupling constants, together with the
    corresponding SM predictions are shown. The second, `high energy' phase of LEP
    operation at collision energies above the threshold for W pair production
    provided essentially one additional high precision SM parameter, $m_W$. This measurement,
    in combination with the FERMILAB measurement of comparable accuracy of the same 
    quantity, is also shown in
    Tables 19 and 20. Other important measurements performed during the second LEP phase
     were of the triple boson WW$\gamma$ and WWZ couplings, but these have less impact
    than the fermion couplings as a test of the core physics underlying the SM:
    a renormalisable quantum field theory incorporating local gauge invariance that is
    spontaneously broken by the Higgs mechanism\footnote{In fact the now experimentally
    well verified~\cite{EWWG} SM predictions for the WW$\gamma$ and WWZ couplings were aleady contained
   in Glashow's original electroweak paper~\cite{Glashow} where they follow at tree level from 
    global SU(2)$_L$ invariance and quantum mechanical mixing of the W$^0$ and B$^0$
    fields. Their values then shed little light on the correctness (or otherwise)
    of either the renormalisabilty of the theory or the Higgs mechanism. A genuine test of 
    local gauge symmetry would be provided by measurements of the strength of
    quadrilinear boson couplings. So far this has not been done.}. They are therefore 
    not discussed further in this paper. In the conventionally used on-shell
    renormalisation scheme~\cite{OSRS} where $m_W$ is traded as an input parameter
    for the much more precisely known Fermi constant, $G_{\mu}$, derived from the
   measured muon lifetime, the predictions of the SM depend (apart from other and
    better known parameters) on three relatively poorly known ones: $m_t$, the 
    electromagnetic couplant at the Z mass scale, $\alpha(m_Z)$, and the mass of the Higgs 
    boson, $m_H$. Because of quantum loop corrections, the right- and left-handed
    couplings, $\glr$ and $\gll$, of the Z to charged lepton pairs, as well as $m_W$, 
    are strongly sensitive to the values of $m_t$ and $m_H$. Actually, it is only the
    ratio $\glr/\gll$ and $m_W$ which are strongly sensitive to $m_H$, so that the measurements
    of these two quantites provide the most stringent limits, from quantum
    corrections, on the value of $m_H$.
    \par In the second phase of LEP, an unsuccessful search was
     performed for directly produced Higgs bosons~\cite{HIGGSMD}, resulting in the 95$\%$
    confidence level (CL) lower limit $ m_H > 114.4$ GeV. The principle aim of the present
    paper is to combine, in a transparent way, this direct limit with the indirect information
     derived from $\glr/\gll$ (or equivalently $A_l$, see below) and $m_W$, to derive combined
     curves of $\overline{{\rm CL}}$, the confidence level that the $m_H$-sensitive
     data agrees with the SM, as a function of $m_H$. The reader might then reasonably hope that the
    result of the paper would be a single curve of $\overline{{\rm CL}}$ versus $m_H$.
    In fact she (or he)  will find eight figures where  $\overline{{\rm CL}}$ is plotted versus
     $m_H$ containing in total 18 different curves. The reason for this complication is that the
    data, though perfectly consistent exprimentally\footnote{That is, good agreement is found 
    between different measurements of the same experimental observables.
    For details see Reference~\cite{EWWG}},
     is not consistent with the SM (see Tables 19 and 20) and depending on the assumptions
    made (SM correct, model-independent analysis, certain data included or excluded) 
     different results  are found for the $\overline{{\rm CL}}$ curves. I have  included
     a number of different possibilities to demonstrate the inconsistency
    of the data with SM predictions. The reader may then choose the curve for which the
    assumptions match best her (or his) own favourite ones.
     \par The structure of the paper is as follows: In the next Section some general remarks 
    concerning the different functions of the science of Statistics in data analysis are made. In
    particular it is pointed out that the choices made until now for the $\chi^2$ estimator
   in global electroweak analyses give an over-optimistic estimate of the level of agreement
    of the data with the SM. In Section 3, the heavy quark asymmetry measurements that are
    {\it prima facie} inconsistent with charged lepton asymmetries, when both are interpreted 
    within the SM, are discussed. In particular, the internal consistency of the data
    and systematic error estimates are examined in some detail. The sensitivities of
    different electroweak observables to $m_t$ and $m_H$ are discussed in Section 4.
     This is the only place in the paper where fit results are shown and discussed.
    It is demonstrated
    that the sensitivity to $m_H$ comes essentially from only the observables $A_l$ 
     and $m_W$. Section 5 describes the algorithm used for combining the direct 
    and indirect measurements of $m_H$. Results for $\overline{{\rm CL}}$ derived from
     $A_l$ and $m_W$, assuming the correctness of the SM, but selecting different data,
    are shown. Also shown in this Section is the sensitivity of $\overline{{\rm CL}}$
    to the values of the parameters $m_t$ and $\alpha(m_Z)$. In Section 6 the alternative
    interpretations of the result of the NuTeV experiment are explained and it is
    pointed out that the interpretation, as required in a model-independent analysis, as a measurement
    of the Z$\nu\overline{\nu}$ coupling,
    instead of $m_W$, is strongly favoured
    by arguments of statistical consistency. In Section 7 a complete set of model
    independent observables is extracted and compared with SM predictions. Constraints are 
    set on the coupling of non-b down-type quarks using the precisely measured
    observable $\Gamma_{had}$. Quantum corrections are extracted for
    different fermion flavours and compared with SM predictions. Finally, in Section 7,
    curves of $\overline{{\rm CL}}$ versus $m_H$ derived from a $\chi^2$ estimator using
    all or selected subsets of the considered observables (including now $m_H$-insensitive
    ones) are shown. In Section 8 values of  $\overline{{\rm CL}}$ obtained as 
    described in Section 7, are compared with the confidence levels of previously
    published global fits to similar data. The confidence levels are seen to be
    very consistent when the purely statistical dilution of the hypothesis testing power
    of the $\chi^2$ estimators of the global fits, as discussed in Section 2, is
    taken into account. Section 8 also contains a critical discussion of $m_H$
  limits determined from $\Delta \chi^2$ plots. Section 9 contains a summary and conclusions,
    including the author's personal choice of three most pertinent
     $\overline{{\rm CL}}$ versus $m_H$ plots. The busy reader
    is encouraged to read this Section first to get a general view of the results and conclusions
    returning later (if still interested) to the earlier Sections for more information
    and supporting arguments.
    \par When the first version of the present paper was almost complete a new experimental 
   value of $m_t$, $178\pm 4.3$ GeV, was announced by the CDF and D0 collaborations~\cite{Newmt}.
    Since the change from the previous value of $174.3 \pm 5.1$ GeV has a dramatic effect on the
     $\overline{{\rm CL}}$ curves, especially for large values of $m_H$, all such curves shown, when the 
    contrary is not explicitly stated, use the new value of $m_t$. However no global fits have yet
    been published by the EWWG and EWPDG using the new value. Therefore, in Section 4 where
    comparisons with the results of EWWG global fits are made, the old value of $m_t$ is used.
    The conclusions of Section 8, where the consistency of the confidence levels found in the 
    present paper with those quoted for global fits is discussed, are unaffected by the change
   in the measured value of $m_t$.

\SECTION{\bf{Statistics: Data Consistency versus Hypothesis Testing}}
 In the context of the analysis of experimental data, Statistics has three quite distinct
 roles to play. These are:
 \begin{itemize}
 \item[(i)] To judge whether different measurements of the same physical quantity 
   are consistent with each other, and to derive an unbiased weighted average value
    of the quantity.
  \item[(ii)] To test the hypothesis that an ensemble of measurements of the same or 
  different physical quantities are consistent with some theory.
  \item[(iii)] In the case of positive answers to the questions implicit in
    (i) and (ii), to determine numerical values of unknown or partially known
    parameters of a theory from the data.
 \end{itemize}
 In previous and current analyses of precision electroweak data performed by
   the LEP and SLD electroweak working groups (EWWG)~\cite{EWWG} and the standard 
  model sub-group of the Particle Data Group (EWPDG)~\cite{EWPDG}, only the functions
  (i) and (iii) above are systematically performed, with little, if any regard for
  (ii). In fact tests of data consistency (comparisons of different measurements of the same
   physical quantity) are performed by the EWWG using the $\chi^2$ estimator, with, in general,
   very satisfactory results~\cite{EWWG}. In the global fit to all data, since the
   point (ii) is not addressed, all relevant data is used for parameter estimation
   in the global fit. In the case that all of the data is in agreement with the SM
  this procedure gives the best, unbiased, estimate of parameter values. However, if
  certain sub-sets of data do not agree with the SM, biased results may be obtained
  using this procedure. In particular, as will be discussed below, the fitted value
  of $m_H$ obtained with the current data set is biased towards higher values
  by about 50 GeV by just such an effect. The EWPDG also do not investigate the level of agreement
  of different data sub-sets with the SM and related biases, being concerned only with
  the function (iii), parameter estimation on the assumption that all 
  data is correctly described by the SM~\cite{erlerpc}. In this case different measuremnts
   of the same physical quantity are included as independent data in the fit without
  any prior consistency checks such as those prrformed by the EWWG. In the case that 
  subsets of data do not agree with the SM, fitted prameters may then be biased in 
   just the same way as in the EWWG global fits. It seems to the present writer 
   that seeking the answer to the question posed in (ii) above should be the 
   principle aim of experimental investigations of the SM, but, as a point of fact,
   this is an avowed goal of neither the EWWG nor the EWPDG.
     \par So what is the answer to the question implicit in (ii) above provided by the
        current electroweak data set?  The nature of the problem is well illustrated by some
    fit results quoted in
   a paper devoted to a
   search for possible evidence of supersymmetry in precision electroweak data~\cite{ACGGR}.
   Fitting, as a preamble, the minimal electroweak standard model to only 
   the $\sin^2\Theta_{eff}^{lept}$ values derived from either leptonic or hadronic asymmetry measurements,
    a $\chi^2$ per degree of freedom ($\chi^2/{\rm d.o.f.}$) of 18.4/4 was obtained corresponding to a 
    CL of 0.001. A fit by the EWWG to the same data set, but using
   instead about 20 observables\footnote{Many of these quantities are actually
   `pseudo-observables', but for brevity the term `observable' will be used throughout this
    paper for extracted physical quantities sensitive to parameters of the electroweak theory.}
    reported a $\chi^2/{\rm d.o.f.}$ of 26.0/15, with a CL
   of 0.04. An analysis of essentially the same year 2000 data set by EWPDG, but fitting the
   SM to more than 40 observables found, for a global fit, a  $\chi^2/{\rm d.o.f.}$ of
   42/37 with a CL of 0.27~\cite{EWPDG1}. The fitted value of $m_H$,
    was very similar in these three different fits since, as discussed below, almost all
   the sensitivity to $m_H$ is found in only two observables, $\sin^2\Theta_{eff}^{lept}$ and
    $m_W$. Thus, for essentially the same fitted value of  $m_H$,
    CLs differing by a factor of up to 270, according to the fit procedure used, were obtained.
    Which (if any) of the different CLs most truly reflects the agreement between the data
    and the SM prediction? The principal aims of the present paper are, firstly, to provide an answer
   to this question, and, secondly, to combine CLs derived from direct and indirect
   experimental limits on $m_H$
   so has to obtain an meaningful overall CL that reflects both the internal
   consistency of different observables and the global level of
   agreement with the SM.
   \par The explanation of the poor CL obtained in the fit to only the leptonic and hadronic
     $\sin^2\Theta_{eff}^{lept}$ values is now well known. As first pointed out in analyses of the
    1996 data set~\cite{Renton,JHF1} The Z-boson b-quark couplings appear to be anomalous
    at about the three standard deviation level. These couplings are quite insensitive, in
    the SM, to $m_H$ and $m_t$, but, due to a correlation effect, when
    heavy quark forward/backward asymmetries are analysed, assuming the correctness of the SM,
    to extract a value of  $\sin^2\Theta_{eff}^{lept}$, the latter is found to correspond
    to a much larger value of $m_H$ than that derived from purely leptonic
     measurements~\cite{JHF2,JHF3}. This leads to barely compatible values of 
      $\sin^2\Theta_{eff}^{lept}$ from leptonic and hadronic (essentially b-quark) data
     and explains the poor CL of the fit to this data to obtain  $m_H$ and $m_t$ mentioned
     above.
    \par More recently, much more precise experimental measurements of $m_W$ have
      become available. These
     are found to favour a value of  $m_H$ even lower than that suggested by the leptonic data,
     thus resulting in a large discrepancy between the $m_H$ value obtained by combining the
     leptonic data and  $m_W$  and that derived from hadronic asymmetries. This problem has been
    has been recently stressed in the literature~\cite{Chanowitz} and is now generally
    appreciated~\cite{Gambino}.
     \par The reason for the factor $\simeq 300$ difference in the CLs of different fits
      is easily understood. 
      The point is that the fit to only the  $\sin^2\Theta_{eff}^{lept}$
     values was essentally performing the function (ii) above, i.e. hypothesis testing, whereas
     the EWWG and EWPDG fits were combining the functions (i) and (ii) with a large weighting
     factor in favour of (i). How this happens will now be explained. In addition to this 
     effect, the hypothesis testing ability of the fit $\chi^2$ is further blunted by the
     inclusion of observables in the fit, that have almost no sensitivity to $m_H$ and
     $m_t$, in both the EWWG and EWPDG analyses.  
     \par Consider a number, $N$, of independent measurements, $Q_i$,  of the same quantity,
      $Q$. The 
      theoretical expectation for $Q$ is $Q_{Thy}$ and the weighted everage value of the
      measurements is $\bar{Q}$. With the assumption of
      uncorrelated experimental errors, three different Pearson $\chi^2$ estimators may be defined,
       as follows:
      \begin{eqnarray}
      \chi^2_{data,WA} & = & \sum_{i=1}^{N}\frac{(Q_i-\bar{Q})^2}{\sigma_i^2} \\
     \chi^2_{data,Thy} & = & \sum_{i=1}^{N}\frac{(Q_i-Q_{Thy})^2}{\sigma_i^2} \\
    \chi^2_{WA,Thy} & = & \frac{(\bar{Q}-Q_{Thy})^2}{\bar{\sigma}^2}
      \end{eqnarray}
         In Eqn(2.3), $\bar{\sigma}$ is the weighted mean error on the quantity  $\bar{Q}$.
       Assuming uncorrelated, Gaussian distributed, errors it is given by the relation:
      \begin{equation}
       \frac{1}{\bar{\sigma}^2}  \equiv \sum_{i=1}^{N} \frac{1}{\sigma_i^2}
      \end{equation}
      where $\sigma_i$ is the estimated RMS uncertainty on  $Q_i$.
       Noting the identity:
        \begin{equation}
     Q_i-Q_{Thy} \equiv (Q_i-\bar{Q})+ (\bar{Q}-Q_{Thy})
      \end{equation}
     Eqn(2.2) may be written as:
     \begin{eqnarray}
        \chi^2_{data,Thy} & = & \sum_{i=1}^{N}\left[\frac{(Q_i-\bar{Q})^2}{\sigma_i^2}
      + \frac{(\bar{Q}-Q_{Thy})^2}{\sigma_i^2}
      + \frac{2(Q_i-\bar{Q})(\bar{Q}-Q_{Thy})}{\sigma_i^2}\right] \nonumber \\
        & = &   \chi^2_{data,WA} +(\bar{Q}-Q_{Thy} )^2\sum_{i=1}^{N} \frac{1}{\sigma_i^2}
        +  2(\bar{Q}-Q_{Thy}) \left[ \sum_{i=1}^{N}\frac{Q_i}{\sigma_i^2}
         - \bar{Q} \sum_{i=1}^{N} \frac{1}{\sigma_i^2}\right] \nonumber \\
         & = &   \chi^2_{data,WA} +  \chi^2_{WA,Thy} +  2(\bar{Q}-Q_{Thy}) \left(
         \sum_{i=1}^{N} \frac{1}{\sigma_i^2}\right)
         \left[\frac{ \sum_{i=1}^{N}\frac{Q_i}{\sigma_i^2}}{\sum_{i=1}^{N} \frac{1}{\sigma_i^2}}
          - \bar{Q}\right] \nonumber \\
          & = &   \chi^2_{data,WA}  +  \chi^2_{WA,Thy}
     \end{eqnarray}

   where, in the third line of Eqn(2.6) the definition of $ \bar{\sigma}$, (2.4) and Eqn(2.3) 
  have been
   used, and in the fourth line the definition of $\bar{Q}$:
     \begin{equation}
      \bar{Q} \equiv 
      \frac{ \sum_{i=1}^{N}\frac{Q_i}{\sigma_i^2}}{\sum_{i=1}^{N}\frac{1}{\sigma_i^2}}     
   \end{equation}
   So, in the simple case of uncorrelated Gaussian errors, the $\chi^2$ for consistency
   of the data with the theory is equal to the simple sum of the $\chi^2$ for consistency
   of the data with its weighted average plus the $\chi^2$
    for consistency
    of the theory with the weighted average. Clearly $\chi^2_{data,WA}$ is a measure only
   of the internal consistency of the data and so the corresponding CL provides an 
    answer only to the question raised in point (i) above.  $\chi^2_{WA,Thy}$ gives,
    providing the CL for $\chi^2_{data,WA}$ is acceptable, an estimate  of probability
    that the data is correctly described by 
    the theory, and so provides the hypothesis test mentioned in (ii) above, as well as estimating
     the values of unknown parameters of the theory (for example, $m_H$ if $Thy =$ SM)
     in accordance with point (iii) above. However, the value
    of $\chi^2_{data,Thy}$, the statistical estimator universally used by both the EWWG and
    the EWPDG, reflects {\it both} the internal consistency of the data {\it and} the level of
    agreement of the data with the theory. If the number of data is very large, the relative 
    contribution of  $\chi^2_{WA,Thy}$ to $\chi^2_{data,Thy}$ becomes very small, since the
    former $\chi^2$ has only one degree of freedom. Under these circumstances, the CL of
     $\chi^2_{data,Thy}$ is not a meaningful indicator of 
      the level of agreement of the data and the theory.
    \par To take a simple example, suppose that there are 40 data and
     that $\chi^2_{data,WA} = 30$ and $\chi^2_{WA,Thy} = 16$, so that, on the assumption
     of uncorrelated Gaussian errors, Eqn(2.6) gives  $\chi^2_{data,Thy} = 46$.
     The corresponding confidence levels are: $\chi^2_{data,WA}/{\rm d.o.f.} = 30/39$,
     CL$ = 0.849$ ;  $\chi^2_{WA,Thy}/{\rm d.o.f.} = 16/1$,  CL$ = 6.3 \times 10^{-5}$;
     $\chi^2_{data,Thy}/{\rm d.o.f.} = 46/40$,  CL$ = 0.28$. Thus the effect of the four
      standard deviation discrepancy observed in  $\chi^2_{WA,Thy}$ is
      diluted to give an innocuous CL of
      0.28 for the statistical estimator $\chi^2_{data,Thy}$.
      \par To take properly into account both the internal consistency of different measurements
       of the same quantity, and the level of agreement of the data with theory, a useful
      statistical procedure is to combine the confidence levels of the appropriate
       $\chi^2$ functions. Since  $\chi^2_{data,WA}$ and  $\chi^2_{WA,Thy}$  are independent
       statistical estimators, the corresponding CLs may be combined
       by use of the formula~\cite{EDJRS1}:
       \begin{equation}
        {\rm CL}(\alpha_1,\alpha_2) = \alpha_1 \alpha_2 [1- \ln(\alpha_1 \alpha_2)] 
       \end{equation}
        where $\alpha_1$ and $\alpha_2$ are the two independent CLs to be combined.
        It follows that in the simple example considered above the combined CL has
       the value $5.8 \times 10^{-4}$ so that the data/theory discrepancy is still well
       in evidence. Note that the combined CL is a factor 493 smaller than the CL
       of  $\chi^2_{data,Thy}$ in this case! In the following the combined CL given by
       Eqn(2.8) will be used to calculate the overall confidence level that the relevant
      data are   consistent and that the data are in agreement with the SM, for different
      values of $m_H$.
      \par  In the above example each datum has the same sensitivity to the parameters 
       of the theory. However, among  the $\simeq 20$ observables included in the global
       electroweak fits performed by the EWWG and the $\simeq 40$ in the similar EWPDG
      fits, the majority are only weakly sensitive to the values of $m_H$ and $m_t$.
       This effect dilutes even further the hypothesis testing power of the the statistical
       estimator $\chi^2_{data,Thy}$ beyond that due to the dominant contribution
       of  $\chi^2_{data,WA}$ discussed above. In the statistical analysis presented below,
       the separate contributions of $\chi^2_{data,WA}$ and $\chi^2_{WA,Thy}$ to 
        $\chi^2_{data,Thy}$ will be extracted to provide separate answers to the questions
       posed in points  (i) and (ii) above. The overall CL will then be calculated 
       according to Eqn(2.8) above. In the case of a small number of sensitive observables
     \footnote{Indeed, for $m_H$, there are only two such observables $A_l$ and $m_W$ as
      discussed in Section 4 below.}it will be found that, unlike in the example discussed above,
     good agreement is found
     between the CL of the total $\chi^2$:  $\chi^2_{data,WA} +\chi^2_{WA,Thy}$ and the combined
     CL calculated using Eqn(2.8). The former CL is then used as a statistical estimator
      for the indirect Higgs mass analysis. 
 
        \par Previous authors~\cite{DD,Erler} have calculated normalised probability density
         functions (PDFs) giving the relative probability of different values of $m_H$, by
         combining direct and indirect limits. Instead, in the present
          paper, the combined CL is found by combining the CLs of the direct and indirect
          measurements in 
         region of overlap using Eqn(2.8). This combined CL gives an
       absolute rather than a relative probability that the SM is consistent with the data
        for any value of $m_H$.  In this way the hypothesis testing aspect of
         the comparison of the data with the SM is addressed. This is not done by the
         normalised PDFs derived in References~\cite{DD,Erler}.
 %$       
 \SECTION{\bf{Heavy Quark Asymmetry Measurements}}
    A discussion of the consistency of the b-quark asymmetry measurements in the
    data up to 1999 may be found in Reference~\cite{JHFDS}. 
    The current LEP and SLD heavy flavour asymmetry measurements are collected in Table 1
    (b-quarks)~\cite{ABCEWWG}
    and Table 2 (c-quarks)~\cite{ABCEWWG}. In Table 1 are reported eight independent
    measurements of the forward/backward asymmetry $\Afbb$ as well as the direct SLD
    measurement of $A_b$ from the 
    forward/backward,left/right asymmetry. Table 2 contains seven LEP measurements 
    of $\Afbc$ and the direct $A_c$ measurement from SLD. For each LEP asymmetry
     measurement the corresponding value of $A_b$ or $A_c$ is estimated using the 
     relation:
    \begin{equation}
 A_{Q} =  \frac{4 A_{FB}^{0,Q }}{3 A_{l}}~~~~(Q=b,c)
   \end{equation}
   where $A_l$ is the LEP+SLD average value of the charged lepton asymmetry
   parameter extracted by assuming charged lepton universality\footnote{The notation 
    follows that of Reference~\cite{JHF2}}:
  \begin{equation}
    A_l =  \frac{2 \vbl \abl}{\vbl^2 + \abl^2} = \frac{2 \rbl}{1+\rbl^2}
  \end{equation}
    where
   \begin{equation}
   \rbl \equiv \frac{\vbl}{\abl} = 1-4\sin^2\Theta_{eff}^{lept}
  \end{equation}
   The value used is~\cite{EWWG}\footnote{Errors are quoted on the least 
    significant digits. e.g. 4.123(32) means 4.123 $\pm$ 0.032. When two errors 
    are quoted, the first is statistical and the second systematic}:
   \begin{equation}
     A_l = 0.1501(16)
    \end{equation}
     The values of $A_b$ and $A_c$ derived in this manner are presented in the last columns of
      Tables 1 and 2.
   The SM predictions for the values of $A_b$ and $A_c$ are 0.935 and 0.668
   respectively, with a negligible dependence on $m_H$ and $m_t$ at the scale 
   of the present experimental errors. Also shown in Tables 1 and 2 are the LEP
   average values of $\Afbb$, $A_b$, $\Afbc$ and $A_c$ as well as the LEP+SLD 
   combined values of $A_b$ and $A_c$. The uncertainties on the LEP average
    values of  $A_b$ and $A_c$ come mainly from those on  $\Afbb$ and  $\Afbc$
    (1.6 $\%$ and 5.0 $\%$) rather than that on $A_l$ (1.1 $\%$). The statistical
    and systematic errors on both the LEP average and the SLD measurements of 
    $A_b$ and $A_c$ are of comparable magnitude.

 \begin{table}
\begin{center}
\begin{tabular}{|c|c|c|} \hline
 Experiment & $\Afbb$ & $A_b$ \\  \hline
 ALEPH leptons & 0.1009(38)(17)  & 0.896(39) \\
 DELPHI leptons & 0.1031(51)(24) & 0.916(51) \\
 L3 leptons & 0.1007(60)(35) & 0.895(70) \\
 OPAL leptons & 0.0983(38)(18) & 0.873(38) \\
 ALEPH inclusive & 0.1015(25)(12) & 0.902(27) \\
 DELPHI inclusive & 0.0984(30)(15) & 0.874(32) \\
 L3 jet-charge & 0.0954(101)(56) & 0.847(103) \\
 OPAL inclusive & 0.1000(34)(18) & 0.888(35) \\
 SLD   & $-$ & 0.925(14)(14) \\  \hline
 LEP average & 0.0997(14)(7) & 0.885(14)(10) \\
 LEP+SLD average & $-$ & 0.902(13) \\  \hline
\end{tabular}
\caption[]{{\sl The LEP and SLD measurements of
 b-quark asymmetry parameters. When two uncertainties are 
   quoted, the first is statistical, the second systematic.}}
\end{center}
\end{table}

 \begin{table}
\begin{center}
\begin{tabular}{|c|c|c|} \hline
 Experiment & $\Afbc$ & $A_c$ \\  \hline
 ALEPH leptons & 0.0733(53)(36)  & 0.651(57) \\
 DELPHI leptons & 0.0724(84)(62) & 0.643(93) \\
 L3 leptons & 0.0832(301)(197) & 0.739(320) \\
 OPAL leptons & 0.0642(51)(37) & 0.570(56) \\
 ALEPH D$^*$ & 0.0696(85)(33) & 0.618(81) \\
 DELPHI D$^*$ & 0.0693(87)(27) & 0.615(81) \\
 OPAL  D$^*$ & 0.0759(109)(57) & 0.674(109) \\
 SLD   & $-$ & 0.670(20)(16) \\  \hline
 LEP average & 0.0706(31)(17) & 0.627(26)(19) \\
 LEP+SLD average & $-$ & 0.653(20) \\  \hline
\end{tabular}
\caption[]{{ \sl The LEP and SLD measurements of
 c-quark asymmetry parameters.When two uncertainties are 
   quoted, the first is statistical, the second systematic.}}
\end{center}
\end{table}      

 \begin{table}
\begin{center}
\begin{tabular}{|c|c|c|c|c|} \hline
   & $\chi^2_{data,WA}/{\rm d.o.f.}$, CL &  $\chi^2_{WA,Thy}/{\rm d.o.f.}$, CL 
   &  $\chi^2_{data,Thy}/{\rm d.o.f.}$, CL & Comb. CL  \\   \hline
  $A_b$ & 3.2/8,~~~0.92 & 6.4/1,~~~ 0.011 & 12.0/9,~~~0.21 & 0.057 \\
  $A_c$ & 3.2/7,~~~0.90 & 0.56/1,~~~ 0.45 & 4.1/8,~~~0.85 & 0.77  \\ \hline
\end{tabular}
\caption[]{{ \sl Different $\chi^2$ estimators and CLs derived from
  the LEP and SLD measurements of $A_b$ and $A_c$.}}
\end{center}
\end{table}

 \par Values of the different $\chi^2$ estimators:  $\chi^2_{data,WA}$, $\chi^2_{WA,Thy}$,
 and  $\chi^2_{data,Thy}$ introduced above for the quantities $Q = A_b, A_c$ are 
  presented in Table 3. The $\chi^2_{data,WA}$ CLs of 0.92 and 0.90 for
  $A_b$ and $A_c$ indicate good internal consistency of the data, but also, possibly, 
  an over-estimate of systematic errors. The $\chi^2_{WA,Thy}$ CLs of $1.11 \times 10^{-2}$
   and 0.45 for $A_b$ and $A_c$ indicate in the former case a 2.5$\sigma$ discrepancy, and,
   in the latter, good agreement with the SM prediction. As in the example discussed above,
   the $A_b$ discrepancy is not evident in the CL of $\chi^2_{data,Thy}$, which takes the value
   0.21. The combined CLs, according to Eqn(2.8), that the $A_b$ and $A_c$ data are both 
   consistent and in agreement with the SM are 0.057 and 0.77 repectively. This would seem to 
   indicate that it is not unreasonable that the deviation of $A_b$ from the SM prediction
   could be due to statistical fluctuation perhaps in combination with some unknown
    systematic effect.
    However, this conclusion requires confidence
   in the estimation of the systematic errors. As will be discussed below, there is some 
   evidence, from the data itself, that the uncorrelated systematic errors may be somewhat
   overestimated, thus reducing from its true value the significance of the observed $A_b$ deviation. 
   \par It is interesting to note that a goodness-of-fit estimator, independant of the
    $\chi^2$ test, is provided by the so-called `Run Test'~\cite{EDJRS2}. For the case
    of the $A_b$ data in Table 1, since all 9 independent measurements constitute a single
    `run' (they are all less than the SM prediction) the corresponding CL is easily
    calculated. Since a single run can occur in only two ways (all data higher than or
    all data lower than the theoretical prediction) the CL is $2/2^9 = 3.9 \times 10^{-3}$.
    Unlike  for the $\chi^2$ test, the CL of the Run Test is insensitive to over-
    or under-estimation of uncorrelated systematic errors. Since the CLs of the Run
    Test, of $\chi^2_{data,WA}$ and of $\chi^2_{WA,Thy}$ are all independent, they may be
    combined into a single CL using the formula that generalises Eqn(2.8) to the case of
    three independent CLs: $\alpha_1$, $\alpha_2$ and  $\alpha_3$~\cite{EDJRS1}:
    \begin{equation}
     {\rm CL}(\alpha_1,\alpha_2,\alpha_3 ) = \alpha_1 \alpha_2 \alpha_3
       [1- \ln(\alpha_1 \alpha_2 \alpha_3 ) + \frac{[\ln(\alpha_1 \alpha_2 \alpha_3 )]^2}{2}] 
     \end{equation}
      The combined CL for the $A_b$ data given by Eqn(3.5) is $2.5 \times 10^{-3}$.
      The single run of the $A_b$ data may be associated with a genuine deviation of the
      data from the SM prediction or a large correlated systematic effect of unknown origin.
      It is argued below that the latter explanation is unlikely. The third possible 
     explanation, a statistical fluctuation, is also unlikely, given the small value
     of the combined confidence level.
     \par As there are $\simeq 10$ independent measurements of both $A_b$ and $A_c$, it is
      possible to compare errors estimated directly from the data, with the calculated 
       statistical and estimated uncorrelated systematic errors on the weighted average values of 
       $A_b$ and $A_c$ shown in Tables 1 and 2 .
       The estimators for the error on the weighted average, $\bar{\sigma}$
      and its RMS uncertainty $\sigma_{\bar{\sigma}}$ are given by the formulae~\cite{RJB}:
       \begin{eqnarray}
       \bar{\sigma} & =  & \sqrt{\frac{\sum_{i}(Q_i-\bar{Q})}{N(N-1)}} \\
       \sigma_{\bar{\sigma}} & = & \frac{\bar{\sigma}}{\sqrt{2N(N-1)}}
       \end{eqnarray}
     These formulae yield values of $\bar{\sigma}$ of 0.0089(22) for  $A_b$,  and
     0.018(5) for $A_c$, to be compared with the estimated total errors on the WA values
     of 0.013 and 0.020 respectively in these quantities. The agreement is good for
     $A_c$  , but for $A_b$ there is an indication, at the 1.9$\sigma$ level, that the
     uncorrelated systematic errors may be overestimated. This is confirmed by calculation of the
     WA statistical error on the LEP+SLD weighted average value of $A_b$, which is
     just 0.0088, in perfect agreement with the value of $\bar{\sigma}$ estimated
     directly from the data. Using this error to calculate $\chi^2_{WA,Thy}$
     gives  $\chi^2/{\rm d.o.f.} = 14.0/1$, CL $= 1.8 \times 10^{-4}$ a 3.7$\sigma$
     effect. Thus, although (see Table 1) the estimated systematic error on the
      LEP+SLD average value of $A_b$ is by no means the dominant one, the significance
     of the apparent deviation of $A_b$ from the SM prediction is very sensitive to it.
      There is some evidence, from the data itself, that the uncorrelated part of
      this systematic error may be over-estimated.
      \par The dominant source of correlated systematic error on both $\Afbb$ and $\Afbc$ 
      arises from the QCD corrections~\cite{Aabetal} 
      of (2.96 $\pm$  0.40)$\%$ for $\Afbb$ and (3.57 $\pm$  0.76)$\%$ for $\Afbc$. This source
      contributes 35$\%$ of the total systematic error on the LEP average value of $A_b$,
      the remaining part being essentially uncorrelated between the different measurements.
      Thus the observed fractional discrepancy, 0.053, between the LEP average value of $A_b$  
      and the SM prediction, is about 1.8 times larger than the
      QCD correction and about 13 times larger than the estimated uncertainty on this
      correction. The latter would have to have been underestimated  by more than an order
      of magnitude in order to explain the observed discrepancy with the SM prediction.
      This seems unlikely. Note, however, that the estimate of systematic error from the data 
      itself using Eqns(3.6) and (3.7) gives no information on such a correlated uncertainty. 
      \par In conclusion, the different measurements of $A_b$ are in very good agreement with 
      each other, but their average value shows a -2.5$\sigma$ deviation from the SM
      prediction. There is some evidence, from the data itself, that uncorrelated systematic
      errors may be over-estimated, thus possibly reducing the observed deviation from
       its true value. The correlated
      systematic error must have been underestimated by a large factor if the origin of
       the  $A_b$ deviation is unknown systematics rather than a breakdown of the SM.
       The measurements of $A_c$, on the other hand are found to be both consistent
       and in good agreement
       (within their much larger errors) with the SM prediction. Also however, as will be
        seen below, all of the hadronic asymmetries show similar fractional deviations from the SM
      parameters favoured by the purely leptonic data. The possiblity of deviations from
      the SM in the c-quark and light quark sectors as large as that observed in the b-quark
      sector is therefore not excluded by the asymmetry data.  

  \SECTION{\bf{Sensitivities of Electroweak Observables to $m_t$ and $m_H$}}
    To justify the restricted choice of observables used below to calculate the $\chi^2$
     estimators for
    the data/SM comparison this Section presents some results of fits to obtain $m_H$,
      or $m_H$ and $m_t$.
     The overall approach used is the `model-independent' one of References~\cite{JHF1,JHF2,JHF3}
    All charged lepton and heavy quark measurements from LEP and SLD are combined to 
    obtain the independent observables: $A_l$, $\sbl$, A$_b$, $\sbb$ , $A_c$ and $\sbc$.
    The $A_f$ ($f=l,b,c$) parameters are defined as in Eqn(3.2) above, with small additional correction
    terms in the case of $A_b$. The quantity, $\sbf$, is defined as $\sbf = \vbf^2+\abf^2$
    and so is proportional to the partial width for $\rm{Z} \rightarrow f \bar{f}$ decays.
    Again, due to the large mass of the b-quark, small corrections are included in this
     case~\cite{JHF2}. The LEP+SLD average values of these observables used in the fits presented 
      below are shown in Table 4. To take properly into account error correlations,
      the directly measured values; $A_b = 0.925(20)$ and $A_c = 0.670(26)$
      from SLD are assigned speparate terms in the $\chi^2$ estimator. Correlations
      between $A_l$, $A_b$ and $A_c$ resulting from Eqn(3.1) are
    included in the  $\chi^2$ error matrix. Also shown in Table 4 are the SM predictions
     for $m_t = 174$ GeV and $m_H = 100$ GeV as well as normalised deviations. This Table
     has the same format and SM predictions as Table 3 of~\cite{JHF2}, with which it may be directly 
      compared.

\begin{table}
\begin{center}
\begin{tabular}{|c|c|c|c|c|c|c|} \cline{2-7}
\multicolumn{1}{c}{ } & \multicolumn{2}{|c}{leptons } & \multicolumn{2}{|c}{ c quarks }
 & \multicolumn{2}{|c|}{ b quarks }  \\ \cline{2-7}
\multicolumn{1}{c|}{ } & $A_l$  & $\sbl$  &  $A_c$ & $\sbc$  & $A_b$ & $\sbb$  \\ \hline
 Meas. & 0.1501(16) & 0.25268(26)  & 0.653(20) & 0.2897(50) &
0.902(13)  & 0.3663(13) \\ \hline
 SM &  0.1467 & 0.25272  & 0.6677 & 0.2882 & 0.9347 & 0.3647 \\ \hline
Dev($\sigma$) & 2.1 & -0.15  & -0.74  &  0.3 & -2.5 & 1.2 \\
 \hline      
\end{tabular}
\caption[]{{ \sl Measured values of $A_f$ and $\sbf$ compared to SM 
 predictions for $m_t =$ 174 GeV, $m_H =$ 100 GeV.  Dev($\sigma$) = (Meas.-SM)/Error. }} 
\end{center}
\end{table}

    \par The sensitivity of different observables to $m_t$ and $m_H$ is presented 
     in Table 5
     \footnote{Note that the  $(\Delta X/\sigma_X)_{m_t}$ entries of $\sbl$ and $\sbb$
           of the similar table in Reference~\cite{JHF2} are incorrect}.
      To take into account both the intrinsic sensitivity and the effect of
     experimental uncertainty, the quantities  $(\Delta X/\sigma_X)_{m_t}$  and
      $(\Delta X/\sigma_X)_{m_H}$ are shown for each observable, $X$, with experimental
      uncertainty $\sigma_X$. The quantity $\Delta X$ in  $(\Delta X/\sigma_X)_{m_t}$ is the
      change in the value of $X$ for a variation of $m_t$ from 164 GeV to 184 GeV, with
      $m_H =$ 120 GeV and
       $\Delta X$ in  $(\Delta X/\sigma_X)_{m_H}$ is the
      change in the value of $X$ for a variation of $m_H$ from 20 GeV to 200 GeV, with
      $m_t = 174.3$ GeV. Most of the sensitivity to $m_t$ resides in the observables
      $A_l$, $\sbl$ and $m_W$, to $m_H$ in  $A_l$ and $m_W$ only. The sensitivity of $\sbl$ (or $\Gamma_l$)
      to $m_t$, but not to $m_H$, has also been noted in a recent paper~\cite{TKRK}.

     \par Also included in Table 5 are the observables:  $\sbn = \vbn^2+\abn^2$ 
      and $\sbnbp$ (to be discussed below) which is similarly defined to $\sbc$ and $\sbb$
      but for non-b quarks. Both these observables have a moderate sensitivity to
      both $m_t$ and $m_H$. 

\begin{table}
\begin{center}
\begin{tabular}{|c|c|c|c|c|}  \hline
 $X$ & $X_{expt}$ & $\sigma_X$ &  $(\Delta X/\sigma_X)_{m_t}$  &  $(\Delta X/\sigma_X)_{m_H}$ \\ \hline
 $A_l$ & 0.1501 & 0.0016 & 3.1 & -5.1 \\
 $\sbl$ & 0.25268 & 0.00026 & 2.4 & -0.11 \\
 $A_c$ & 0.653 & 0.020 & 0.10 & -0.18 \\
 $\sbc$ & 0.2897 & 0.0050 & 0.18 & -0.077 \\
 $A_b$ & 0.902 & 0.013 & 0.012 & 0.051 \\
 $\sbb$ & 0.3663 & 0.0013 & -0.12 & -0.22 \\
 $\sbn$ & 0.5014 & 0.0015 & 0.77 & -0.55 \\
 $\sbnbp$ & 1.3211 & 0.0043 & 0.93 & 0.32 \\
 $m_W$ & 80.426 & 0.034 & 3.5 &  -3.1 \\
 \hline      
\end{tabular}
\caption[]{{ \sl Sensitivities of different measured quantites to $m_t$ and $m_H$ (see text).}} 
\end{center}
\end{table}
 
      \par As pointed out above, if $\Afbb$ is used as observable to estimate, via
       quantum corrections, $m_H$, a very different value is obtained from that favoured
      by $A_l$ or $m_W$. This is due to the linear dependence of  $\Afbb$ on
      $A_l$ (see Eqn(3.1) above) and the 2.5$\sigma$ deviation of $A_b$ from the 
       SM prediction discussed in the previous Section\footnote{This effect is
        particularly transparent in Fig.1 of Reference~\cite{JHF3} or in
        Fig 15.1 of Reference~\cite{EWWG}}. Assuming the correctness of the SM,
        `hadronic' values of $A_l$ may be extracted from the measurements of
       $\Afbb$ and $\Afbc$ by substituting the SM predictions for $A_b$ ,$A_c$
       (which are essentially independent of $m_H$ and $m_t$) into Eqn(3.1). 
        Another, independent `hadronic' value of $A_l$ may be derived from 
        the value of $\sin^2\Theta_{eff}^{lept}$ obtained from the SM analysis
       the quark anti-quark charge asymmetry, $Q_{FB}^{had}$~\cite{EWWG}.
        These different `hadronic' determinations of $A_l$, obtained by assuming
        the correctness of the SM, are presented,
        together with the `leptonic' value from Table 4, in Table 6. Note that the
         `leptonic' value of $A_l$, although derived  assuming 
        charged lepton universality does {\it not} assume the correctness
       of the SM, only that the process ${\rm Z} \rightarrow l \bar{l}$ 
      is described by some effective vector and axial vector couplings
      so that Eqn(3.1) is obeyed.
 
 \begin{table}
\begin{center}
\begin{tabular}{|c|c|c|c|c|c|c|} \hline
   Source & leptons & b-quarks & c-quarks & $Q_{FB}^{had}$ & hadronic mean  & overall mean \\
  \hline
    $A_l$ & 0.1501(16) & 0.1422(23) & 0.1423(72) & 0.1401(95) & 0.1421(21) & 0.1472(13) \\ \hline
\end{tabular}
\caption[]{{ \sl Different experimental determinations, derived assuming the correctness of the SM,
   of the leptonic asymmetry parameter $A_l$.}}
\end{center}
\end{table} 

\begin{table}
\begin{center}
\begin{tabular}{|l|c|c|} \hline
 \multicolumn{1}{|c|}{ Fitted quantities} &  $m_H$ [GeV] &  $\chi^2/{\rm  d.o.f}$,  CL \\ \hline
   $A_l(lept)$, $\sbl$,$A_b$, $\sbb$, $A_c$, $\sbc$, $m_W$ & $97_{-24}^{+31}$ & 14.7/8, 0.065 \\
    $A_l(all)$,  $m_W$ & $97_{-24}^{+32}$ & 1.99/1,  0.16 \\
   $A_l(lept)$, $\sbl$,$A_b$, $\sbb$, $A_c$, $\sbc$, $m_W$  $m_W({\rm NuTeV})$ 
     & $112_{-27}^{+35}$ &  23.3/9, 0.0056 \\
    $A_l(all)$,  $m_W$,  $m_W({\rm NuTeV})$ & $113_{-28}^{+36}$ & 10.6./2,  0.0050 \\ 
   $A_l(lept)$,  $m_W$  & $53_{-18}^{+22}$ & 0.20/1,  0.66 \\
  $A_l(lept)$,  $m_W$,  $m_W({\rm NuTeV})$  & $66_{-20}^{+28}$ & 10.9/2,  0.0043 \\
   $A_l(had)$,  $m_W$  & $154_{-47}^{+65}$ & 8.0/1,  0.0047 \\
  $A_l(had)$,  $m_W$,  $m_W({\rm NuTeV})$  & $196_{-58}^{+79}$ & 14.6/2,  0.00068 \\ \hline
\end{tabular}
\caption[]{{ \sl Results of $m_H$ fits to different sets of observables.}}
\end{center}
\end{table}

\begin{table}
\begin{center}
\begin{tabular}{|c|c|c|c|} \cline{3-4} 
\multicolumn{2}{c}{ } &\multicolumn{1}{|c}{ all data except NuTeV}
    &\multicolumn{1}{|c|}{ all data} \\  \hline
  This & $m_H$ & $102_{-35}^{+53}$ & $107_{-37}^{+58}$ \\
  paper & $m_t$ &  $175.0_{-4.2}^{+4.4}$ & $173 .7_{-4.3}^{+4.5}$ \\
 &  $\chi^2/{\rm d.o.f}$,  CL & 14.7/8, 0.065 & 23.3/9, 0.0056 \\ \hline  
  EWWG & $m_H$ & $91_{-36}^{+55}$ & $96_{-38}^{+60}$ \\
 Ref.~\cite{EWWG} & $m_t$ &  $175.3_{-4.3}^{+4.4}$ & $174.3_{-4.4}^{+4.5}$ \\
        &  $\chi^2/{\rm d.o.f}$,  CL & 16.7/14, 0.27 & 25.4/15, 0.045 \\ \hline         
\end{tabular}
\caption[]{{ \sl Global electroweak fits for $m_H$ and $m_t$. }} 
\end{center}
\end{table}

        \par The `leptonic' value of $A_l$ in Table 6 is derived from $e$, $\mu$ and $\tau$
       forward/backward charge asymmetries and from $\tau$-polarisation measurements.
       The hadronic ones from quark forward/backward charge asymmetries. In fact the
       $A_l$ derived uniquely from  $\tau$-polarisation measurements:
       $A_l(\tau-poln) = 0.1465(33)$ lies almost exactly  mid-way between the SLD ALR and 
       LEP $\Afbl$ weighted average of $0.1513(19)$ and the value of $A_l(had)$
      quoted in Table 6. It is 1.3$\sigma$ below the former and 1.1$\sigma$ above the
     latter. In the 1996 data set~\cite{EWWG96} the difference between  $A_l(\tau-poln)$
    and the ALR, $\Afbl$ average was much larger, 2.5$\sigma$, so that the inclusion
    (or not) of the $\tau$-polarisation data had a large effect on the value of $A_l$ extracted
    using Eqn(3.1). This was discussed in some detail in Reference~\cite{JHF1}. In another
    paper discussing the same 1996 data set~\cite{TA} it was pointed out that, considering also the
    $\tau$-polarisation data as `hadronic' (because of the predominantly hadronic final states),
   the value of  $\sin^2\Theta_{eff}^{lept}$ derived from the leptonic ALR and $\Afbl$ 
   measurements was found to differ by more than 3$\sigma$ from that given by the `hadronic'
    ones, i.e. $\tau$-polarisation and quark asymmetry measurements. The situation is much improved
   in the current (essentially final) LEP+SLC data. Excluding the  $\tau$-polarisation
   measurements gives a minor change in the WA $A_b$ value: $A_b(\tau-poln~out) = 0.898(13)$ to be compared
    with the value quoted in the last row of Table 1. The deviation of $A_b$  from the SM prediction
   is only  increased from 2.5$\sigma$ to 2.8$\sigma$, instead of the $\simeq 1\sigma$ increase
   found in the 1996 data set~\cite{JHF1}.

        \par Because of its strong dependence on $m_H$ and $m_t$ then, unlike
      in the case of $A_b$ and $A_c$, no definite SM prediction exists for 
       $A_l$. However it is of interest to compare the `leptonic' value
       of $A_l$, $A_l(lept)$ with the different `hadronic'
       values $A_l(had)$. The following $\chi^2$ values and confidence
        levels are obtained:    
        $\chi^2_{had,WA}/{\rm d.o.f.}$ = 0.047/2, CL= 0.977;
        $\chi^2_{all,WA}/{\rm d.o.f.}$ = 9.0/3, CL= 0.029: 
        $\chi^2_{had,lept}/{\rm d.o.f.}$ = 9.2/1, CL= 0.0024.
      Thus the three hadronic determinations are very consistent with each other,
      whereas the hadronic and leptonic determinations differ by 3 standard
      deviations. This poor overall consistency of the different values of $A_l$,
      extracted assuming the correctness of the SM for the quark couplings, 
       must be taken
      into account when assessing the overall level of agreement of the 
      data with the SM.\footnote{Indeed, the consistency of the three different `hadronic'
      estimates of  $A_l$ is much better than expected. Because of the large
     statistical uncertainties of the b- and c-quark data this is most likely due
     to a chance co-incidence rather than any over-estimation of systematic errors.}
         It is important to stress that this mismatch is not the result of any inconsistency
       evident in the experimental data themselves, but rather the result of interpreting 
     the data according to the SM prediction.
       \par The following strategy is now followed for fits to obtain
        limits on $m_H$: In a first step, fits similar to those previously
      presented in References~\cite{JHF2,JHF3} are performed to the the entire
     LEP+SLD data set contributing to the six observables of Table 4, 
      as well as LEP+FERMILAB
      combined direct  measurement of $m_W$: $m_W = 80.426(34)$ GeV. Other fits are done including
     also the indirect determination of $m_W$: $m_W(\rm{NuTeV}) = 80.136(83)$ GeV by NuTeV~\cite{NuTeV1}.
      Only $m_H$ is varied in the fits, the other important parameters: $m_Z$,
      $m_t$, $\alpha(m_Z)$ and $\alpha_s(m_Z)$  being fixed at their measured values of 91.1875 GeV,
      174.3 GeV, 0.007755 and 0.118 respectively. The effect of variation of the second and third of  
      these parameters, within their experimental uncertainies, on the CL for agreement of
       the data with the SM, will be discussed in Section 5 below. The values of other 
      fixed parameters are specified in References~\cite{JHF1,JHF2,JHF3}.
       \par In the fits, a numerical parameterisation, accurate at the per mil level,
        of the effective weak mixing angle given the two-loop ZFITTER 5.10
       program~\cite{ZFITTER} was used\footnote{The formula is valid at the quoted accuracy
       for $m_H \ge 40$ GeV. For lower values of $m_H$, small corrections are made to
       the constant term and the coefficient of $\ln m_H$.}:
   \begin{equation}
       \sin^2\Theta_{eff}^{lept} = 0.233657-8.42\times 10^{-8}m_t^2-3.86 \times 10^{-8}\ln m_t
          +5.00\times 10^{-4}\ln m_H
   \end{equation}
     where $m_t$ and $m_H$ are expressed in GeV units.
    The overall normalisation factors $\rho_f$ ($f = l,\nu,u,d,b$) for fermionic
    widths of the Z are given by a numerical parametrisation similar to Eqn(4.1)
     of the entries in Table 2 of Reference~\cite{Hollik}. For $m_W$, the parameterisation
     of Reference~\cite{ACFW} was used. 
       \par The fits for $m_H$ are then repeated using only the `$m_H$-sensitive' observables
      $A_l(all)$ and $m_W$ where $A_l(all)$ is the weighted average of the leptonic and hadronic
      values of $A_l$ given in the last column of Table 6. Similar fits are performed including also
      $m_W({\rm NuTeV})$. The results of this comparison are shown in the first four rows of
      Table 7. In can be seen that essentially the same range of Higgs masses is obtained
      whether fits are made to the complete set of electroweak observables or only to the
      $m_H$-sensitive ones  $A_l$  and $m_W$. The fit results presented in the fifth and sixth 
      rows of Table 7 demonstrate that very low values of $m_H$, with best fit values
     incompatible with the 95$\%$ direct lower limit of 114.4 GeV, are found when fitting only
     the $m_H$-sensitive observables $A(lept)_l$ and $m_W$. As shown
    in the last two rows of Table 7, much higher values of
     $m_H$ are found when $A(lept)_l$ is replaced by  $A(had)_l$ in the fits. This is a consequence 
     of the deviation of the measured value of $A_b$ from the SM expectation, and the strong
    correlation between $A_l$ and $A_b$ resulting from Eqn(3.1), when $\Afbb$ is measured.
     In all cases inclusion of the NuTeV $m_W$ measurement results in slightly
     higher fitted values for $m_H$ and reduces all confidence levels, by about 
     an order of magnitude, to values less than 0.01.
      \par As another cross-check of both the fitting procedure and the $m_H$, $m_t$
       sensitivity of different observables, simultaneous fits to $m_H$ and $m_t$ were
      performed including also in the $\chi^2$ estimator the directly measured value of $m_t$ from
       FERMILAB: $ m_t = $ 174.3(5.1) GeV. The results of these fits, both including and excluding
     the indirect NuTeV $m_W$ measurement, are presented in Table 8, where they may be
     compared with the results of similar fits from the most recent EWWG report~\cite{EWWG}.
      Slightly lower fitted values of $m_H$ are found in the latter, probably due to the 
     inclusion of other observables such as $\Gamma_Z$, and $Q_W(Cs)$ from
      atomic parity-violating experiments, that have some sensitivity to $m_H$, in the 
       EWWG fits.
      The uncertainties on both $m_H$ and $m_t$ found in the two sets of fits are very
      similar. In fact slightly more precise values of $m_H$ are obtained in the fits
      of the present paper. This, in combination with the results shown in the first four
      rows of Table 7, shows that the restriction to $A_l$ and $m_W$ entails no significant
      loss of sensitivity in the indirect determination of $m_H$.
      The dilution effect discussed in Section 1 of the hypothesis-testing power of the
      $\chi^2$ estimator, due to the inclusion of unaveraged equivalent obervables, or
      additional `noise' observables, that are insensitive to $m_H$ and $m_t$, is evident
      in the $\chi^2/{\rm d.o.f}$s and CLs of the fits that are also presented in Table 8.
      The EWWG fits have a CL that is a factor of 8(4) times larger than those of
      the present paper for the fits including(excluding) the NuTeV  $m_W$ measurement.
      A more detailed discussion of the global EWWG fits is found in Section 8
      below. 

     \SECTION{\bf{Combining Confidence Levels of Direct and Indirect Limits on $m_H$}}

\begin{figure}[htbp]
\begin{center}\hspace*{-0.5cm}\mbox{
\epsfysize10.0cm\epsffile{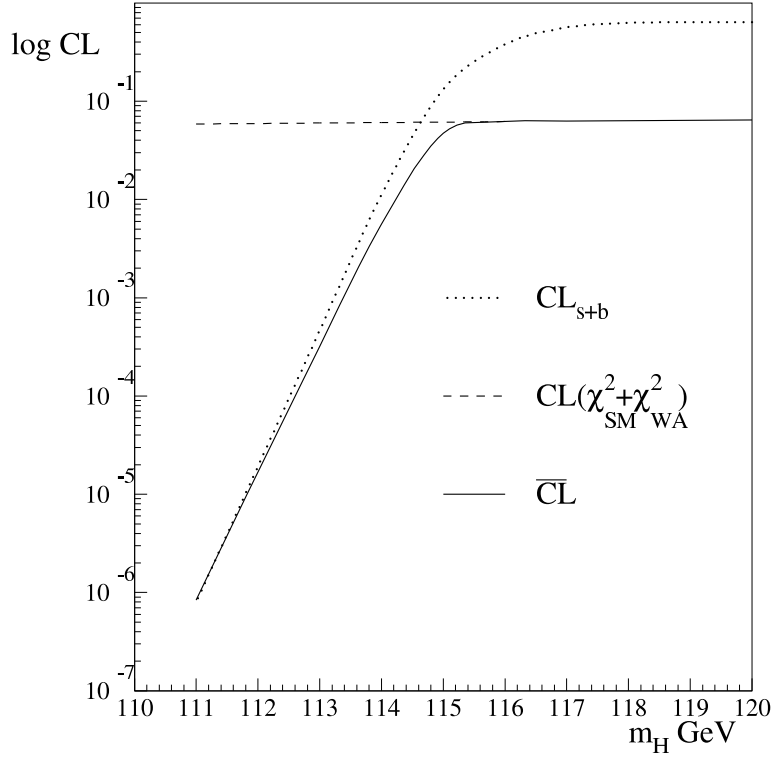}}
\caption{{\sl Illustration of the combination of direct (${\rm CL}_{s+b}$) and indirect
 $\rm{ CL(\chi^2_{SM}+\chi^2_{WA})}$ $m_H$ confidence levels using Eqn(2.8). The observables
 used to calculate  $\rm{ CL(\chi^2_{SM}+\chi^2_{WA})}$ are $A_l(all)$ and $m_W$.}}  
\label{fig-fig1}
\end{center}
 \end{figure}

\begin{figure}[htbp]
\begin{center}\hspace*{-0.5cm}\mbox{
\epsfysize10.0cm\epsffile{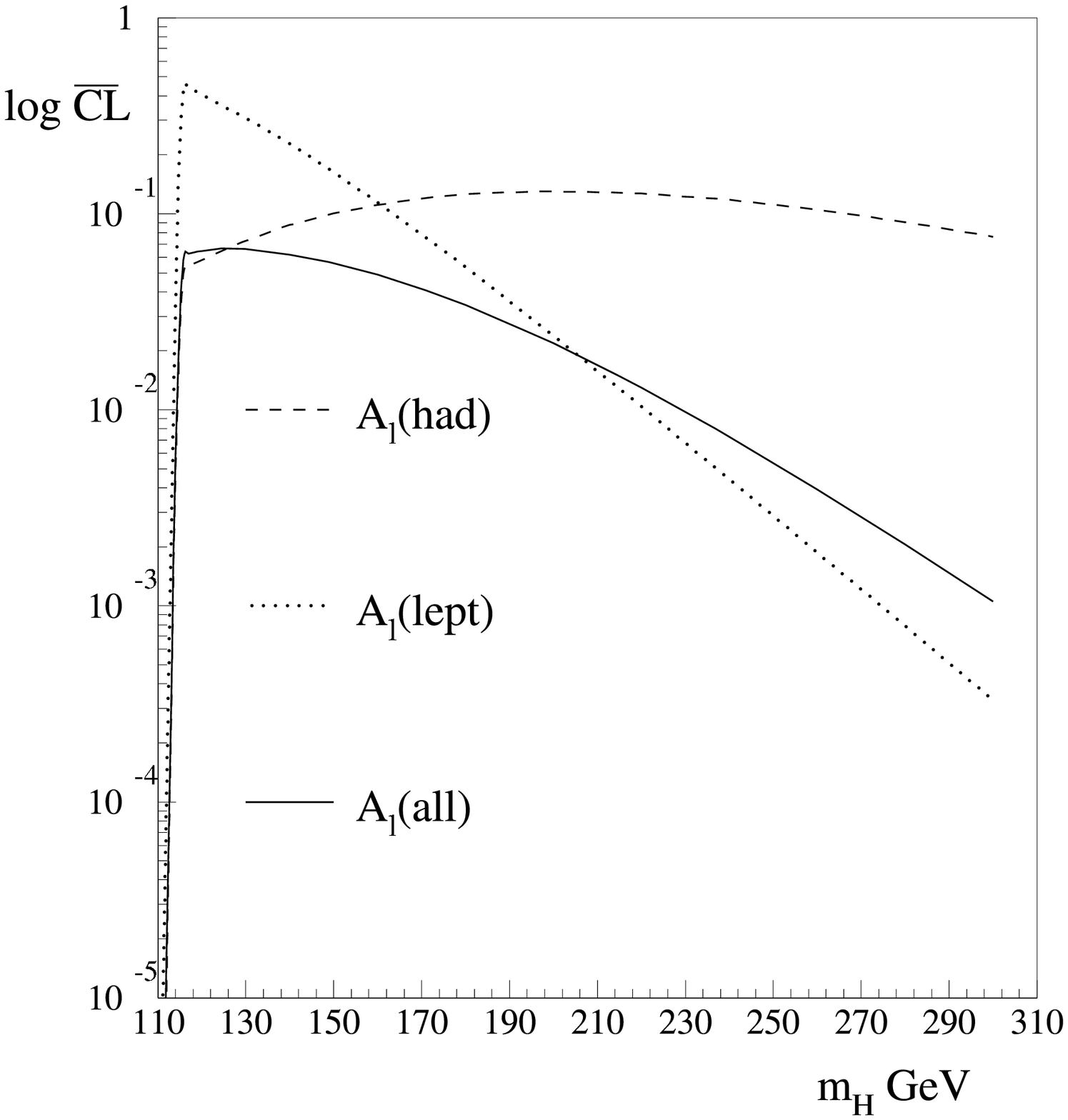}}
\caption{{ \sl Combined $m_H$ confidence levels. The observables
 used to calculate  $\rm{ CL(\chi^2_{SM}+\chi^2_{WA})}$ are $A_l$ and $m_W$.}} 
\label{fig-fig2}
\end{center}
 \end{figure}

\begin{figure}[htbp]
\begin{center}\hspace*{-0.5cm}\mbox{
\epsfysize10.0cm\epsffile{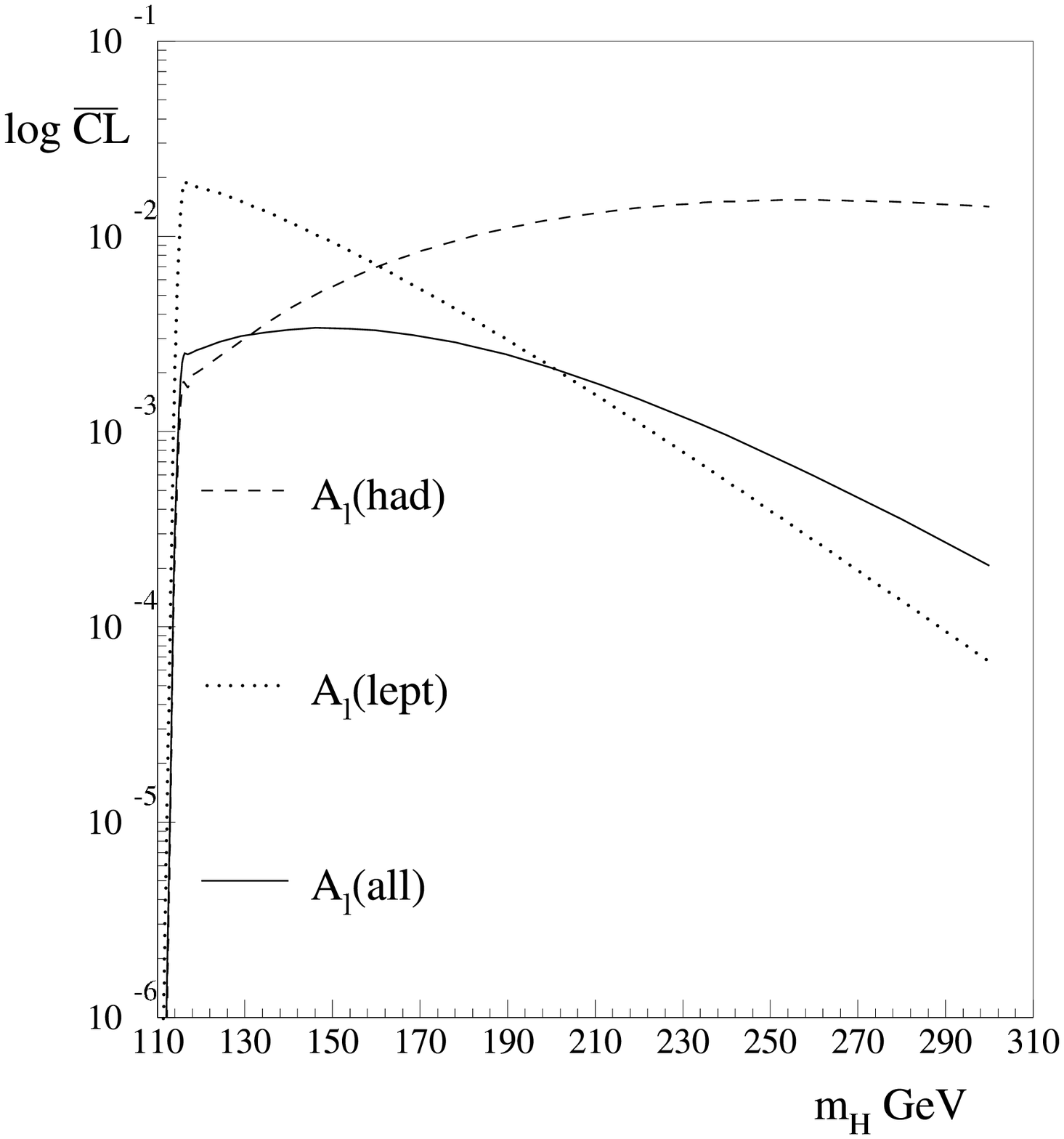}}
\caption{{ \sl Combined $m_H$ confidence levels. The observables
 used to calculate  $\rm{ CL(\chi^2_{SM}+\chi^2_{WA})}$ are $A_l$, $m_W$ and $m_W({\rm NuTeV})$.}}  
\label{fig-fig3}
\end{center}
 \end{figure}

%\begin{table}
%\begin{center}
%\begin{tabular}{|c||c||c|c|c|c|c|} \hline
% Observables & $m_H$ [GeV] & 120 & 160 & 200 & 240 & 280 \\ \hline \hline
%  $A_l(all)$, $m_W$ & CL($\chi^2_{SM}$) & 0.28  & 0.074 & 0.012 & 0.0017 & 0.00022 \\  \cline{2-7}
% $\chi^2_{WA}/{\rm d.o.f.} = 9.0/3$  & CL($\chi^2_{SM}+\chi^2_{WA}$) & 0.040  & 0.014 & 0.0032 & 0.00057 &
%  9.5$\times 10^{-5}$ \\  \cline{2-7}
%   CL = $0.029$  & CL($Comb$) & 0.046  & 0.015 & 0.0032 & 0.00054 &
%  8.3$\times 10^{-5}$ \\  \hline \hline
%  $A_l(lept)$, $m_W$ & CL($\chi^2_{SM}$) & 0.053 & 0.0044 & 0.00036 & 3.2 $\times 10^{-5}$ & 
%  3.0 $\times 10^{-6}$  \\  \cline{2-7}
% $\chi^2_{WA}/{\rm d.o.f.} = 1.6/2$  & CL($\chi^2_{SM}+\chi^2_{WA}$) & 0.112 0 & 0.014 & 0.0016 & 
% 1.7 $\times 10^{-4}$ &  1.9$\times10^{-5}$ \\  \cline{2-7}
%   CL = $0.45$  & CL($Comb$) & 0.112 0 & 0.014 & 0.0016 & 
% 1.7 $\times 10^{-4}$ &  1.9$\times10^{-5}$ \\ \hline \hline
%  $A_l(had)$, $m_W$ & CL($\chi^2_{SM}$) & 0.015 & 0.0018 & 0.014 &  0.0084 & 0.0043  \\  \cline{2-7}
% $\chi^2_{WA}/{\rm d.o.f.} = 0.047/2$  & CL($\chi^2_{SM}+\chi^2_{WA}$) & 0.075 & 0.09 & 0.073 & 
%  0.047 &  0.027 \\  \cline{2-7}
%   CL = $0.98$  & CL($Comb$) & 0.075 & 0.09 & 0.072 & 
%  0.047 &  0.027 \\ \hline
%\end{tabular}
%\caption[]{{ \sl Confidence levels as a function of $m_H$ for different sets of observables. }} 
%\end{center}
%\end{table}

\begin{table}
\begin{center}
\begin{tabular}{|c||c||c|c|c|c|c|} \hline
 Observables & $m_H$ [GeV] & 120 & 160 & 200 & 240 & 280 \\ \hline \hline
  $A_l(all)$, $m_W$ & CL($\chi^2_{SM}$) & 0.51  & 0.35 & 0.13 & 0.047 & 0.0074 \\  \cline{2-7}
 $\chi^2_{WA}/{\rm d.o.f.} = 9.0/3$  & CL($\chi^2_{SM}+\chi^2_{WA}$) & 0.065  & 0.049 & 0.022 & 0.0073 &
  0.0021 \\  \cline{2-7}
   CL = $0.029$  & CL($Comb$) & 0.076  & 0.057 & 0.024 & 0.0104 &
  0.0020 \\  \hline \hline
  $A_l(lept)$, $m_W$ & CL($\chi^2_{SM}$) & 0.30 & 0.054 & 0.008 & 0.0011 & 
  0.00017  \\  \cline{2-7}
 $\chi^2_{WA}/{\rm d.o.f.} = 1.6/2$  & CL($\chi^2_{SM}+\chi^2_{WA}$) & 0.41 0 & 0.11 & 0.024 & 
  0.0044 &  0.00079 \\  \cline{2-7}
   CL = $0.45$  & CL($Comb$) & 0.41 0 & 0.11 & 0.024 & 
  0.0044 & 0.00079 \\ \hline \hline
  $A_l(had)$, $m_W$ & CL($\chi^2_{SM}$) & 0.11 & 0.024 & 0.029 &  0.026 & 0.019  \\  \cline{2-7}
 $\chi^2_{WA}/{\rm d.o.f.} = 0.047/2$  & CL($\chi^2_{SM}+\chi^2_{WA}$) & 0.058 & 0.111 & 0.130 & 
  0.118 &  0.091 \\  \cline{2-7}
   CL = $0.98$  & CL($Comb$) & 0.058 & 0.111 & 0.130 & 
  0.118 &  0.091 \\ \hline
\end{tabular}
\caption[]{{ \sl Confidence levels as a function of $m_H$ for different sets of observables.
   The values of  $\chi^2_{WA}/{\rm d.o.f.}$ and CL refer to $A_l$.}} 
\end{center}
\end{table}

%\begin{table}
%\begin{center}
%\begin{tabular}{|c||c||c|c|c|c|c|} \hline
% Observables & $m_H$ [GeV] & 120 & 160 & 200 & 240 & 280 \\ \hline \hline
%  $A_l(all)$, $m_W$, & CL($\chi^2_{SM}$) & 0.014 & 0.0069 & 0.0020 & 4.2 $\times 10^{-4}$ 
%  &  7.9 $\times 10^{-5}$ \\  \cline{2-7}
%  $m_W({\rm NuTeV})$ & CL($\chi^2_{SM}+\chi^2_{WA}$) & 0.0032  & 0.0017 &  5.5$\times 10^{-4}$ & 
%  1.4 $\times 10^{-4}$ & 3.0 $\times 10^{-5}$ \\  \cline{2-7}
%  & CL($Comb$) & 0.0036  & 0.0019 &  6.1 $\times 10^{-4}$ &  1.5 $\times 10^{-4}$ &
%  3.2 $\times 10^{-5}$ \\  \hline \hline
%  $A_l(lept)$, $m_W$,  & CL($\chi^2_{SM}$) & 0.003 & 5.0 $\times 10^{-4}$  & 6.9 $\times 10^{-5}$
%  & 9.2 $\times 10^{-6}$ &  1.2 $\times 10^{-6}$  \\  \cline{2-7}
%  $m_W({\rm NuTeV})$ & CL($\chi^2_{SM}+\chi^2_{WA}$) & 0.0083 & 0.0017 & 2.7 $\times 10^{-4}$ & 
% 4.2 $\times 10^{-5}$ &  6.5 $\times10^{-6}$ \\  \cline{2-7}
%  & CL($Comb$) & 0.010 & 0.0021 &  3.5 $\times 10^{-4}$ & 
% 5.5 $\times 10^{-5}$ &  8.6 $\times10^{-6}$ \\ \hline \hline
%  $A_l(had)$, $m_W$, & CL($\chi^2_{SM}$) & 9.0 $\times 10^{-4}$ & 0.0019 & 0.0022 &  0.0019 & 0.0013
%   \\  \cline{2-7}
%  $m_W({\rm NuTeV})$ & CL($\chi^2_{SM}+\chi^2_{WA}$) & 0.0055 & 0.011 & 0.012 & 
%  0.011 &  0.0064 \\  \cline{2-7}
%  & CL($Comb$) & 0.0071 & 0.014 & 0.016 & 
%  0.012 &  0.0082 \\ \hline
%\end{tabular}
%\caption[]{{ \sl Confidence levels as a function of $m_H$ for different sets of observables. }} 
%\end{center}
%\end{table}

\begin{table}
\begin{center}
\begin{tabular}{|c||c||c|c|c|c|c|} \hline
 Observables & $m_H$ [GeV] & 120 & 160 & 200 & 240 & 280 \\ \hline \hline
  $A_l(all)$, $m_W$, & CL($\chi^2_{SM}$) & 0.012 & 0.015 & 0.0089 & 0.0036 
  &  0.0012 \\  \cline{2-7}
  $m_W({\rm NuTeV})$ & CL($\chi^2_{SM}+\chi^2_{WA}$) & 0.0027  & 0.0033 & 0.0021 & 
  0.00096 &  0.00036 \\  \cline{2-7}
  & CL($Comb$) & 0.0030 & 0.0037 & 0.0024 &  0.00107 &
  0.00039 \\  \hline \hline
  $A_l(lept)$, $m_W$,  & CL($\chi^2_{SM}$) & 0.0071 &  0.0025  & 0.00066
  &  0.00013 &  3.3 $\times 10^{-5}$  \\  \cline{2-7}
  $m_W({\rm NuTeV})$ & CL($\chi^2_{SM}+\chi^2_{WA}$) & 0.018 & 0.0072 & 0.0022 & 
  0.00056 & 0.00014 \\  \cline{2-7}
  & CL($Comb$) & 0.022 & 0.0089 & 0.0027 & 
 0.00065 & 0.00018  \\ \hline \hline
  $A_l(had)$, $m_W$, & CL($\chi^2_{SM}$) & 0.00031 & 0.0012 & 0.0022 &  0.0029 & 0.0029
   \\  \cline{2-7}
  $m_W({\rm NuTeV})$ & CL($\chi^2_{SM}+\chi^2_{WA}$) & 0.0021 & 0.007 & 0.012 & 
  0.015 &  0.015 \\  \cline{2-7}
  & CL($Comb$) & 0.0028 & 0.0091 & 0.016 & 
  0.020 &  0.019 \\ \hline
\end{tabular}
\caption[]{{ \sl Confidence levels as a function of $m_H$ for different sets of observables. }} 
\end{center}
\end{table}

%\begin{table}
%\begin{center}
%\begin{tabular}{|c|c|c|c|} \cline{2-4} 
%\multicolumn{1}{c}{ } &\multicolumn{1}{|c}{ $A_l(All)$ }
%    &\multicolumn{1}{|c|}{ $A_l(lept)$ } & \multicolumn{1}{c|}{ $A_l(had)$} \\  \hline
%\multicolumn{1}{|c}{$m_H$ (GeV) } &\multicolumn{1}{|c|}{ } & \multicolumn{1}{c|}{ }
% &  \multicolumn{1}{c|}{ } \\ \cline{1-1}
%111 & $8.4 \times 10^{-7}$  &  $6.3 \times 10^{-6}$ & $6.6 \times 10^{-7}$ \\
%113 & $3.2 \times 10^{-4}$  &  $2.1 \times 10^{-3}$ & $2.6 \times 10^{-4}$ \\
%115 & $0.047$  &  $0.23$ & $0.041$ \\
%140 &  $0.062$   &  $0.23$ & $0.088$ \\
%180 &  $0.034$  &  $0.054$ & $0.13$ \\
%220 & $0.013$  &  $0.010$ & $0.13$ \\
%260 &  $0.0039$  &  $0.0019$ & $0.11$ \\
%300 &  $0.0011$  &  $0.00033$ & $0.08$ \\ \hline          
%\end{tabular}
%\caption[]{{ \sl Combined confidence levels $\overline{CL}$ for consistency with the 
%    SM as a function of $m_H$. Observables used in the 
%    $\chi^2$ estimator: $A_l$ and $m_W$. }} 
%\end{center}
%\end{table}

\begin{table}
\begin{center}
\begin{tabular}{|c|c|c|c|} \cline{2-4} 
\multicolumn{1}{c}{ } &\multicolumn{1}{|c}{ $A_l(All)$ }
    &\multicolumn{1}{|c|}{ $A_l(lept)$ } & \multicolumn{1}{c|}{ $A_l(had)$} \\  \hline
\multicolumn{1}{|c}{$m_H$ (GeV) } &\multicolumn{1}{|c|}{ } & \multicolumn{1}{c|}{ }
 &  \multicolumn{1}{c|}{ } \\ \cline{1-1}
111 & $8.4 \times 10^{-7}$  &  $6.3 \times 10^{-6}$ & $6.6 \times 10^{-7}$ \\
113 & $3.2 \times 10^{-4}$  &  $2.1 \times 10^{-3}$ & $2.6 \times 10^{-4}$ \\
115 & $0.047$  &  $0.23$ & $0.041$ \\
140 &  $0.062$   &  $0.23$ & $0.088$ \\
180 &  $0.034$  &  $0.054$ & $0.13$ \\
220 & $0.013$  &  $0.010$ & $0.13$ \\
260 &  $3.9 \times 10^{-3}$  &  $1.9 \times 10^{-3}$ & $0.11$ \\
300 &  $1.1 \times 10^{-3}$  &  $3.3 \times 10^{-4}$ & $0.077$ \\ \hline          
\end{tabular}
\caption[]{{ \sl Combined confidence levels $\overline{CL}$ for consistency with the 
    SM as a function of $m_H$. Observables used in the 
    $\chi^2$ estimator: $A_l$ and $m_W$. }} 
\end{center}
\end{table}

%\begin{table}
%\begin{center}
%\begin{tabular}{|c|c|c|c|} \cline{2-4} 
%\multicolumn{1}{c}{ } &\multicolumn{1}{|c}{ $A_l(All)$ }
%    &\multicolumn{1}{|c|}{ $A_l(lept)$ } & \multicolumn{1}{c|}{ $A_l(had)$} \\  \hline
%\multicolumn{1}{|c}{$m_H$ (GeV) } &\multicolumn{1}{|c|}{ } & \multicolumn{1}{c|}{ }
% &  \multicolumn{1}{c|}{ } \\ \cline{1-1}
%111 & $5.2 \times 10^{-8}$  &  $1.8 \times 10^{-7}$ & $7.0 \times 10^{-8}$ \\
%113 & $2.1 \times 10^{-5}$  &  $6.8 \times 10^{-5}$ & $3.0 \times 10^{-5}$ \\
%115 & $3.6 \times 10^{-3}$  &  $9.9 \times 10^{-3}$ & $4.9 \times 10^{-3}$ \\
%140 &  $2.6 \times 10^{-3}$   &  $3.9 \times 10^{-3}$  &  $8.3 \times 10^{-3}$ \\
%180 &  $1.0 \times 10^{-3}$  &  $6.8 \times 10^{-4}$ & $0.012$ \\
%220 & $2.8 \times 10^{-4}$  &  $1.1\times 10^{-4}$ & $0.012$ \\
%260 &  $6.4 \times 10^{-5}$  &  $1.6 \times 10^{-5}$ &  $9.1 \times 10^{-3}$ \\
%300 &  $1.3 \times 10^{-5}$  &  $2.5 \times 10^{-6}$ & $6.4 \times 10^{-3}$  \\ \hline          
%\end{tabular}
%\caption[]{{ \sl Combined confidence levels $\overline{CL}$ for consistency with the 
%    SM as a function of $m_H$. Observables used in the 
%    $\chi^2$ estimator: $A_l$, $m_W$ and $m_W({\rm NuTeV})$.  }} 
%\end{center}
%\end{table}

\begin{table}
\begin{center}
\begin{tabular}{|c|c|c|c|} \cline{2-4} 
\multicolumn{1}{c}{ } &\multicolumn{1}{|c}{ $A_l(All)$ }
    &\multicolumn{1}{|c|}{ $A_l(lept)$ } & \multicolumn{1}{c|}{ $A_l(had)$} \\  \hline
\multicolumn{1}{|c}{$m_H$ (GeV) } &\multicolumn{1}{|c|}{ } & \multicolumn{1}{c|}{ }
 &  \multicolumn{1}{c|}{ } \\ \cline{1-1}
111 & $3.6 \times 10^{-8}$  &  $3.0 \times 10^{-7}$ & $2.6 \times 10^{-8}$ \\
113 & $1.6 \times 10^{-5}$  &  $1.1 \times 10^{-4}$ & $1.1 \times 10^{-5}$ \\
115 & $2.8 \times 10^{-3}$  &  $0.017$ & $2.2 \times 10^{-3}$ \\
140 &  $3.3 \times 10^{-3}$   &  $0.012$  &  $4.2 \times 10^{-3}$ \\
180 &  $2.8 \times 10^{-3}$  &  $4.0 \times 10^{-3}$ & $9.8 \times 10^{-3}$ \\
220 & $1.5 \times 10^{-3}$  &  $1.1\times 10^{-3}$ & $0.014$ \\
260 &  $6.0 \times 10^{-4}$  &  $2.8 \times 10^{-4}$ &  $0.015$ \\
300 &  $2.1 \times 10^{-4}$  &  $6.6 \times 10^{-5}$ & $0.014$  \\ \hline          
\end{tabular}
\caption[]{{ \sl Combined confidence levels $\overline{CL}$ for consistency with the 
    SM as a function of $m_H$. Observables used in the 
    $\chi^2$ estimator: $A_l$, $m_W$ and $m_W({\rm NuTeV})$.  }} 
\end{center}
\end{table}
 
       The combined result of the direct searches for the Standard Model Higgs Boson by the LEP
       Collaborations ALEPH, DELPHI, L3 and OPAL is given in Fig.9 of Reference~\cite{HIGGSMD}.
       This shows the confidence level ratio: ${\rm CL}_{s} \equiv {\rm CL}_{s+b}/{ \rm CL}_b$
       as a function 
       of $m_H$. ${\rm CL}_{s+b}$ is the confidence level of the signal-plus-background hypothesis
        and ${\rm CL}_b$ that of the background-only hypothesis. Inspection of the figure shows that,
        at percent level accuracy: ${\rm CL}_s = 10^{-6},0.05,0.08$ for $m_H = 111,114.4,120$ GeV,
        respectively. For the present study it is preferred to work directly with ${\rm CL}_{s+b}$,
        which is similar to the $\chi^2$ confidence level given by comparing the SM to
        Z-decay data, to obtain indirect $m_H$ limits. As shown in Fig.7 of
         Reference~\cite{HIGGSMD}, the value of ${\rm CL}_b$ is about 0.8 in the region
         110 GeV $< m_H <$ 120 GeV, of interest for the present study. This gives the estimates:
          ${\rm CL}_{s+b} = 10^{-7},0.04,0.64$ for $m_H = 111,114.4,120$GeV, respectively. 
         \par The following numerical parameterisation of ${\rm CL}_{s+b}$ is used:
          \begin{eqnarray}
                111~\rm{GeV} <&m_H& < 114.4~\rm{GeV} \nonumber \\
          \log{\rm CL}_{s+b}& = &1.382 m_H(\rm{GeV})-159.51 \\
                114.4~\rm{GeV} \le &m_H& < 120~\rm{GeV}\nonumber \\ 
   \log{\rm CL}_{s+b}& = &-0.1938-\left(\frac{120-m_H(\rm{GeV})}{5.3945}
   \right)^{4.968}
           \end{eqnarray}
      As shown in Fig.1, the function of Eqn(5.2) has the same value and first derivative
     as that of Eqn(5.1), at the matching point $m_H = 114.4$ GeV, and vanishing first derivative
    at $m_H = 120$ GeV, where ${\rm CL}_{s+b} = 0.64$. Allowing for the overall scale factor
   of $0.8$, the parameterisation of Eqns(5.1) and (5.2) describes well the experimentally
    determined curve of  ${\rm CL}_{s+b}/{\rm CL}_b$ in Fig.9 of  Reference~\cite{HIGGSMD}.
    It should be noted,
   however, that the precise shape of ${\rm CL}_{s+b}$ has only a small effect on the final confidence level
    curves to be presented below. The direct search excludes, with a CL of $\le 10^{-3}$, the
    possibility  that the SM Higgs boson exists with mass of less than 113 GeV, and gives
   essentially no information for $m_H > 115$ GeV. Thus the region where it is of interest to
   combine ${\rm CL}_{s+b}$ with indirect  confidence levels covers only
   a narrow range of $m_H$. 
   \par In order to define the confidence level for agreement of Z-decay data with the SM, the $\chi^2$
    of the data/SM comparision is simply calculated as a function of $m_H$, setting $m_t$ and 
     $\alpha(m_Z)$ to the measured values given above. Therefore no fit to the data is necessary.
     The sensitivity of the CL curves to the assumed values of $m_t$ and  $\alpha(m_Z)$ is discussed 
     below.     
      In this Section only the '$m_H$-sensitive' observables $A_l$ and $m_W$ are included in the $\chi^2$
      estimator, where the W mass is either the directly measured value from LEP and FERMILAB
    or the indirectly determined NuTeV value. $A_l$ is determined either by using all
    asymmety data ($A_l(all)$), lepton data only ($A_l(lept)$) or only hadronic data
     ($A_l(had)$). The corresponding values are presented in Table 6 above. To take into
     account the internal consistency of the different data sets the values of $\chi^2_{all,WA}$,
     $\chi^2_{lept,WA}$,  or  $\chi^2_{had,WA}$ are added to the $\chi^2$ of the SM comparison:
     $\chi^2_{WA,SM}$ in each case. As shown below, almost identical CLs are found using 
     either $\chi^2_{X,WA}+\chi^2_{WA,SM}$ ($X=all, lept, had$) or by combining the CLs of
    $\chi^2_{X,WA}$ and $\chi^2_{WA,SM}$ using Eqn(2.8). The former CL is then combined with
     ${\rm CL}_{s+b}$ using Eqn(2.8) to yield the direct plus indirect confidence level curves shown below.
     The values of $\chi^2/{\rm d.o.f.}$ for  $\chi^2_{had,WA}$ and  $\chi^2_{all,WA}$ are given 
     above; that for the leptonic data: $\chi^2_{lept,WA}/{\rm d.o.f.} = 1.6/2$, CL $= 0.45$, 
    given by combining the $A_l$ values obtained from lepton forward/backward asymmetries and
    tau polarisation measurements from LEP and the $A_{LR}$ measurement from SLD,
    is taken from Reference~\cite{EWWG}.
     \par Some typical CLs for the indirect $m_H$ analysis, obtained as described above, 
     are presented in Table 9 (observables considered: $A_l$, $m_W$) and Table 10
     (observables considered: $A_l$, $m_W$, $m_W(\rm{NuTeV}$)). In all cases good agreement is found
     between
     CL($\chi^2_{SM}+\chi^2_{WA}$) and  CL($Comb$) calculated using Eqn(2.8), where the abbreviations
     $\chi^2_{SM} \equiv \chi^2_{WA,SM}$  $\chi^2_{WA} \equiv \chi^2_{X,WA}$ have been introduced.
      \par The combination of  CL($\chi^2_{SM}+\chi^2_{WA}$) (indirect measurements) and ${\rm CL}_{s+b}$  
       (direct measurements) for different values of $m_H$ is illustrated in Fig.1. Since ${\rm CL}_{s+b}$
      provides little information on $m_H$ for $m_H > 114.4$ GeV (the 95$\%$ CL lower limit of
     the direct search),  CL($\chi^2_{SM}+\chi^2_{WA}$) is combined with  ${\rm CL}_{s+b}$ provided
     that  CL($Comb$) is less than CL($\chi^2_{SM}+\chi^2_{WA}$). In the contrary case 
      CL($\chi^2_{SM}+\chi^2_{WA}$) alone is used. Thus the algorithm used to obtain the combined
      confidence level, $\overline{{\rm CL}}$, is:
      \begin{itemize}
      \item Calculate $\alpha_3$ from  $\alpha_1 = \rm{ CL(\chi^2_{SM}+\chi^2_{WA})}$ and 
        $\alpha_2 = {\rm CL}_{s+b}$
              according to Eqn(2.8).
      \item If $\alpha_3 <$  CL($\chi^2_{SM}+\chi^2_{WA}$), set $\overline{{\rm CL}} =$ 
       CL($\alpha_1$ $\alpha_2$).
        \item  If  $\alpha_3 \ge$  CL($\chi^2_{SM}+\chi^2_{WA}$), set $\overline{{\rm CL}} =$ 
        CL($\chi^2_{SM}+\chi^2_{WA}$).
       \end{itemize}
       In Fig.1 the dotted curve shows  ${\rm CL}_{s+b}$, the dashed curve  CL($\chi^2_{SM}+\chi^2_{WA}$)
       and the solid curve $\overline{{\rm CL}}$.
       \par Curves of  $\overline{{\rm CL}}$ calculated in this manner are shown in Figs.2 and 3.
         When $A_l(lept)$ and $A_l(all)$ are used, the general 
         shape, with a sharp peak above, but close to, the direct lower limit and a rapid fall-off for higher
        values of $m_H$ is similar to that of the PDFs presented in References~\cite{DD,Erler}. However
       $\overline{{\rm CL}}$, unlike the PDFs, gives an estimate of the absolute probability that the
       data is consistent
       with the SM for a given value of $m_H$. An exception to this behaviour is provided by the 
       data sets $A_l(had)$, $m_W$ and  $A_l(had)$, $m_W$, $m_W({\rm NuTeV})$ where the maximum of
      $\overline{{\rm CL}}$ occurs at much higher values of $m_H$ and on average much larger
      values of $\overline{{\rm CL}}$ are obtained. The confidence level curves in Figs.2 and 3 are 
    presented in numerical form in Tables 11 and 12, in the same format as the similar 
    curves  considered in Section 7 below. The latter use all precision observables  , 
    rather than only the $m_H$-sensitive ones as in Figs. 2 and 3.  Comparison of the two sets of curves then 
    shows the effect of `non $m_H$-sensitive' observables on the level of agreement with the 
    SM prediction.
       \par The effect on the $\overline{{\rm CL}}$ curves of variation of the value of $m_t$ by plus
       or minus the experimental error around the measured value is shown in Fig.4 and Table 13. The
       effect of a similar variation of $\alpha(m_Z)$ is shown in Fig.5 and Table 13.  For large values
       of $m_H$ this variation
      of $m_t$ changes the values of $\overline{{\rm CL}}$ by many orders of magnitude. Even so, in 
      the case of the data set $A_l(lept)$, $m_W$, shown in Figs.4 and 5,  which can be argued (see below) to be likely to 
      give the most reliable estimate of  $m_H$, $\overline{{\rm CL}}$ is still only, at best, 0.01,
       at $m_H = 300$ GeV. Change of $\alpha(m_Z)$ within the current experimental errors can change
        $\overline{{\rm CL}}$ by up to an order of magnitude, but the effect is much less dramatic than
       for $m_t$. Clearly a much improved measurement of $m_t$ is needed to significantly improve the
       indirect limits on $m_H$.
  %\begin{figure}[htbp]
%\begin{center}\hspace*{-0.5cm}\mbox{
%\epsfysize10.0cm\epsffile{smstatf4c.eps}}
%\caption{{ \sl Dependence of combined $m_H$ confidence levels on the value
% of $m_t$. The observables
% used to calculate  $\rm{ CL(\chi^2_{SM}+\chi^2_{WA})}$ are $A_l(All)$ and $m_W$.}} 
%\label{fig-fig4}
%\end{center}
% \end{figure}

\begin{figure}[htbp]
\begin{center}\hspace*{-0.5cm}\mbox{
\epsfysize10.0cm\epsffile{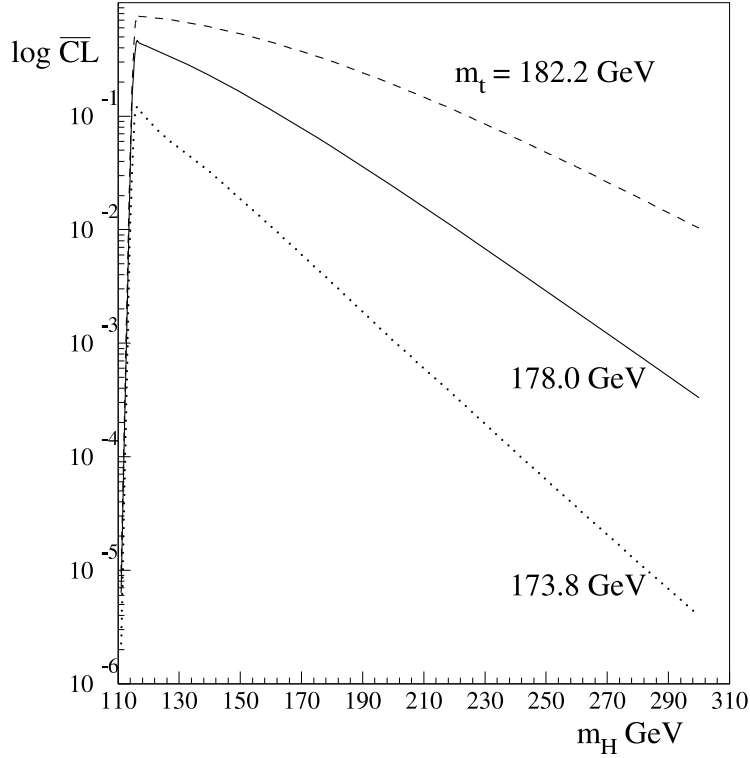}}
\caption{{ \sl Dependence of combined $m_H$ confidence levels on the value
 of $m_t$. The observables
 used to calculate  $\rm{ CL(\chi^2_{SM}+\chi^2_{WA})}$ are $A_l(lept)$ and $m_W$.
  $\alpha(m_Z) = 0.007755$ is assumed.}}  
\label{fig-fig4}
\end{center}
 \end{figure}

%\begin{figure}[htbp]
%\begin{center}\hspace*{-0.5cm}\mbox{
%\epsfysize10.0cm\epsffile{smstatf6c.eps}}
%\caption{{ \sl Dependence of combined $m_H$ confidence levels on the value
% of $\alpha(m_Z)$. The observables
% used to calculate  $\rm{ CL(\chi^2_{SM}+\chi^2_{WA})}$ are $A_l(All)$ and $m_W$.}} 
%\label{fig-fig6}
%\end{center}
% \end{figure}

\begin{figure}[htbp]
\begin{center}\hspace*{-0.5cm}\mbox{
\epsfysize10.0cm\epsffile{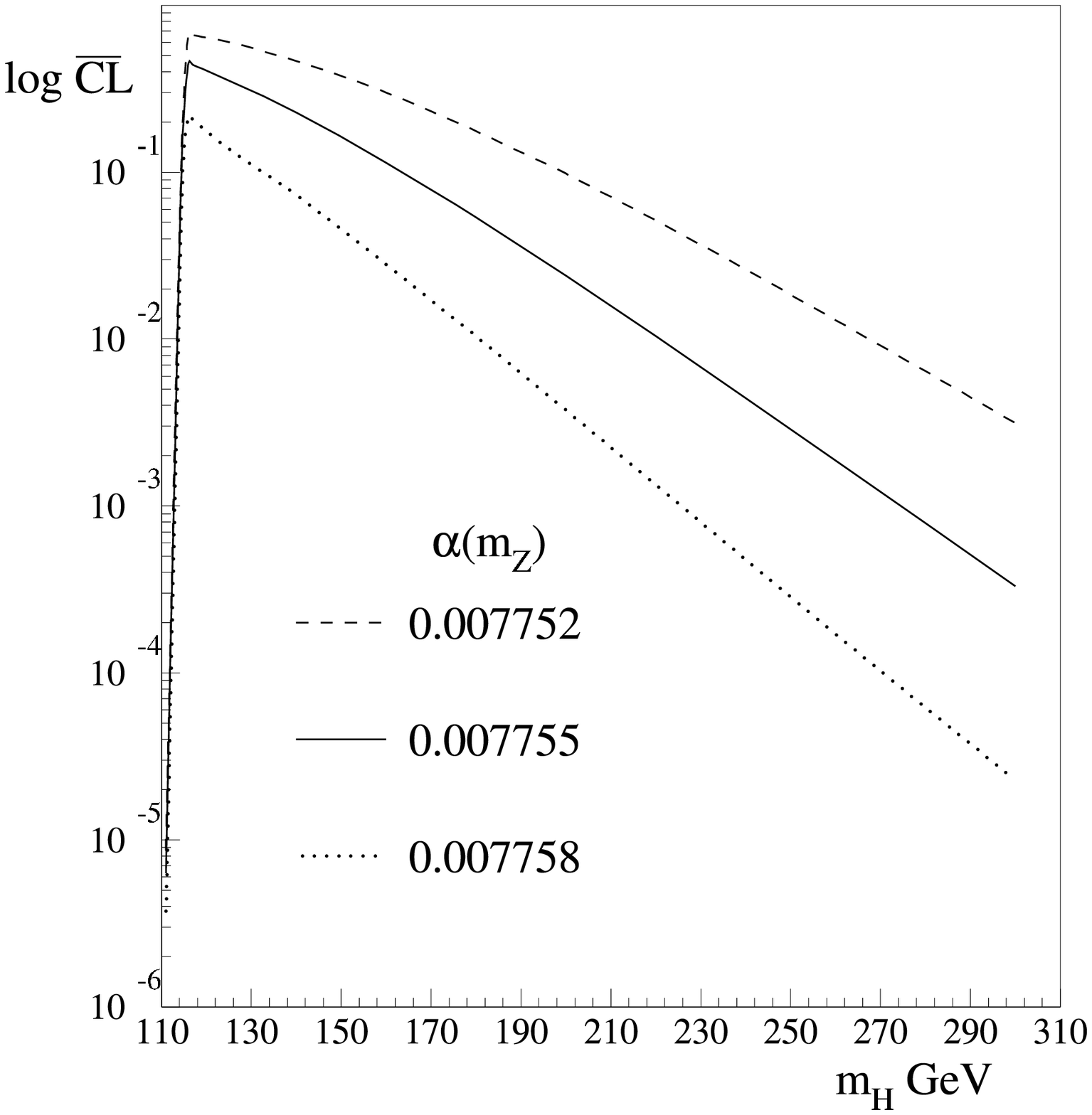}}
\caption{{ \sl Dependence of combined $m_H$ confidence levels on the value
 of  $\alpha(m_Z)$. The observables
 used to calculate  $\rm{ CL(\chi^2_{SM}+\chi^2_{WA})}$ are $A_l(lept)$ and $m_W$.
 $m_t = 178$ GeV is assumed.}}  
\label{fig-fig5} 
\end{center}
\end{figure}

\begin{table}
\begin{center}
\begin{tabular}{|c|c|c||c|c|} \cline{2-5}
 \multicolumn{1}{c}{ } & \multicolumn{2}{|c|}{ $\alpha(m_Z) = 0.007755$} 
  &  \multicolumn{2}{c|}{$m_t = 178$ GeV}  \\  \hline
 \multicolumn{1}{|c}{$m_H$ GeV } &  \multicolumn{2}{|c|}{ $m_t = 173.8$ GeV~~~~182.2 GeV}
  &  \multicolumn{2}{c|}{ $\alpha(m_Z) =$ 0.007752~~~~0.007758}  \\  \hline \hline
  111 &~~~~~ $2.1 \times 10^{-6}$  & $9.4 \times 10^{-6}$ &~~~~~~~~~$8.6 \times 10^{-6}$ & 
  $3.5 \times 10^{-6}$ \\
  113 &~~~~~ $6.9 \times 10^{-4}$  & $3.2 \times 10^{-3}$ &~~~~~~~~~ $2.9 \times 10^{-3}$ &
  $1.1 \times 10^{-3}$ \\
 115 & $0.085$  & $0.33$ & $0.30$ &  $0.13$ \\
 140 & $0.032$  & $0.61$ & $0.46$ &  $0.073$ \\
 180 &  $3.4 \times 10^{-3}$  & $0.30$ & $0.18$ &  $0.010$ \\
 220 &  $3.5 \times 10^{-4}$  & $0.11$ & $0.051$ &  $1.3 \times 10^{-3}$  \\
 260 &  $3.6 \times 10^{-5}$  & $0.036$ & $0.013$ &  $1.7 \times 10^{-4}$  \\
 300 &  $4.0 \times 10^{-6}$  & $0.010$ &  $3.1 \times 10^{-3}$ &  $2.3 \times 10^{-5}$  \\
 \hline
\end{tabular}
\caption[]{{ \sl Combined confidence level curves  $\overline{{\rm CL}}$ for variation
  of $m_t$ and  $\alpha(m_Z)$ by plus or minus one standard deviation around
  their measured values. Observables used in the 
    $\chi^2$ estimator: $A_l(lept)$ and $m_W$. }} 
\end{center}
\end{table}

  \par The analysis presented in this Section has many similarities with that of Reference~\cite{Chanowitz}
       where confidence levels taking into account both direct and indirect information on $m_H$
       were derived. However, the present writer has doubts about the mathematical
     correctness of the method used in Reference~\cite{Chanowitz}.
     It consists of combining the confidence level given by the $\chi_{min}^2$
      and $d.o.f.$ of a fit with that derived from
       ~$\Delta \chi^2 = \chi^2(m_H = 114~{\rm GeV}) - \chi^2_{min}$~ of the {\it  same} fit,
      assuming them to be independent. Certainly, the formula used to combine the confidence
     levels, simple multiplication instead of Eqn(2.8) above, is incorrect\footnote{The present author
       made the same mistake in Reference~\cite{JHF1}}. The approach used is essentially to replace the
      measured ${\rm CL}_{s+b}$ curve by a $\theta$-function at $m_H = 114$ GeV. As can be seen in Fig.1,
      this is quite a good approximation. The global fits used in Reference~\cite{Chanowitz} took no account
      of the dilution of the hypothesis testing power of the $\chi^2$ estimator
      resulting from the use of unaveraged and insensitive
      observables, as discussed in Section 2 above and, in more detail, in Section 8 below. For instance there
     is the statement: `The global SM fit was excellent in 1998 and has now (2002) become poor'.
     This is not at all true of the contribution to  the $\chi^2$  of the $m_H$-sensitive observables, which
      is similarly high for both data sets. In fact, the high confidence level for the 1998 global fit
      is a consequence of an anomalously {\it low} contribution from the non  $m_H$-sensitive observables
       ~\cite{JHF3}. Also, although the NuTeV measurement was discussed, together with $A_b$, as a 
       possible source of anomaly relative to the SM prediction, only the published interpretation
       as a measure of $\sin^2\theta^{on-shell}_W$ or $m_W$ was considered.
      Also the correctness of the SM prediction for the quark couplings to the Z
      was assumed to define different values of $\sin^2\Theta_{eff}^{lept}$ (equivalent to $A_l$ in
     the present paper). The same two assumptions have been made in the analysis presented in this
      Section. In the analysis of model-independent observables presented in Section 7 below
       the alternative interpretation of the NuTeV result, discussed in the following 
       Section, as a measurement of the Z$\nu \overline{\nu}$ coupling is also used, where no {\it a priori}
      assumptions are made concerning the couplings of the Z to fermion pairs.

     \par An important
     point stressed in  Reference~\cite{Chanowitz}, first pointed out in  Reference~\cite{JHF3},
      is that, regardless of how the $A_b$ anomaly is interpreted (statistical fluctuation,
      unknown systematic effect or new physics) the most reliable estimate of
      $m_H$ must be that derived from the charged lepton asymmetries and $m_W$ that
     give consistent predictions for this quantity. Inclusion of the $\Afbb$ measurements in the
     fit results in a positive $\simeq 50$ GeV bias on the 95$\%$ CL upper limit on $m_H$, 
      due to the $A_l$-$A_b$ correlation resulting from Eqn(3.1). Although fits 
     with and without the `anomalous' NuTeV measurement are routinely presented by the
     EWWG, fits excluding the (equally `anomalous') hadron asymmetry data that, would provide the most 
    reliable estimate of $m_H$, have (to my best knowledge) never been shown in the periodical
     updates of the status of electroweak measurements, such as Reference~\cite{EWWG}, produced
     by this working group. 
    I agree with almost all of the general conclusions of Reference~\cite{Chanowitz},
    in particular, that if the lepton asymmetry data is correct then, regardless of the status
    of the hadronic asymmetry data, the SM provides only a poor global description of the data. 
    I would also remark that the confidence
   levels of the global fits quoted, although small, will become even smaller when 
   corrected for the dilution effects discussed in Section 2 above.
   Use of the correct formula for combining confidence levels will, on the other hand,
   give higher combined confidence levels. The analysis of Reference~\cite{Chanowitz}
   gives, however, no hint of the very low values of $\overline{{\rm CL}}$ for large values of $m_H$
   apparent in Figs.2 and 3 and Tables 11 and 12. Finally, in connection with Reference~\cite{Chanowitz},
   as discussed in Section 8 below, the use of $\Delta \chi^2$  to provide confidence levels for parameter 
   estimation is of doubtful validity when, as is the case for the current electroweak data, the 
   absolute confidence level derived from $\chi^2_{min}$ shows that the model containing the
   parameter of interest does not adequately describe the data.
   
 \SECTION{\bf{Alternative Interpretations of the NuTeV Experiment}}
   The publication of the results of the NuTeV experiment~\cite{NuTeV1}
 gives an estimation of the value of the on-shell weak mixing angle:
  \begin{equation}
  \sin^2\theta^{on-shell}_W = 0.2277(13)(9)
  \end{equation}
  that may be translated directly into a W-mass measurement via
  the defining relation of the on-shell renormalisation scheme:
   \begin{equation}
  \sin^2\theta^{on-shell}_W \equiv 1-\frac{m_W^2}{m_Z^2}
 \end{equation}
  This interpretation however requires that the $Z \nu \overline{\nu}$
  coupling is assigned its standard model value. As discussed in
  Reference~\cite{NuTeV2} and shown, for example, in Fig.1 of
  Reference~\cite{MG}, the experiment actually measures a
  quantity that is sensitive both to $\sin^2\theta^{on-shell}_W$
 and the strength of the $Z \nu \overline{\nu}$ coupling, which
  may be specified by a parameter $\rho_0$ such that:
    \begin{equation}
   \sbn = \vbn^2 + \abn^2 = \rho_0^2 \sbn({\rm SM})
 \end{equation}
  Assuming the SM value ($\rho_0 = 1$) gives a prediction for
  $\sin^2\theta^{on-shell}_W$:
    \begin{equation}
  \sin^2\theta^{on-shell}_W = 0.22733(135)(93)-8.8 \times 10^{-8}(
   m_t[{\rm GeV}]^2-175^2)+3.2 \times 10^{-4}\ln(m_H[{\rm GeV}]/150) 
  \end{equation} 
  Assuming instead the SM value of $\sin^2\theta^{on-shell}_W$
  the experiment provides a measurement of $\rho_0$~\cite{NuTeV2}:
     \begin{equation}
  \rho_0 = 0.9942(13)(16)+2.4 \times 10^{-8}(
   m_t[{\rm GeV}]^2-175^2)-1.6 \times 10^{-4}\ln(m_H[{\rm GeV}]/150) 
  \end{equation} 
  Choosing the values  $m_t = 175$ GeV and $m_H = 150$ GeV consistent with the measured 
  value of $m_W$, (6.5) gives a measurement of the
   model-independent parameter $\sbn$:
    \begin{equation}
   \sbn({\rm NuTeV})= 0.4992(21)
    \end{equation}
 which may be compared with the LEP measurement quoted in Table 5:
  \begin{equation}
  \sbn({\rm LEP})= 0.5014(15)
  \end{equation} 
  Since: 
 \begin{equation}
  \sbn({\rm LEP})-  \sbn({\rm NuTeV})= 0.0022(26)
  \end{equation} 
  the two measurements are quite consistent
   ($\chi^2_{WA,\sbn}/d.o.f. = 0.78/1$, CL = 0.38) and yield the
  weighted average value:
   \begin{equation}
  \sbn({\rm LEP+NuTeV})= 0.5006(12)
  \end{equation}
   Alternatively assuming that $\rho_0 =1$ and using Eqns(6.2) and (6.4) to
   obtain $m_W$ gives:
  \begin{eqnarray}
  \sbn({\rm NuTeV}) & = &  \sbn({\rm SM})= 0.5050 \\
  m_W({\rm NuTeV}) & = & 80.136(83)~{\rm GeV} 
  \end{eqnarray} 
  In this case the assumed, SM,  value of $\sbn$ differs from the LEP
  measurement by 2.4$\sigma$ ($\chi^2/d.o.f. = 5.76/1$, CL = 0.016) and
  also 
 \begin{equation}
 m_W({\rm LEP+FERMILAB})- m_W({\rm NuTeV}) = 0.290(90)
  \end{equation}
 the 3.2$\sigma$ discrepancy  ($\chi^2/d.o.f. = 10.4/1$, CL = 0.0013)
 mentioned in Reference~\cite{NuTeV1}. 
 Using Eqn(2.8) to combine the data consistency CLs for
  $\sbn$ and $m_W$ yields an overall CL of $2.5 \times 10^{-4}$. Thus
  on the assumption that the NuTeV measurement is correct, the alternative
  interpretation of the experiment is strongly favoured statistically
  as the ratio of data consistency CLs of the two interpretations
  is $\simeq 1.5 \times 10^3$.  
   \par For both interpretations the SM prediction is unfavoured.
   For the standard one ($m_W$ measurement and $\rho_0 =1$) the CL
  of the SM comparison is that just quoted:  $2.5 \times 10^{-4}$.
 For the alternative interpretation ($\rho_0\ne 1$) it is found that:
 \begin{equation}
 \sbn({\rm LEP+NuTeV})-  \sbn({\rm SM}) = -0.0044(12)
 \end{equation}
  or a 3.7$\sigma$ ($\chi^2/d.o.f. = 13.4/1$, CL $ = 2.5 \times 10^{-4}$) 
   deviation from the SM prediction. The alternative interpretation 
   thus shows exactly the same deviation from the SM as the one
   proposed in Reference~\cite{NuTeV1}.
     \par A number of theoretical assumptions must be made in order to
   derive Eqns(6.4) and (6.5) from the experimental quantities: 
 \[ R_{\nu}^{exp} = \frac{\sigma( \nu Fe \rightarrow \nu X)}{\sigma( \nu Fe \rightarrow \mu^- X)},
  ~~~ R_{\overline{\nu}}^{exp} = \frac{ \sigma( \overline{\nu} Fe \rightarrow \overline{\nu} X)}
    {\sigma( \overline{\nu} Fe \rightarrow \mu^+ X)} \]
 actually measured by the NuTeV experiment. A recent concise review of the situation may be 
 found in Reference~\cite{Gambino} in which citations of related work can be found. The most important
   and extensively discussed assumption concerns the supposed symmetry of the strange sea momentum
  distribution in a nucleon. A recent analysis by the CTEQ Collaboration~\cite{CTEQ} presented
  in Reference~\cite{Gambino} finds some evidence for a positive asymmetry of the strange quark sea:
   \begin{equation}
  s^- = \int_{0}^{1} x(s(x)-\overline{s}(x))dx = 0.002(1)
   \end{equation}
   It is pointed out in Reference~\cite{Gambino} that an asymmetry of 0.002
 has the effect of reducing by 42$\%$ the discrepancy between the measured value
  of $\sin^2\theta^{on-shell}_W$ derived from the direct $m_W$ measurements, and the value
  of the same quantity found from Eqn(6.4). The alternative, and statistically favoured,
   interpretation of the NuTeV experiment as a measurement of $\rho_0$, was not considered
   in Reference~\cite{Gambino}, but in view of the linear correlation between 
    $\sin^2\theta^{on-shell}_W$ and  $\rho_0$ provided by the measurement\footnote{See, for example, 
      Figure 1 of Reference~\cite{MG}.}, it is reasonable to suppose that, in the alternative 
    interpretation, the strange quark sea asymmetry will reduce the deviation of $\rho_0$ (or,
     equivalently $\sbn$) from the SM expectation by the same fraction. Thus the estimated
     value of  $\sbn$, correcting for the effect of the asymmetry of Eqn(6.14) is
      0.5015(21), which agrees prefectly with LEP measurement in Eqn(6.7). The LEP+NuTeV
     weighted average becomes 0.5014(12), which still lies 3.0$\sigma$ below the SM expectation.
     The apparent anomaly in the Z$\nu \overline{\nu}$ coupling is therefore reduced, but
     not removed, by
     the estimated effect of a strange quark sea asymmetry on the NuTeV results. 
   \par It remains true however that because of the  many systematic effects, detailed in
     Reference~\cite{Gambino}, the results of the NuTeV experiments are less `sure' than the measurement
    of the related quantity $\Gamma_{inv}$ at LEP. Because of this some authors~\cite{GAMG} prefer to adopt
    a conservative position and exclude the NuTeV results completely from global
    electroweak analyses. In contrast, the present paper takes a strictly neutral position
    on the question of the reliability, or otherwise, of the NuTeV results.    
   In the following Section then, CLs as a function of $m_H$ will be
    calculated using all available LEP and SLD data on
    the assumption 
   of either of the two possible interpretations of the NuTeV
   experiment, or by excluding the experiment. The CLs obtained
   are later compared in Section 8 below with those obtained from the global EWWG and EWPDG
   fits.

 \SECTION{\bf{Model-Independent Observables Compared to SM Predictions}}
   Following the model-independent approach of References~\cite{JHF1,JHF2,JHF3} essentially all 
   precision information on the SM provided, to date, by the LEP, SLC and FERMILAB
    experimental programs, as well as by the NuTeV experiment, is contained in the values of the 
   nine observables listed in Table 14. The other important quantities $m_t$ and
   $\alpha(m_Z)$ are not included as they are here considered as
    input parameters for the SM prediction rather than measurements which provide
    a test of the SM. All information from leptonic forward/backward charge asymmetry
    and $\tau$-polarisation measurements as well as the SLC ALR measurement is
    condensed into the single parameter $A_l(lept)$, equivalent to  $\sin^2\Theta_{eff}^{lept}$.
    The quantity $\sbl$ is derived from the width of the Z for decay into charged leptons
    on the assumption of charged-lepton universality. The quantities $A_c$ and $A_b$ 
    are derived using Eqn(3.1) from the c- and b-quark  forward/backward charge asymmetries 
    measured at LEP as well as from the SLC measurement of forward/backward-left/right asymmetries
    of c and b quarks. $\sbc$ and $\sbb$ are obtained from the Z decay widths into c and b quarks, more
    conventionally expressed in terms of the ratios: $R_Q = \Gamma_Q/\Gamma_{had},~(Q=c,b)$,
     using the relation:
  \begin{equation}
\sbQ = \sqrt{\frac{2 \pi}{3}}\frac{R_Q \Gamma_Z}{G_{\mu} M^2_Z}
\frac{\sqrt{R_l \sigma^0_h}}{C^{QED}_Q C^{QCD}_Q}~~~(Q=c,b),
\end{equation}
  The QED and QCD correction factors are given in Reference~\cite{JHF1}. 
    The observable $\sbn$ is given by the invisible width, $\Gamma_{inv}$, of the Z boson,
    determined from the Z-boson total width, $\Gamma_Z$, the hadronic width, $\Gamma_{had}$
    and the leptonic width, $\Gamma_l$ via the relation:
    \begin{equation}
     \Gamma_{inv} = \Gamma_Z - \Gamma_{had} - 3 \Gamma_l
    \end{equation}
    Since the hadronic width of the Z is quite precisely measured: $\Gamma_{had} = 1.7444(20)$ GeV,
    the measurements of $R_b$ or $\sbb$ can be used, in combination with the former, to  extract the quantity:
     \begin{equation}
     \sbnbp = \sum_{q=u,c}[(\vbq)^2+(\abq)^2]C_u^{QED}+  \sum_{q=d,s}[(\vbq)^2+(\abq)^2]C_d^{QED}
   \end{equation}
    The subscript `nb' here stands for `non-b' quarks. As will be discussed below, the measurements
    of $\Gamma_{had}$ and $\sbnbp$ provide much more stringent constraints on the possible values
     of the couplings of non-b down-type quarks to the Z than
     the existing direct measurements of
     these couplings, which have large experimental errors.
 
     \par The experimental errors on the observables listed in Table 14 are largely uncorrelated
     between the observables, which facilitates calculation of a $\chi^2$ estimator for global SM comparisons.
     Correlations exist between: $A_l$, $A_c({\rm LEP})$ and $A_b({\rm LEP})$ due to the use of Eqn(3.1)
     to obtain $A_c({\rm LEP})$ and $A_b({\rm LEP})$. Because of the small uncertainty on $\Gamma_{had}$
   the errors on $\sbb$ and $\sbnbp$ are strongly anticorrelated. Weaker correlations exist between
    $\sbc$ and $\sbb$. In view of the relatively poor precision of the $\sbc$ measurement in comparison with
    those of $\sbb$ and $\sbnbp$, and the  correlations between these three observables,
    the contribution of the former is omitted from the $\chi^2$ estimator used in the
    global comparisons with the SM shown below.
    As previously mentioned, to take properly into account correlations, the direct SLC measurements
   of $A_c$ and $A_b$ are assigned separate terms from the LEP measurements in the $\chi^2$
    estimator. In Table 14, however, the weighted average LEP+SLC values of  $A_c$ and $A_b$
    are quoted. 
     \par The value of $\sbn$ given in Table 14 is the LEP+NuTeV weighted average, i.e. the
      `alternative' interpretation of the NuTeV experiment is taken. This is mandatory in
    a model-independent analysis where no {\it a priori} assumptions concerning the couplings
    of the leptons or quarks to the Z (with the exception of charged lepton universality, well 
      respected by experiment) is made. The NuTeV measurement may be compared, in this sense, 
       with the LEP measurements of $\AfbQ$ (q=b,c). The latter depend, via Eqn(3.1), 
     on $A_l$ (defined by the values of the Z charged-lepton couplings) and $A_Q$ 
    (defined by the values of the Z heavy-quark couplings), Since $A_l$ is independently measured,
    $A_Q$ can then be extracted from the measured value of $\AfbQ$. Similarly the NuTeV
    result depends, in a correlated way, on the values of $ \sin^2\theta^{on-shell}_W$ (equivalent to $m_W$)
    and $\rho_0$ (equivalent to the ${\rm Z} \nu \overline{\nu}$ coupling). Since $m_W$
    (and hence $\sin^2\theta^{on-shell}_W$)  is precisely
     determined at LEP and FERMILAB the correlated value of $\rho_0$, and so also $\sbn$,
     can be extracted in a similar fashion to $A_Q$ from $\AfbQ$.

%\begin{table}
%\begin{center}
%\begin{tabular}{|c|c|c|c|c|} \hline  
%$X$ & $X_{expt}$ &  $X_{expt}/\sigma_X$ ($\%$) & $X_{SM}$ & ($X_{expt}-X_{SM})/\sigma_X$ \\  \hline
% $A_l(lept)$ & 0.1501(16) & 1.07 & 0.1472 & 1.81  \\  \hline
%$ $\sbl$ & 0.25268(26) & 0.103 & 0.25276 & -0.31 \\   \hline
% $A_c$ & 0.653(20) & 3.06 & 0.668 & -1.06 \\   \hline
% $\sbc$ & 0.2897(50) & 1.73 & 0.2883 & 0.00 \\  \hline
% $A_b$ & 0.902(13) & 1.44 & 0.9347 &  -2.51  \\  \hline
% $\sbb$ & 0.3663(13) & 0.49 & 0.3647 &  1.15 \\  \hline
% $\sbn$ & 0.5006(12) & 0.24 & 0.5047 & -3.4 \\  \hline
% $\sbnbp$ & 1.3211(43) & 0.33 & 1.3211 & 0.00 \\  \hline
% $m_W$ & 80.426(34) & 0.042 & 80.37 & 1.65 \\  \hline         
%\end{tabular}
%\caption[]{{ \sl model-independent electroweak observables.
%   The SM predictions correspond  to $m_t = 174.3$ GeV, $m_H = 120$ GeV and
%   $\alp = 0.007755$.}} 
%\end{center}
%\end{table}

\begin{table}
\begin{center}
\begin{tabular}{|c|c|c|c|c|} \hline  
$X$ & $X_{expt}$ &  $X_{expt}/\sigma_X$ ($\%$) & $X_{SM}$ & ($X_{expt}-X_{SM})/\sigma_X$ \\  \hline
 $A_l(lept)$ & 0.1501(16) & 1.07 & 0.1481 & 1.05  \\  \hline
 $\sbl$ & 0.25268(26) & 0.103 & 0.25277 & -0.35 \\   \hline
 $A_c$ & 0.653(20) & 3.06 & 0.668 & -1.06 \\   \hline
 $\sbc$ & 0.2897(50) & 1.73 & 0.2884 & 0.26 \\  \hline
 $A_b$ & 0.902(13) & 1.44 & 0.9347 &  -2.51  \\  \hline
 $\sbb$ & 0.3663(13) & 0.35 & 0.3648 &  1.15 \\  \hline
 $\sbn$ & 0.5006(12) & 0.24 & 0.5050 & -3.7 \\  \hline
 $\sbnbp$ & 1.3211(43) & 0.33 & 1.3218 & -0.16 \\  \hline
 $m_W$ & 80.426(34) & 0.042 & 80.394 & 0.94 \\  \hline         
\end{tabular}
\caption[]{{ \sl Model-independent electroweak observables.
   The SM predictions correspond  to $m_t = 178$ GeV, $m_H = 120$ GeV and
   $\alp = 0.007755$.}} 
\end{center}
\end{table}

%\begin{table}
%\begin{center}
%\begin{tabular}{|c|c|c|c|} \hline
%  Coupling & Expt value & SM  &(Exp-SM)/$\sigma$ \\  \hline
% $\vbl$ & -0.03783(41) & -0.03709 & -1.8 \\ 
% $\abl$ & -0.50125(26) & -0.50128 & 0.12 \\
% $|\vbn| = |\abn|$ & 0.50068(75) & 0.50240 & -2.3  \\ 
% $\vbc$ & 0.1875(68) & 0.1919  & -0.65 \\
% $\abc$ & 0.5045(50) & 0.5015  & 0.60 \\
% $\vbb$ & -0.3232(78) & -0.3435 & 2.6 \\
% $\abb$ & -0.5133(50) & -0.4983  & -3.0 \\ \hline     
%\end{tabular}
%\caption[]{{ \sl Effective coupling constants of the Z boson to lepton 
%  neutrino and heavy quark pairs.
%  The SM predictions correspond to $m_t$ = 174.3 GeV, $m_H$ = 120 GeV and
%   $\alp$ = 0.007755.}} 
%\end{center}
%\end{table}

\begin{table}
\begin{center}
\begin{tabular}{|c|c|c|c|} \hline
  Coupling & Expt value & SM  &(Exp-SM)/$\sigma$ \\  \hline
 $\vbl$ & -0.03783(41) & -0.03734 & -1.2 \\ 
 $\abl$ & -0.50125(26) & -0.50137 & 0.46 \\
 $|\vbn| = |\abn|$ & 0.5003(6) & 0.50251 & -3.7  \\ 
 $\vbc$ & 0.1875(69) & 0.1921  & -0.67 \\
 $\abc$ & 0.5045(50) & 0.5015  & 0.60 \\
 $\vbb$ & -0.3232(78) & -0.3435 & 2.6 \\
 $\abb$ & -0.5133(50) & -0.4983  & -3.0 \\ \hline     
\end{tabular}
\caption[]{{ \sl Effective vector and axial-vector coupling constants of the Z boson to lepton 
  neutrino and heavy quark pairs.
  The SM predictions correspond to $m_t$ = 178 GeV, $m_H$ = 120 GeV and
   $\alp$ = 0.007755.}} 
\end{center}
\end{table}

\begin{table}
\begin{center}
\begin{tabular}{|c|c|c|c|} \cline{2-4} 
\multicolumn{1}{c}{ } &\multicolumn{1}{|c}{ Coupling}
    &\multicolumn{1}{|c|}{ Value }  &\multicolumn{1}{|c|}{ (Exp-SM)/$\sigma$}\\  \hline
     & $\gbr$ & 0.0774 & $-$ \\
  SM    &   &   &   \\
    & $\gbl$ & -0.4209 & $-$ \\ \hline
  1996  & $\gbr$ & 0.1098(101) &  3.2 \\
  data   &   &   &   \\
  \cite{EWWG96}   & $\gbl$ & -0.4155(30) & 1.8 \\ \hline
  1998  & $\gbr$ & 0.1050(90) &  3.1 \\
  data   &   &   &   \\
  \cite{EWWG98}  & $\gbl$ & -0.4159(24) & 2.1 \\ \hline
  2003  & $\gbr$ & 0.0951(63) &  2.8 \\
  data   &   &   &   \\
  \cite{EWWG}  & $\gbl$ & -0.4182(16) & 1.7 \\ \hline      
\end{tabular}
\caption[]{{ \sl History of measurements of $\gbr$ and $\gbl$.
  The SM predictions correspond to $m_t$ = 178
 GeV, $m_H$ = 120 GeV and
   $\alp$ = 0.007755.}} 
\end{center}
\end{table}

\begin{table}
\begin{center}
\begin{tabular}{|c|c|c|c|} \hline 
$X$ & 1996 & 2003 & $[X(2003)-X(1996)]/ \sigma_X(2003)$ \\  \hline
$\Afbb$(LEP) & 0.0979(23) & 0.0997(16) & 1.1 \\
 $\sbb$(LEP) & 0.3676(24) & 0.3663(13) & -1.0 \\
  $A_b$(SLC) & 0.863(49) & 0.925(20) & 3.1 \\
  $A_l$(LEP) & 0.1466(33) & 0.1482(26) & 0.62  \\
  $A_l$(SLC) & 0.1543(37) & 0.1513(21) & -1.4  \\
  $A_l$(LEP+SLC) & 0.1501(24) & 0.1501(16) & 0.0  \\   \hline         
\end{tabular}
\caption[]{{ \sl Observables in the 1996 and 2003 data sets contributing
 to measurements of the b-quark effective coupling constants.}} 
\end{center}
\end{table}

   \par The experimental values of the observables in Table 14 are compared 
     with the SM prediction for $m_t = 178$ GeV, $m_H = 120$ GeV and
    $\alp = 0.007755$. This choice of $m_H$ (just above the experimental 
     lower limit) is near to the maxima of the $A_l(lept)$ and $A_l(all)$ curves
       of $\log\overline{{\rm CL}}$ plotted in Figs.2 and 3. Note that the model-independent
    analysis corresponds to only the $A_l(lept)$ curves in Figs.2-6. Only by making the
    stronger assumption of the SM values of the Z couplings to quarks, is it possible to
     derive  $A_l(had)$ and  $A_l(all)$.
  \par The agreement with the SM predictions shown in Table 14 is not completely
   satisfactory. The largest deviations are for $\sbn$ ( -0.84$\%$ and $3.7\sigma$) and $A_b$
  (-3.5$\%$ and $2.5\sigma$). The positive 1.05$\sigma$ and 0.94$\sigma$ deviations 
   of both $A_l$ and $m_W$
    respectively
   reflect the fact that the central values of $m_H$ preferred by these observables
     \footnote{ These fitted values of $m_H$ are for $\alp = 0.007755$ and $m_t = 178$ GeV}:
   \[  m_H = 72.5_{-24.2}^{+36.4}~~~~~{\rm fit~of}~A_l(lept)~{\rm only} \] 
  \[  m_H = 65.3_{-37.8}^{+59.9}~~~~~{\rm fit~of}~m_W~{\rm only} \]
     are incompatible with the direct lower limit of $m_H = 114.4$ GeV.
    \par The effective vector and axial vector couplings of charged leptons,
     neutrinos, c quarks and b quarks, that may be directly derived from
    the observables $\sbf$ and $A_f$ ($f = l,\nu, Q$) are presented in Table 15,
    in comparison with SM predictions. The agrement with the SM is satisfactory
    for the charged leptons and c quarks, but the neutrino couplings show a 
    $3.7\sigma$, $\vbb$ a $2.5\sigma$ and $\abb$ a $2.9\sigma$ deviation.
    As previously pointed out~\cite{JHF1,JHF2} the apparently anomalous 
    behaviour of the b-quark couplings is essentially found in the right-handed
    effective coupling, $\gbr$ rather than the left-handed one, $\gbl$ where:
    \begin{eqnarray}
     \gbr & = & \frac{\vbb-\abb}{2} \\
     \gbl & = & \frac{\vbb-\abb}{2} 
    \end{eqnarray}
  It is interesting to consider the history of this apparent anomaly, which is illustrated in Table 16.
  The most significant deviation (42$\%$ and 3.2$\sigma$) was seen in the 1996
  data set~\cite{EWWG96}. In the current, essentially final, data set the size of the effect is 
  reduced to 32$\%$ but still has a significance of 2.8$\sigma$. The left-handed coupling is now
  slightly more consistent with the SM (1.7$\sigma$ deviation) as compared to 1998 (2.1$\sigma$ deviation) and 
  1996 (1.8$\sigma$ deviation). The experimental error on $\gbr$ is reduced by $\simeq 40\%$ in the current
  data set as compared to that of 1996. The sources of these changes are made clear by the 
  entries of Table 17 in which are presented the model-independent observables used  to calculate
   the b-quark couplings, as derived from the 1996 and 2003 data sets. Also shown are the shifts 
   of the observables in units of the 2003 experimental errors. It can be seen that the most
   important change occurs in the direct measurement of $A_b$ from SLC. The value of $A_l$(LEP+SLC)
   used to extract  $A_b$(LEP) from $\Afbb$ is the same for the two data sets.

   \par To date, only a few authors~\cite{CCM,CTW,HV} have proposed new physics interpretations 
    of the measured b-quark couplings. Other authors~\cite{ACGGR,GAMG} have argued that there
    is unlikely to be a new physics interpretation of the observed anomaly. The present writer
   finds all the reasons given for this conclusion to be either simply wrong, or unconvincing.
   It was argued that: `the sensitivity of $\Afbb$ to $A_b$ is small, because $A_l$ is small'.
   In fact, the size of $A_l$ is irrelevant. What are important are the {\it relative precisions}
   with which it and $\Afbb$ are known. Since the errors on $A_l$ and $\Afbb$ are uncorrelated
    it follows from Eqn(3.1) that:
    \begin{equation}
    \frac{\sigma(A_b)}{A_b} = \sqrt{\left( \frac{\sigma(A_l)}{A_l}\right)^2
      +\left( \frac{\sigma(\Afbb)}{\Afbb}\right)^2} 
    \end{equation}
     where $\sigma(X)$ is the experimental error on $X$. As shown in Tables 1 and 6, 
      $A_l$ and $\Afbb$ have relative uncertainies of 1.2$\%$ and 1.6$\%$ respectively. Since
     the observed deviation from the SM is 3.6$\%$, the contribution to the uncertainity
     on $A_b$ due to that on $A_l$ is essentially negligible in comparison with the the observed
   deviation from the SM prediction. It is further argued in Reference~\cite{ACGGR}
   that a new physics interpretation is disfavoured because a significant deviation
   is seen only in $\Afbb$ from LEP and not in $R_b$ (equivalent to $\sbb$) or $A_b$(SLC). But, 
   in the case of a deviation from the SM only in the right-handed coupling, no significant
   change is expected in $R_b$\footnote{ Since $\sbb \simeq 2[(\gbr)^2+(\gbl)^2]$ and,
     with the SM values of the couplings: $(\gbr)^2= 0.005$, $(\gbl)^2= 0.18$,
     a 100$\%$ deviation of $\gbr$ from the SM prediction changes $\sbb$ (or $R_b$) by only
     8$\%$.}. Also the measured value of $A_b$(SLC) lies 1.5$\sigma$ above  $A_b$(LEP)
     and 0.5$\sigma$ below the SM prediction. Furthermore $\chi^2_{A_b,WA}/d.o.f. = 2.32/1$,
     CL=0.13. There is therefore no strong evidence for any incosistency between the measured
    values of $A_b$(SLC) and $A_b$(LEP). Both values are included in the weighted average
     that differs by 2.5$\sigma$ from the SM prediction. In Reference~\cite{ACGGR} 
    it is stated that: `One concludes that most probably the observed discrepancy is due
     to a large statistical fluctuation and/or an experimental problem'. The discussion in
     Section 3 above reaches just the opposite conclusion. In fact the statement just quoted
    tacitly implies (without any justification) that all the LEP experiments have seriously
    underestimated the systematic arrors of their $\Afbb$ measurements; indeed, as discussed
    in Section 3 above, by at least an order of magnitude. This may be possible, but hardly
    seems likely. The total estimated experimental error on the LEP value of $\Afbb$ is
    is largely statistical. In contrast the $R_b$ measurement has a statistical error of
     0.20$\%$ to be compared with a systematic one of 0.22$\%$, so that the estimated systematic
    contribution is much more important than that for $\Afbb$. Indeed from general experimental
    considerations, asymmetries such as  $\Afbb$ are expected to have smaller
    systematic uncertainities
    than quantities such as $R_b$ where absolute experimental detection efficencies play a
     role. Contrary to what is implied in Reference~\cite{ACGGR} then, there is no
    objective reason to suppose that the $\Afbb$ measurement should be less reliable than
    the $R_b$ one. As discussed in Section 3, the hypothesis that the $A_b$ deviation is
    a purely statistical effect has a CL of $\simeq 10^{-3}$ and so, though not completely
    excluded, is very unlikely. It remains true however that a correlated systematic error 
   of unknown origin in the LEP $\Afbb$ measurements is expected to produce an anomaly
    predominantly in the right-handed effective coupling. Assuming SM values for the couplings
   and that the -3.5$\%$ discrepancy in $A_b$ is of systematic origin, the derived values of the
    couplings: $\gbr = 0.09486$ and  $\gbl = -0.4187$ are in good agreement with the measured
    values in Table 16. Another argument~\cite{JHF2} in favour of an unknown systematic origin
    for the $A_b$ discrepancy is to note the good agreement of the measured value of $\sbb$
    with the SM prediction shown in Table 14. This agreement requires the presence of
    of large, $m_t^2$ dependent, quantum corrections originating in the strong breaking
    of quark flavour symmetry in the third generation of SM fermions. Neglect of these
     corrections gives a prediction of 0.3707 for $\sbb$, differing from the measured
    value by 3.7$\sigma$. In the case of a new physics explanation of the $A_b$
    discrepancy, the appearence of the expected quantum corrections from the SM
    for $\sbb$ must be regarded as fortuitous. Thus, although there are no
    objective experimental reasons to doubt the correctness of the $A_b$ measurement,
    a systematic effect of unknown origin cannot be excluded. The good agreement of
     $\sbb$ with the SM prediction and the large observed deviation in the right-handed 
    coupling are consistent with this hypothesis. A purely statistical fluctuation
    is very unlikely. The effect could also be explained by new physics. There are
    no good reasons for the statement in Reference~\cite{ACGGR} that: `It is well
    known that this ($A_b$) discrepancy is not likely to be explained by some
    new physics effect in the $b\overline{b}$Z vertex'. In fact all the explanations mentioned
    above (including new physics) remain open possibilities. Only better
    experimental data can decide between them.

\begin{table}
\begin{center}
\begin{tabular}{|c|c|c|} \hline
 $\Gamma_{had}$(thy) definition & $\Gamma_{had}$ [GeV] &
 [$\Gamma_{had}$(thy)-$\Gamma_{had}$(expt)]/$\sigma$(expt) \\   \hline
  SM & 1.7439 & -0.4  \\
  b $\rightarrow$ b(meas) & 1.7451 & 0.35 \\
  b,d $\rightarrow$ b(meas) & 1.7414 & -1.50 \\
  b,d,s $\rightarrow$ b(meas) & 1.7377 & -3.4 \\  \hline
 $\Gamma_{had}$(expt) & 1.7444(20) &   $-$ \\  \hline     
\end{tabular}
\caption[]{{ \sl Constraints on the Z couplings to d-type quarks 
  from the LEP average measurement of $\Gamma_{had}$.}} 
\end{center}
\end{table}

    \par The next question that obviously arises is whether there is any evidence that
   the couplings of non-b quarks may also deviate from the SM predictions. Direct 
    measurements of the light quark couplings have been performed by the DELPHI~\cite{DELPHIlq}
    and OPAL~\cite{OPALlq} Collaborations. For example, OPAL found:
    \[ \overline{g}_{\rm{d,s}}^L = -0.44_{-0.09}^{+0.13}~~,~~
        \overline{g}_{\rm{d,s}}^R = 0.13_{-0.17}^{+0.15} \]
      in good agreement with the SM predictions of -0.424 and 0.077 respectively. However,
    the very large uncertainties on the measurements preclude obtaining any useful information
    concerning deviations at the few $\%$ level. such as that observed for the parameter $A_b$.
     In fact the observed deviations of $A_b$ and $A_c$ from the SM predictions by factors of 1.036 and
    1.023 are of comparable size. Since the relative error on $A_c$ is two times larger than that
    on $A_b$, only for the latter is a possibly significant deviation from the SM observed.
     The similar qualitative behaviour of  $A_b$ and $A_c$ with respect to the SM prediction
    can be seen in Fig.15.1 of Reference~\cite{EWWG}. The LEP measurement of $Q_{FB}^{had}$ can be
    used to extract an average value of $A_{\overline{q}}$ (averaged over all quark flavours) that
    is 1.07(7) times the SM prediction for this quantity. Thus, as previously mentioned, the 
    different quark charge asymmetry measurements do not exclude deviations of $A_q$ ($q = u,d,s,c$)
    as large as that observed for $A_b$. The present data are not, however, sufficiently precise
    to give any positive evidence for such an effect.
    \par As pointed out in Reference~\cite{JHF4}, much stronger constraints on the non-b quark
    couplings are provided by the LEP average value of the hadronic width, $\Gamma_{had}$, of the 
     Z boson~\cite{EWWG}:
      \[ \Gamma_{had} = 1.7444(20)~{\rm GeV} \]
      This value is in excellent agreement with the SM prediction\footnote{Unless otherwise stated,
      all SM predictions are for $m_t = 178$ GeV, $m_H = 120$ GeV and $\alp = 0.007755$.} of 1.7439 GeV
      (0.03 $\%$, 0.25$\sigma$ deviation). One may also note in Table 14 the almost perfect agreement
     of the $\sbnbp$ measurement with the SM prediction. The small relative uncertainty of 0.11$\%$
     on $\Gamma_{had}$ allows significant constraints to be placed on different hypotheses concerning
     the size of the Z$q \overline{q}$ couplings. Using the relation:
     \begin{equation}
       \Gamma_{had} = \frac{\sqrt{2} G_{\mu} m_Z^3}{4 \pi} \sum_{q}^{u,d,s,c,b}
         \sbq C_{q}^{QED} C_{q}^{QCD}
      \end{equation}
       the SM predictions for the $d$ and $s$ quarks may be replaced by the central values of the
       measured b-quark couplings from Table 14. The results given by replacing, in Eqn(7.7), the
       SM predictions for (i) $b$ quarks, (ii) $b$ and $d$ quarks and (iii) $b$, $d$  and $s$ quarks 
      by the measured b-quark couplings from Table 14 are presented in Table 18, in comparison
       with the measured value of $\Gamma_{had}$. It can be seen that, although the prediction
     is little changed for case (i), case (iii) is excluded by the measured value of $\Gamma_{had}$
     at the 3.4$\sigma$ level. 
     \par Since (see Table 14) the measured value of $\sbc$ agrees well with the SM prediction,
      the measured value of  $\Gamma_{had}$ will provide no useful constraints if the procedure
       used in Table 18 is repeated fou u-type quarks. The effective coupling constants
      $\vbc$ and $\abc$ also agree well with the SM predictions.
     \par The experimental situation concerning measurements of right-handed and left-handed
        Z-fermion pair couplings and the W boson mass is summarised in Tables 19 and 20.
       In Table 19 the couplings of charged leptons, c quarks, b quarks and neutrinos are
     compared with SM predictions. Similar comparisons are made in Table 20, varying the values
   of $m_t$ and $m_H$ in the SM predictions. In this case, for clarity, only deviations from the
    SM predictions are tabulated. Tables 19 and 20 contain, in concise form, essentially all the
     precision information on the SM derived from the experimental programmes of LEP and SLD as
     well as the main contributions of FERMILAB to the same subject (essentially measurements of $m_t$, 
      $m_W$ and the Z$\nu \overline{\nu}$ coupling) during the same period. 
     \par Looking at this comparison, it is difficult to conclude that the level of agreement
     with the SM is good. For $m_H = 120$ GeV, i.e. around the maximum value of  $\overline{{\rm CL}}$,
     as in Table 19, $\gbr$ and $\gnl$ show deviations of 2.8 and -3.7 standard deviations
     respectively. All of $\glr$, $\gll$ and $m_W$ show negative deviations around the one standard
     deviation level as a consequence of the low values of $m_H$ (inconsistent with the experimental 
     direct lower limit) favoured by the measured values of these quantities. For $m_H = 300$ GeV
       (see Table 20) five out of the eight EW parameters show deviations around three standard 
     deviations. Those associated with $\gbr$ and $\gnl$ are almost independent of $m_H$ and 
     vary only weakly with $m_t$. Another feature is that increasing (decreasing) the value of $m_t$
     improves (worsens) the agreement for $\glr$ and $\gll$ ($m_W$) for all values of $m_H$. 
     In any case, the SM still fares badly for $m_H = 300$ GeV and above. This is aleady
    apparent in Figs.2 and 3, and will also be evident in the combined confidence level curves based
     on all precision data to be discussed below.
     \par Perhaps the most important aspect of the SM that has been tested by recent precision 
     measurements is the renormalisability of the theory. This enables quantum loop corrections
    involving fermions, weak bosons and the Higgs boson to be calculated and compared to experiment,
     just as precise measurements of similar effects in QED, such as the Lamb shift of Hydrogen and the
     anomalous magnetic moments of the electron and muon, were important to establish the essential
     correctness of the theory for the description of such quantities. The model-independent
   observables shown in Table 14 can be used to isolate the effect of quantum corrections in the coupling
    of charged  leptons, c quarks, b quarks and neutrinos to the Z boson. For this only the values of
    $A_l(lept)$, $A_c$ and $A_b$ are required for charged leptons and heavy quarks, and that of $\sbn$
     for neutrinos. The $A$ parameters are all simple mappings (see Eqn(3.2)) of the ratio:
     $r =v/a$ of the vector and axial-vector coupling constants. At  tree-level in the SM the
     following relations hold:
     \begin{eqnarray}
     r_l & = & 1-4s_W^2  \\  
     r_c & = & 1-\frac{8}{3}s_W^2  \\  
     r_b & = & 1-\frac{4}{3}s_W^2  \\
     g_{\nu}^L & = & \frac{1}{2}
     \end{eqnarray}
     where $s_W^2 = \sin^2\theta^{on-shell}_W$ as defined in Eqn(6.2). In the presence of 
     quantum corrections  $r_f \rightarrow \overline{r}_f$, and  $s_W^2 
      \rightarrow (\overline{s}_W^2)^f$ ($f = l,c,b$) in Eqns(7.8)-(7.10) and $g_{\nu}^L  \rightarrow
      \gnl$ in Eqn(7.11). Thus parameters, $\delta_{Quant}^f$, that measure directly the effect of
   quantum corrections can be introduced according to the definitions:
    \begin{eqnarray}
    \delta_{Quant}^l & \equiv & \frac{(\overline{s}_W^2)^l - s_W^2}{s_W^2} =
        \frac{1- \overline{r}_l}{4 s_W^2}-1  \\
    \delta_{Quant}^c & \equiv & \frac{(\overline{s}_W^2)^c - s_W^2}{s_W^2} =
        \frac{3(1- \overline{r}_c)}{8 s_W^2}-1  \\
    \delta_{Quant}^b & \equiv & \frac{(\overline{s}_W^2)^b - s_W^2}{s_W^2} =
        \frac{3(1- \overline{r}_b)}{4 s_W^2}-1  \\
  \delta_{Quant}^{\nu} & \equiv & (\gbl-\frac{1}{2})/ (\frac{1}{2}) = 2\gbl-1 
   \end{eqnarray}
    In the absence of quantum corrections all the $\delta_{Quant}^f$ parameters vanish.
     The experimental values of these parameters derived from the entries of Table 14, as well as
    the corresponding SM predictions, are presented in Table 21. 
    \par It can be seen from this table that the expected size of the corrections in the SM
    is $\simeq 4-5 \%$ for  $l$, $c$ and $b$, and an order of magnitude lower for $\nu$. 
     By far the most significant measurement of quantum corrections (44 standard deviations from zero!)
    is that of $\delta_{Quant}^l$. Good agreement with the SM prediction (1.2$\sigma$ deviation) is
     found, at least for $m_H = 120$ GeV, as used in the SM predictions in Table 21. As discussed
     previously, in connection with $\glr$ and $\gll$ (see Table 20), the agreement worsens considerably 
     for higher values of $m_H$. For example, for $m_H = 300$ GeV, $\delta_{Quant}^l$(SM) = 0.04385,
     a deviation of 3.5$\sigma$ from the measured value.
     For c quarks the expected quantum correction is only a 1.7$\sigma$
    effect. Good agreement with the SM prediction is found in this case. For b quarks the expected
     quantum correction is unmeasurable with present data (expected effect 0.64$\sigma$ from zero)
     whereas a 3.4$\sigma$ effect is actually observed, differing by 2.7$\sigma$ from the SM
     prediction. This is indicative, as discussed previously, of either a large and unknown correlated
     systematic effect (perhaps in combination with a statistical fluctuation) in the LEP
      $\Afbb$ measurements, or of new physics at tree-level, or, indeed some combination of
     the two. A purely statistical fluctuation is very unlikely. The situation is exactly the
     opposite for the neutrino couplings. The SM predicts a large (4.2$\sigma$) quantum 
     correction, while in the data only a 0.5$\sigma$ effect is seen. The discrepancy
     with the SM amounts to -3.7$\sigma$ in this case. Rather than a tree-level effect
     giving a large apparent quantum correction, as for the b quarks, it seems that the
     SM quantum corrections are effectively `turned off' in the case of the the Z$\nu
     \overline{\nu}$ couplings! Theoretical interpretations of this apparent coupling
    suppression have been made in References~\cite{LOTW,LORTW}.

     \par The main conclusion to be drawn from the results presented in Table 21 is that
     only for the charged leptons and c quarks is the present data reasonably consistent with the SM.
     This implies that since the c-quark couplings are almost completely insensitive to the values of
     $m_t$ and $m_H$ (see Table 5) information on these parameters, via quantum corrections, can
    only be reliably obtained from the charged lepton couplings. This further implies (see Tables 11
    and 12) that the maximum values of   $\overline{{\rm CL}}$ of 23$\%$ or 1.7$\%$ are obtained
     at $m_H = 115$ GeV when the NuTeV $m_W$ measurement is excluded or included respectively.
      Much lower confidence levels of 0.054 and 0.0011 are obtained, under the same conditions,
     for $m_H = 220$ GeV, the presently quoted 95$\%$ CL upper limit on $m_H$ from the
     EWWG~\cite{EWWG}. In fact, even lower CLs will be found for a global analysis based on
    all model-independent observables shown in Table 14, which will now be discussed.

\begin{table}
\begin{center}
\begin{tabular}{|c|c|c|c|} \hline
  SM parameter & Expt value & SM  &(Exp-SM)/$\sigma$ \\  \hline
 $\glr$ & 0.23171(25)  & 0.23202 & -1.2 \\
  $\gll$ & -0.26954(23)  & -0.26935 & -0.83 \\   \hline
 $\gcr$ & -0.1585(48)  & -0.1547 & -0.79 \\
  $\gcl$ & 0.3460(36)  & 0.3468 & -0.22 \\   \hline
 $\gbr$ & 0.0951(63)  & 0.0774 & 2.8 \\
  $\gbl$ &  -0.4182(16)  &  -0.4209 & 1.7 \\   \hline
 $\gnl$ & 0.5003(6) & 0.50251 & -3.7 \\  \hline
 $m_W$[GeV] & 80.426(34) &  80.394 & 0.94  \\ \hline     
\end{tabular}
\caption[]{{ \sl Measured values of precision electroweak parameters compared to
 SM predictions for  $m_t = 178$ GeV, $m_H = 120$ GeV and $\alp = 0.007755$.}} 
\end{center}
\end{table}

 %^&
\begin{table}
\begin{center}
\begin{tabular}{|c|c|c|c|c|c|c|c|c|c|} \cline{2-10}
 \multicolumn{1}{c}{ } & \multicolumn{3}{|c|}{ $m_H = 120$ GeV} 
  &  \multicolumn{3}{c|}{ $m_H = 200$ GeV} &   \multicolumn{3}{c|}{ $m_H = 300$ GeV}   \\  \hline
 \multicolumn{1}{|c}{$m_t$ [GeV]} &  \multicolumn{3}{|c|}{173.8~~178.0~~182.2}
  &  \multicolumn{3}{c|}{173.8~~178.0~~182.2}  &  \multicolumn{3}{c|}{173.8~~178.0~~182.2}  \\  \hline \hline
 \multicolumn{1}{|c}{ $\glr$ } &  \multicolumn{3}{|c|}{ -1.6~~-1.2~~-0.82}
  &  \multicolumn{3}{c|}{ -2.4~~-2.1~~-1.7}  &  \multicolumn{3}{c|}{ -3.1~~-2.8~~-2.4}  \\
 \multicolumn{1}{|c}{ $\gll$ } &  \multicolumn{3}{|c|}{ -1.6~~-0.83~~0.03}
  &  \multicolumn{3}{c|}{-3.0~~-2.1~~-1.3}  &  \multicolumn{3}{c|}{ -4.0~~-3.2~~-2.4}  \\ \hline
 \multicolumn{1}{|c}{ $\gcr$} &  \multicolumn{3}{|c|}{ -0.77~~-0.79~~-0.80}
  &  \multicolumn{3}{c|}{ -0.75~~-0.76~~-0.77}  &  \multicolumn{3}{c|}{ -0.72~~-0.73~~-0.75}  \\
 \multicolumn{1}{|c}{ $\gcl$} &  \multicolumn{3}{|c|}{-0.19~~-0.22~~-0.27}
  &  \multicolumn{3}{c|}{-0.11~~-0.16~~-0.2}  &  \multicolumn{3}{c|}{ -0.06~~-0.10~~-0.15}  \\ \hline
\multicolumn{1}{|c}{ $\gbr$ } &  \multicolumn{3}{|c|}{ 2.81~~2.82~~2.82}
  &  \multicolumn{3}{c|}{2.80~~2.80~~2.81}  &  \multicolumn{3}{c|}{ 2.79~~2.79~~2.80}  \\
\multicolumn{1}{|c}{ $\gbl$ } &  \multicolumn{3}{|c|}{ 1.71~~1.70~~1.64}
  &  \multicolumn{3}{c|}{  1.60~~1.57~~1.54}  &  \multicolumn{3}{c|}{1.52~~1.49~~1.46}  \\ \hline
\multicolumn{1}{|c}{$\gnl$ } &  \multicolumn{3}{|c|}{ -3.5~~ -3.7~~-3.9}
  &  \multicolumn{3}{c|}{-3.3~~-3.5~~-3.7}  &  \multicolumn{3}{c|}{ -3.2~~-3.4~~-3.6}  \\ \hline
\multicolumn{1}{|c}{ $m_W$} &  \multicolumn{3}{|c|}{ ~1.7~~~0.94~~~0.18}
  &  \multicolumn{3}{c|}{ 2.6~~~1.9~~~1.1}  &  \multicolumn{3}{c|}{ 3.4~~~2.7~~~1.9}  \\ \hline

%  $\glr$ & -1.6 & -1.2 & -0.82 & -2.4 & -2.1 & -1.7 & -3.1 & -2.8 & -2.4 \\
%  $\gll$ & -1.6 & -0.83 & 0.03 & -3.0 & -2.1 & -1.3 & -4.0 & -3.2 & -2.4 \\ \hline
%  $\gcr$ & -0.77 & -0.79 & -0.80 & -0.75 & -0.76 & -0.77 & -0.72 & -0.73 & -0.75 \\
%  $\gcl$ & -0.19 & -0.22 & 0.27 & -0.11 & -0.16 & -0.20 & -0.06 & -0.10 & -0.15 \\ \hline
%  $\gbr$ & 2.81 & 2.82 & 2.82 & 2.80 & 2.80 & 2.81 & 2.79 & 2.79 & 2.80 \\
%  $\gbl$ & 1.71 & 1.70 & 1.64 & 1.60 & 1.570 & 1.54 & 1.52 & 1.49 & 1.46 \\ \hline
% $\gnl$ & -3.5 & -3.7 & -3.9 & -3.3 &  -3.7 & -3.7 & -3.2 & -3.4 & -3.6 \\ \hline
% $m_W$ & 1.7 & 0.94 & 0.18 & 2.6 &  1.9 & 1.1 & 3.4 & 2.7 & 1.9 \\ \hline
\end{tabular}
\caption[]{{ \sl Values of deviations (Expt-SM)/$\sigma$ for precision
 electroweak parameters. SM predictions with $\alp = 0.007755$.}} 
\end{center}
\end{table}

 \begin{table}
\begin{center}
\begin{tabular}{|c||c|c|c|c|c|} \hline 
 $f$ & $\delta_{Quant}^f(Expt)$ & $\delta_{Quant}^f(Expt)/\sigma_f$ &
 $\delta_{Quant}^f(SM)$ & $\delta_{Quant}^f(SM)/\sigma_f$ & $[Expt-SM]/\sigma_f$ \\  \hline  \hline 
 $l$ & 0.04064(92) & 44.2 & 0.04118 & 44.8 & -0.59 \\
  $c$ & 0.060(25) & 2.4 & 0.0418 & 1.67 & 0.76 \\
  $b$ & 0.249(74) & 3.4 & 0.049 & 0.65 & 2.7 \\
  $\nu$ & 0.0006(12) & 0.5 & 0.00502 & 4.2 & -3.7 \\   \hline  
\end{tabular}
\caption[]{{ \sl Quantum correction parameters for different fermion
 flavours f.
 SM predictions for  $m_t = 178$ GeV, $m_H = 120$ GeV and $\alp = 0.007755$.}} 
\end{center}
\end{table}

\begin{table}
\begin{center}
\begin{tabular}{|c|c|c|c|} \cline{2-4} 
\multicolumn{1}{c}{ } &\multicolumn{1}{|c}{ NuTeV out }
    &\multicolumn{1}{|c|}{ NuTeV  $\sbn$ meas.} & \multicolumn{1}{c|}{ NuTeV  $m_W$ meas.} \\  \hline
\multicolumn{1}{|c}{$m_H$ (GeV) } &\multicolumn{1}{|c|}{ } & \multicolumn{1}{c|}{ }
 &  \multicolumn{1}{c|}{ } \\ \cline{1-1}
111 & $4.9 \times 10^{-7}$  &  $4.9 \times 10^{-8}$ & $3.2 \times 10^{-8}$ \\
113 & $2.0 \times 10^{-4}$  &  $2.1 \times 10^{-5}$ & $1.4 \times 10^{-5}$ \\
115 & $0.031$  &  $3.8 \times 10^{-3}$ &  $2.6\times 10^{-3}$ \\
140 &  $0.051$   & $5.1 \times 10^{-3}$ & $4.2 \times 10^{-3}$ \\
180 &  $0.053$  &  $5.9 \times 10^{-3}$ &  $6.4 \times 10^{-3}$ \\
220 & $0.041$  & $5.6 \times 10^{-3}$ &  $6.4 \times 10^{-3}$ \\
260 & $0.026$  & $2.9 \times 10^{-3}$  &  $5.0 \times 10^{-3}$ \\
300 &  $0.015$   &  $1.6\times 10^{-3}$ &  $3.4 \times 10^{-3}$ \\ \hline          
\end{tabular}
\caption[]{{ \sl Combined confidence levels $\overline{CL}$ for consistency with the 
    SM as a function of $m_H$. All data, as in Fig.6. }} 
\end{center}
\end{table}

 \begin{table}
\begin{center}
\begin{tabular}{|c|c|c|c|} \cline{2-4} 
\multicolumn{1}{c}{ } &\multicolumn{1}{|c}{ NuTeV out }
    &\multicolumn{1}{|c|}{ NuTeV  $\sbn$ meas.} & \multicolumn{1}{c|}{ NuTeV  $m_W$ meas.} \\  \hline
\multicolumn{1}{|c}{$m_H$ (GeV) } &\multicolumn{1}{|c|}{ } & \multicolumn{1}{c|}{ }
 &  \multicolumn{1}{c|}{ } \\ \cline{1-1}
111 & $1.3 \times 10^{-6}$  &  $7.4 \times 10^{-8}$ & $4.7 \times 10^{-8}$ \\
113 & $4.5 \times 10^{-4}$  &  $2.8 \times 10^{-5}$ & $1.9 \times 10^{-5}$ \\
115 & $0.061$  &  $4.7 \times 10^{-3}$ &  $3.3 \times 10^{-3}$ \\
140 &  $0.042$   & $2.2 \times 10^{-3}$ & $2.0 \times 10^{-3}$ \\
180 &  $0.010$  &  $5.3 \times 10^{-4}$ &  $7.4 \times 10^{-4}$ \\
220 &   $2.0 \times 10^{-3}$  & $1.0 \times 10^{-4}$ &  $2.1 \times 10^{-4}$ \\
260 & $3.5 \times 10^{-4}$  & $1.9 \times 10^{-5}$  &  $5.3 \times 10^{-5}$ \\
300 & $6.0 \times 10^{-5}$   &  $3.4 \times 10^{-6}$ &  $1.2 \times 10^{-5}$ \\ \hline          
\end{tabular}
\caption[]{{ \sl Combined confidence levels $\overline{CL}$ for consistency with the 
    SM as a function of $m_H$. Lepton data only, as in Fig.7. }} 
\end{center}
\end{table}
 
   \begin{figure}[htbp]
\begin{center}\hspace*{-0.5cm}\mbox{
\epsfysize10.0cm\epsffile{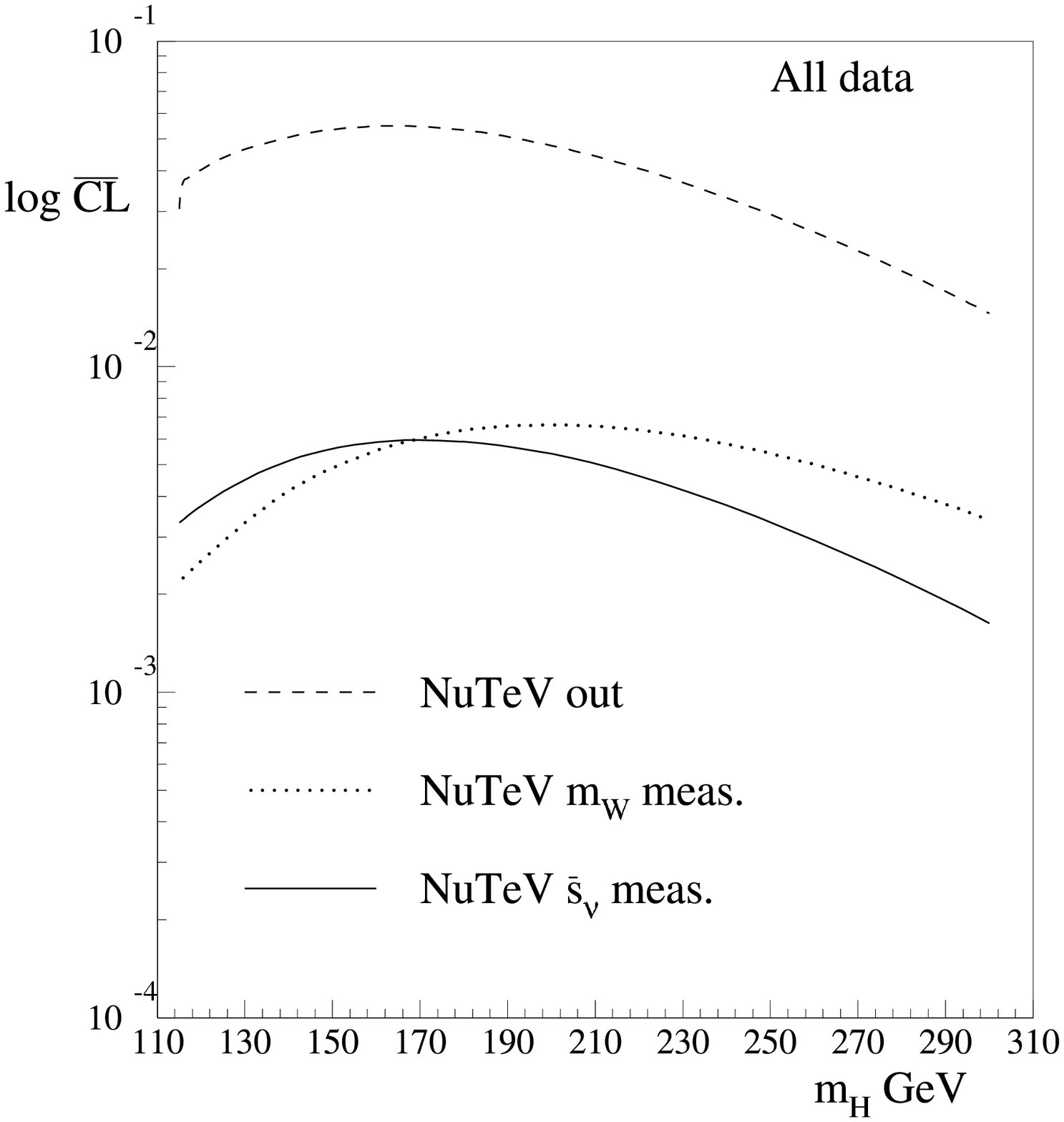}}
\caption{{ \sl Combined $m_H$ confidence levels. ${\rm CL}_{s+b}$ is combined with
  CL($\chi^2_{SM}$) using Eqn(2.8). $\chi^2_{SM}$ is calculated from all model-independent
  variables: $A_l(lept)$, $\sbl$, $A_c$, $A_b$, $\sbb$, $\sbnbp$, $\sbn$, $m_W$ and
  $m_W$(NuTeV).}} 
\label{fig-fig6}
\end{center}
 \end{figure}
 
   \begin{figure}[htbp]
\begin{center}\hspace*{-0.5cm}\mbox{
\epsfysize10.0cm\epsffile{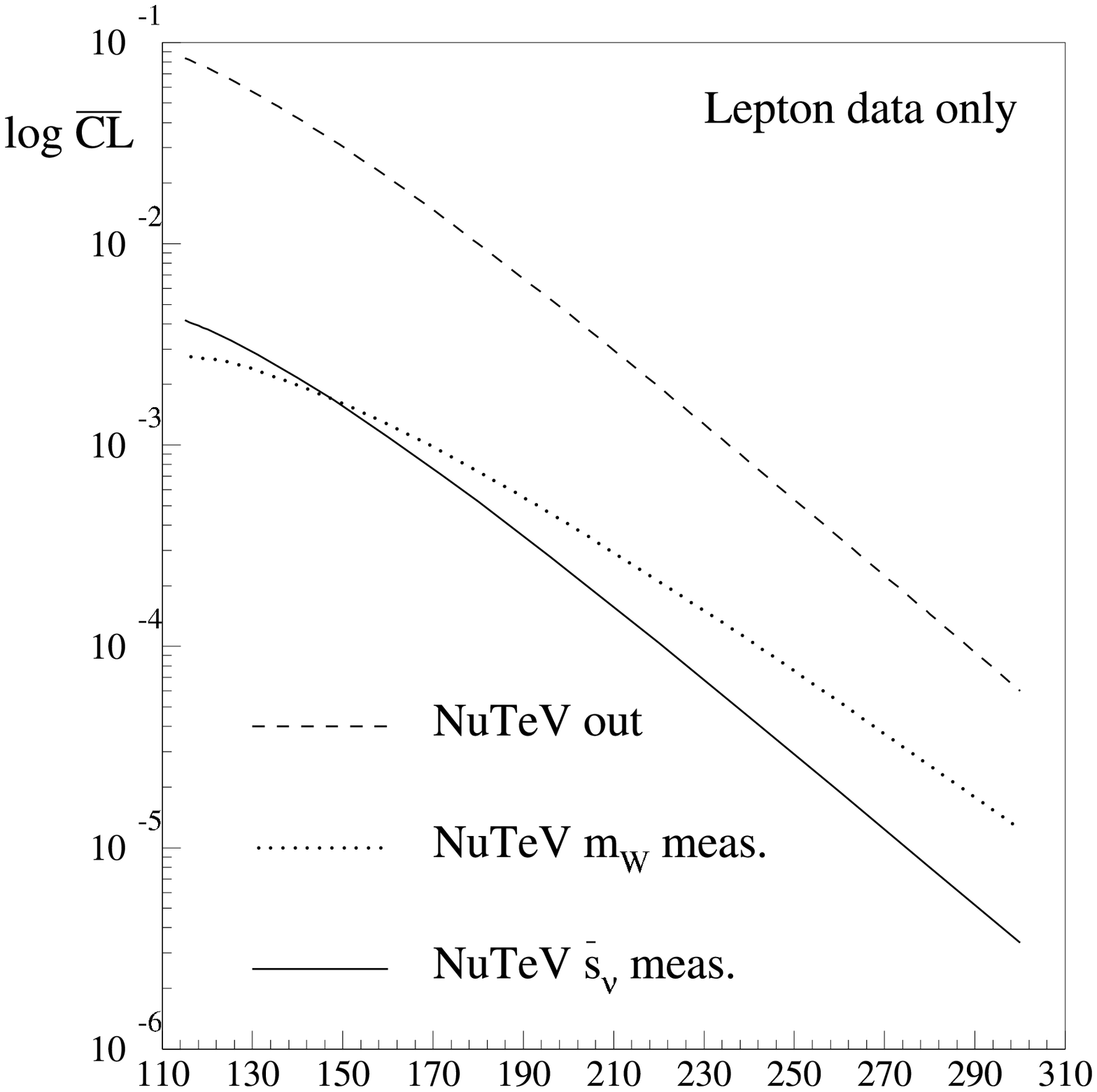}}
\caption{{ \sl Combined $m_H$ confidence levels. ${\rm CL}_{s+b}$ is combined with
  CL($\chi^2_{SM}$) using Eqn(2.8). $\chi^2_{SM}$ is calculated from the leptonic
  model-independent variables: $A_l(lept)$, $\sbl$ $\sbn$, as well as $m_W$ and
  $m_W$(NuTeV).}} 
\label{fig-fig7}
\end{center}
 \end{figure} 
 
     \par In order to test the overall level of agreement of the precision 
     data with the SM, $\chi^2$ estimators are calculated from the observables shown in
    Table 14. The corresponding confidence level CL($\chi_{SM}^2$) is then 
    combined with that, ${\rm CL}_{s+b}$, of the direct Higgs search using Eqn(2.8) to give 
    $m_H$ confidence level curves,  $\overline{{\rm CL}}$. Since all equivalent
    and statisically compatible measurements are combined in order to extract the 
    model-independent observables, there is, in this case, no contribution of the type
    $\chi_{X,WA}^2$ to take into account the degree of statistical compatiblity of 
    different measurements of the same quantity $X$. In view of the possibility of new physics
     or poorly understood systematic effects for the Z$b\overline{b}$ couplings, and of the
    overall status, as well as the different possible interpretations of the NuTeV experiment,
     the above procedure is repeated for different selections and interpretations of the data:
     the results are presented in Tables 22 and 23 and Figs.6 and 7.
      
     \par In Table 22 and Fig.6, all the observables in Table 14, except $\sbc$, are included
  in the $\chi^2$ estimator. In the case that the NuTeV result is excluded, $\sbn$ is assigned
  the LEP-only value of Table 5 and Eqn(6.7). The `NuTeV $\sbn$ meas.' curves use the $\sbn$ value quoted
   in Table 14 and Eqn(6.9) (LEP+ NuTeV average). For the interpretation of the NuTeV result as
    an $m_W$ measurement, $m_W$(NuTeV) is included in $\chi^2$ and  $\sbn$ is set to the 
     LEP-only value. The number of degrees of freedom of the `NuTeV out',`NuTeV $\sbn$ meas.' 
     and `NuTEV $m_W$ meas.' estimators are 10, 10 and 11 respectively.
 
     \par In the case of new physics, or unknown systematic effects, in the $A_b$ measurement,
     it is clearly of interest to test only the level of agreement of the leptonic sector with 
    the SM. For this, combined $m_H$ confidence level curves are also derived using only
    the leptonic observables: $A_l(lept)$, $\sbl$ and $\sbn$ as well as $m_W$ and, possibly
    $m_W$(NuTeV). The `NuTeV out',`NuTeV $\sbn$ meas.' 
     and `NuTeV $m_W$ meas.' cases are considered as previously, giving now 4, 4 and 5 degrees
    of freedom, respectively, for $\chi_{SM}^2$. The results for the `Lepton data only'
    case are presented in Table 23 and Fig.7.
   
   \par It can be seen from Table 22 and Fig.6 that, when all data is included, the maximum of
      $\overline{{\rm CL}}$ occurs around $m_H = 180$ GeV, and has a value $\simeq 0.05$ in the
    case that the NuTeV experiment is excluded and  $\simeq 0.0065$ when it is included,
    independently of the interpretation ($\sbn$ or $m_W$ measurement) of the experiment.
    For $m_H = 300$ GeV, the CLs are much lower, being 0.015, 0.0016 and 0.0034 for the
    `NuTeV out',`NuTeV $\sbn$ meas.' and `NuTeV $m_W$ meas.' cases respectively. In the case that
    only the leptonic Z-decay  data is considered, the maximum value of  $\overline{{\rm CL}}$
    occurs just above the direct lower limit with the values $\simeq 0.06$ (NuTeV out) and
    $\simeq 0.004$ (NuTeV included). For $m_H = 300$ GeV, very low values of
    $\overline{{\rm CL}}$ are found: $6.0 \times 10^{-5}$,  $3.4 \times 10^{-6}$
     and $1.2 \times 10^{-5}$ for the three cases considered previously.

  \SECTION{\bf{Comparison with the EWWG and EWPDG Global Fits}}  
%$#@
    The conclusions presented above concerning the global level of agreement with the SM, and the possible
   value of $m_H$ may seem somewhat at variance with those drawn from the global fits of the EWWG~\cite{EWWG} 
   and EWPDG~\cite{EWPDG}. Indeed, the quoted $\chi^2$ confidence levels of the latter are much higher.
   The EWWG quotes confidence levels of 4.5$\%$ for `all data' and 28$\%$ when NuTeV is excluded.
   These fits obtain (see Table 2) $m_H = 96_{-35}^{+60}$ GeV and  $m_H = 91_{-36}^{+55}$ GeV
    respectively. These central fit values are completely inconsistent with the experimental
    direct lower limit on $m_H$. Already for $m_H = 111$ GeV, ${\rm CL}_{s+b} = 10^{-6}$.
   In addition the fitted values of $m_H$ are strongly biased towards higher values, as
   discussed above, due to the inclusion of the `anomalous' quark asymmetry data in the fits. The latest
   global EWPDG fit, which did not include the NuTeV datum, found $m_H = 98_{-35}^{+51}$ GeV
 and $\chi^2/d.o.f. = 47.3/38$, CL$= 14\%$. Some reasons for the higher confidence levels found for
   the global EWWG and EWPDG fits have been put forward in Section 2 above, so it is interesting,
 in the light of that discussion, to examine in detail the contributions of different types of
  observables to the $\chi^2$ of these fits.
   \par The 20 observables inclused in the EWWG fit may be classified as follows:

   \begin{itemize}
     \item[ ] Measured: $\Delta_{had}^{(5)}(m_Z^2)$, $m_Z$, $m_t$.
      \item[ ] $m_H$-sensitive: $A_{FB}^{0,l}$, $A_l(P_{\tau})$,$\sin^2\Theta_{eff}^{lept}(Q^{had}_{FB})$,
       $A_l(SLD)$,$\Afbb$, $\Afbc$,$m_W$,$\sin^2\Theta_{W}(\nu{\cal N})$. 
      \item[ ] Others:  $\sigma^0_h$,$\Gamma_{{\rm Z}}$, $R_l^0$, $R_b^0$,$R_c^0$,$A_b$,$A_c$
       $\Gamma_{W}$, $Q_W(Cs)$.
     \end{itemize}

 The relative contributions to the total $\chi^2$ of the different types of observable are summarised in the
 first three rows of Table 24. The quantities $\sum\Delta(\chi')^2$ are calculated from the squares
 of the 'pulls' given in the last column of Table 16.1 of Reference~\cite{EWWG}. Also shown in
  Table 18 are the `pseudo-confidence levels', CL', for each  $\sum\Delta(\chi')^2$ assuming that
  the effective number of degrees of freedom, $d.o.f.'$, is equal to the number of observables
  of each type. Since there are five fitted parameters: $m_Z$, $M_t$, log($m_H$), $\alp$ and $\alps$,
  the true total number of degrees of  freedom is 15 rather than 20, so that the quoted values of CL'
 should be considered as upper limits on the true CL to be associated with each type of observable.
  This procedure neglects correlations between the fitted parameters, but is adequate to show
  the very different contributions of the different observable types to the overall $\chi^2$. 
  The `Measured' observables, i.e. those that are identical to fitted parameters, are seen to
  provide an essentially vanishing contribution to $\chi^2$, since the fitted parameters are
  completely determined by the corresponding observables. Thus three extra degrees of freedom
  are obtained, free of charge, which considerably improve the CL of the SM comparison\footnote{
   For example, calculating CL' by summing the values of $\sum\Delta(\chi')^2$ corresponding 
   to each observable type gives  $(\chi')^2/d.o.f.' = 26.64/20$, CL'$ = 0.15$  when the Measured 
   observables are included (see Table 24) and  $(\chi')^2/d.o.f.' = 26.60/17$, CL'$ = 0.064$ when they are
   excluded.}. The Measured parameters should be fixed in the fit, not
   treated as observables to be fitted. The effective confidence level, CL' of the 8 `$m_H$-sensitive'
   observables is a factor of 21 smaller than the CL' of all observables (see Table 24) and a factor
   6.4 smaller than the EWWG global fit CL. The value of CL' for the $m_H$-sensitive observables
   is similar to the maximum value 0.0064 of  $\overline{{\rm CL}}$ for the analysis of all 
    model-independent observables in the last column of Table 22 and Fig.6. The `Other' observables
 give a rather low contribution to the total  $(\chi')^2$ , as compared to the expectation from the 
   number of degrees of freedom, which give a further improvement to the CL of the global fit
  beyond that expected from the inclusion of observables that are only weakly sensitive to $m_t$
   and $m_H$.  Excluding the NuTeV datum $\sin^2\Theta_W(\nu {\cal N})$ from the group
  of $m_H$-sensitive observables gives $\sum \Delta(\chi')^2 = 12.6$, $d.o.f' = 7$, CL'$ =0.08$
  which is similar  to the maximum value of  $\overline{{\rm CL}}$ of
  0.053 for the corresponding analysis of model-independent observables in Fig 6 and the 
  first column of Table 22\footnote{Actually the EWWG and EWPDG fits use the old $m_t$ value
   174.3(5.2) GeV rather than 178.0(4.3) as used in Tables 22 and 23. This, however, 
    has only a minor effect on the maximum values of $\overline{{\rm CL}}$. Using the old value 
   of $m_t$, maximum values of  $\overline{{\rm CL}}$ of 0.0080 [0.057] are found when the
    NuTeV experiment is included [excluded], to be compared with the corresponding 
     numbers 0.0064 [0.053] quoted above using the new $m_t$ measurement.}
 Thus the difference between the confidence
  levels found in the present paper and those of the global EWWG fits can be 
  largely understood as a consequence of the extra degrees of freedom associated with the
  `Measured' and `Other' observables which are only weakly sensitive to
  the crucial unknown parameter, $m_H$, of the SM.
  \par Classifying the 42 observables used in the latest global EWPDG fit~\cite{EWPDG}
   in the same fashion as done above for the EWWG fit gives:
   \begin{itemize}
     \item[ ] Measured: $m_Z$, $m_t$.
      \item[ ] $m_H$-sensitive: $A_{FB}^{0,e}$, $A_{FB}^{0,\mu}$, $A_{FB}^{0,\tau}$, $\Afbb$, $\Afbc$,
            $A_{FB}^{0,s}$,$\sin^2\Theta_{eff}^{lept}(Q^{had}_{FB})$,$A_e(1)$,$A_e(2)$,$A_e(3)$,
       $A_{\mu}$,$A_{\tau}(1)$,$A_{\tau}(2)$, $m_W({\rm FERMILAB})$, $m_W({\rm LEP})$   
      \item[ ] Others: $\Gamma_{{\rm Z}}$, $\Gamma_{had}$,  $\Gamma_{inv}$,  $\Gamma_{l^+l^-}$,
        $\sigma^0_h$, $R_e^0$, $R_{\mu}^0$, $R_{\tau}^0$, $R_b^0$,$R_c^0$,$A_b$,$A_c$,$A_s$,
           $R^-$,$\kappa^{\nu}$, $R^{\nu}(1)$, $R^{\nu}(2)$, $g^{\nu e}_V(1)$,$g^{\nu e}_V(2)$,
      $g^{\nu e}_A(1)$,$g^{\nu e}_A(2)$, $Q_W(Cs)$,$Q_W(Tl)$,$\Gamma(b \rightarrow s \gamma)/
  \Gamma(c \rightarrow e \nu)$, $(g_{\mu}-2-\alpha/\pi)/2$
     \end{itemize}
   The difference with respect to the EWWG fit is that a wider range of observables are included
   as well as several different measurements of the same quantity, indicated, for example,as
   $A_e(1)$,$A_e(2)$..., and that charged lepton universality has not been used to reduce the number
    of observables. It could be argued that some of the `Other' observables such as 
  $\Gamma_{{\rm Z}}$, $\Gamma_{had}$ and $\Gamma_{inv}$ are actually quite `$m_H$-sensitive',
    but the choice of the latter type of observable has been restricted, for purposes of comparison,
     to correspond as closely
     as possible to that made above for the EWWG fit. The values of
     $\sum \Delta(\chi')^2$, $d.o.f.'$ and CL' for the three classes of observables in the
     EWPDG fit are presented in the fifth, sixth and seventh rows of Table 24. There are only two Measured
    observables, $m_t$ and $m_Z$, since the fitted parameters are: $m_t$, $m_Z$, $m_H$
    and $\alp$. Unlike for the EWWG fit $\alps$ is treated as a fixed rather than a fitted
   parameter. The high value of CL' for the Measured observables shows that, as in the case of
   the EWWG fit, it is more appropriate to treat  $m_t$ and $m_Z$ as fixed parameters in the fit.
   As for the EWWG fit, the value of CL' for the $m_H$-sensitive observables is much less
   than the global fit confidence level. The value of CL' for
   the `Other' observables gives no indication for a possible over-estimation of systematic
  errors as in the EWWG fit. However, the observables $\Gamma_{inv}$ and $(g_{\mu}-2-\alpha/\pi)/2$
   not included in the EWWG fit show quite large deviations from the SM prediction. Removing these
   observables from the `Others' set gives: $\sum \Delta(\chi')^2 = 14.4$, $d.o.f.' = 23$, CL'$ =0.91$, again
  indicating a possible overestimation of systematic errors for the remaining observables,
  leading to a confidence level for the global fit that gives an optimistic estimate
  of the level of agreement with the SM. The lower confidence level of the global EWPDG
  fit as compared to  the EWWG one, in spite of the inclusion of many
    unaveraged observables, is also explained by the inclusion of these
  `discrepant' observables in the former fit. Excluding them gives $\chi^2/d.o.f. = 37.7/36$,
 CL = 0.39 which is larger than that of the EWWG fit (0.28), as expected.  
  \par Thus the lower confidence levels found in the global analysis of the
   present paper as compared to those quoted by the EWWG and EWPDG are fully explained
 in terms of the dilution of the hypothesis testing power of the latter fits due to
 the inclusion of unaveraged or insensitive observables in the $\chi^2$ estimator.

 \begin{table}
\begin{center}
\begin{tabular}{|c|c|c|c|c|} \cline {2-5}
\multicolumn{1}{c}{ } & \multicolumn{1}{|c}{ Observable Type } &  \multicolumn{1}{|c}{$\sum\Delta(\chi')^2$}
 &  \multicolumn{1}{|c}{ $d.o.f.'$} &   \multicolumn{1}{|c|}{ CL'} \\ \hline 
   & Measured & 0.04 & 3 & 0.998 \\
  EWWG  & $m_H$-sensitive & 21.0 & 8 & 0.0071 \\
   & Others & 5.6 & 9 & 0.78 \\
   & All & 26.6 & 20 & 0.15 \\   \hline 
  & Measured & 0.05 & 2 & 0.975 \\
  EWPDG  & $m_H$-sensitive & 25.0 & 15 & 0.050 \\
   & Others & 25.1 & 25 & 0.78 \\
   & All & 50.2 & 42 & 0.18 \\   \hline  
\end{tabular}
\caption[]{{ \sl Contributions of different types of observables to the 
   $\chi^2$ of the latest global EWWG~\cite{EWWG} and EWPDG~\cite{EWPDG} 
 fits. See text for definitions of the quantities shown.}} 
\end{center}
\end{table}

  \par A final remark concerning the CLs quoted for the EWWG and EWPDG fits is that, as previously
  mentioned, these numbers correspond to central fitted values of $m_H$ that are grossly inconsistent
  with direct the experimental lower limit of 114.4 GeV. Replacing the fitted values by this
  limit will evidently yield lower CLs. Because of the relatively large uncertainy on the
  fitted value of $m_H$, the expected reduction in the CL will be less than that due to correcting
  for the statistical dilution effects discussed above. It  still remains however
  a meaningless exercise to quote CLs for fits with $m_H \simeq 90-100$ GeV, values that
  are excluded by the direct search result with a CL (see Fig.1) very much less than $10^{-7}$.

  \par The latest EWWG report~\cite{EWWG}, quotes a 95 $\%$ confidence level upper limit
  on $m_H$ of 219 GeV. This estimate is based on the famous `blue band'\footnote{
    So-called becuse of the coloured version shown in dozens of electroweak review
    talks over the last decade} plot
   (Fig 16.5 of Reference~\cite{EWWG}) which shows the quantity: $\Delta \chi^2 = \chi^2- \chi^2_{min}$
    for the global fit as a function of $m_H$. Choosing an appropriate fixed value of
     $\Delta \chi^2$ to define a confidence limit for $m_H$ is a valid procedure only in the case
     that the confidence level derived from the value of $\chi^2$
      of the fit to the $m_H$-sensitive variables is sufficiently
    high. Inspection of Figs 2 and 3 and Tables 11 and 12 shows that this is hardly
   the case for the actual electroweak data set. It was argued in the previous Section
  that most reliable estimate of $m_H$ is that derived using only charged lepton
  asymmetry data and $m_W$.
 In this case the the maximum value of $\overline{{\rm CL}}$ is 0.017
   when the NuTev $m_W$ measurement is included, 0.23, if it is excluded.
   Thus the  $\Delta \chi^2$ estimator for $m_H$ is acceptable only
   for the case of a fit to $A_l(lept)$ and $m_W$ excluding the NuTeV measurement.  However,
  at the 95 $\%$ confidence level EWWG upper limit of $m_H = 220$ GeV,
  the corresponding values of $\overline{{\rm CL}}$ are $1.1 \times 10^{-3}$ and
  0.054 
  with the implication that if a Higgs boson existed with this
  mass, and including, as in the final EWWG fit, the NuTeV measurement,
    the SM would be excluded by the data with a confidence
  level only slightly larger than $10^{-3}$! This is not at all the message that one might naively
  draw from the corresponding EWWG numbers. A confidence level of 4.5$\%$ for a global fit 
   gives the impression that the data is, in general, not too badly described by the
   SM. In fact as a brief inspection of Tables 19-23, which contain all relevant 
   information, show, this is hardly the case, so that a meaningful estimate
   of $m_H$ cannot be derived from a $\Delta \chi^2$ plot based on such a global fit.
   In the model-independent analysis of all precision data in the 
   previous Section, the level of the discrepancy with the SM may be even
   larger than if only the $m_H$-sensitive observables
   $A_l$ and $m_W$ are considered. Referring to Figs 6 and 7 and Tables 22
    and 23 and taking the ' NuTeV $\sbn$ meas.' curves,
    mandatory in a model-independent analysis, gives values of
    $\overline{{\rm CL}}$ of 0.0038 (0.0056) for $m_H = 115~(220)$ GeV for the
   `All data' case and 0.0047 ($1.0 \times 10^{-4}$) for the most reliable
    `Lepton only' data. The reader must judge for herself (or himself) whether, in
    these circumstances, the SM does, or does not, provide an adequate 
    description of the current data. Only in the case that it does, the `blue band' plot,
     derived from a global fit to this data, provides 
     a meaningful upper limit on $m_H$.
 
 \SECTION{\bf{Summary and Conclusions}}
 It has been demonstrated in this paper that the the statistical estimator,  $\chi^2_{data,SM}$
 universally employed by the  EWWG and EWPDG to judge the level of overall agreement of precision
  electroweak data with the SM predictions typically yields a confidence level an order 
  of magnitude higher than estimators chosen to test specifically this level of agreement
  rather than the internal consistency of different measurements. The reasons for this
  are discussed in Section 2. While, as shown in Section 4, the number of independent
  observables sensitive to the most important poorly known parameters $m_t$ and $m_H$ is very 
  small, large numbers of observables (20 for EWWG, 42 for EWPDG) are used to construct
  $\chi^2_{data,SM}$. In these circumstances the corresponding CL reflects more the internal 
  consistency of measurements of different observables than the level of agreement of the essential
  `refined' parameters with the SM. Further dilution of the hypothesis testing power of
   $\chi^2_{data,SM}$ results from the inclusion of many observables only weakly sensitive to
   $m_t$ and $m_H$ as well as fit parameters identical to measured quantities that give anomalously
   low contributions to the global $\chi^2$.
   \par In Section 3 the internal consistency of different heavy quark asymmetry measurements is
    discussed in detail. It is found to be very good. There is a hint, from the data itself, that
    uncorrelated systematic errors may be overestimated. Thus the observed 2.5$\sigma$ deviation 
   of the LEP+SLD average value of $A_b$ may be an underestimation. Combination
   of the statistically independent $\chi^2$ and Run Tests for the LEP and SLD measurements
    of $A_b$ yields a CL of $2.1 \times 10^{-3}$ for a pure statistical fluctuation, which therefore
    seems unlikely. Attributing the $A_b$ deviation to a correlated systematic error of 
    unknown origin in the LEP $\Afbb$ measurements would require an effect that is 1.8 times 
    larger than the estimated QCD correction, which is the dominant source of correlated systematic
    uncertainty on this quantity,
  and 13 times larger than the estimated uncertainty on this
   correction. Even given the inevitable theoretical uncertainties associated with QCD effects
   this again seems unlikely. Thus there is no sound experimental reason to doubt, as suggested by
    some authors~\cite{ACGGR,GAMG}, the validity of the b-quark asymmetry measurements
   and the possible evidence they provide for new physics beyond the SM.
    \par It is shown in Section 4 that essentially all information from quantum loop effects
    on the values of  $m_t$ and $m_H$ is provided by only three observables:
     $A_l(lept)$, $\sbl$ and $m_W$ (see Table 5). For  $m_H$ there are only two strongly
      sensitive observables: $A_l(lept)$ and $m_W$. It is demonstrated by fitting only the 
      $m_H$-sensitive observables and also complete sets of model-independent observables
     that indeed the fitted value of $m_H$ is essentially determined by the former set only.
     In an approach similar to that followed in Reference~\cite{Chanowitz} it is assumed, in
     Section 4, that the SM is valid so that different values of $A_l$ may be extracted
     from purely leptonic data ($A_l(lept)$) and from quark forward/backward asymmetries
      ($A_l(had)$). As previously noted~\cite{Chanowitz}, for the case of the equivalent 
    observable $\sin^2\Theta_{eff}^{lept}$, these two estimates differ by 3.0$\sigma$ showing
    that the SM interpretation of the data is also inconsistent at this level. This is a simple
    consequence of the observed anomalous behaviour of $A_b$ and the possibly similar 
     behaviour of other Z${\rm q \overline{q}}$ couplings. Also, as previously noticed, much
   larger values of $m_H$ are favoured by $A_l(had)$ than by $A_l(lept)$. This is because
    (see Table 5) in the SM only $A_l$, not $A_b$, is sensitive to $m_H$ and the measured 
   forward backward asmmetry $\Afbb$ is, as shown in Eqn(3.1), proportional to
   $A_l A_b$. Whatever the explanation of the $A_b$ anomaly, the most reliable estimate
   of $m_H$ that can be derived from the present data is therefore that provided by $A_l(lept)$.
   Inclusion of $A_l(had)$, assuming the correctness of the SM, gives a calculable bias towards
   higher values of $m_H$ resulting from the $A_l$-$A_b$ correlation in the quark
   forward/backward asymmetry.
   \par In Section 5 a combined confidence level, $\overline{{\rm CL}}$, as a function of
    $m_H$ is derived from the directly measured~\cite{HIGGSMD} confidence level curve
     ${\rm CL}_{s+b}$  and the CL curve derived from the $\chi^2$ estimator using the
     $m_H$-sensitive observables $A_l$ and $m_W$ (see Fig.1). As in Section 4 the correctness
    of the SM is assumed and different  $\overline{{\rm CL}}$ curves are calculated for
     $A_l(lept)$,  $A_l(had)$ and  $A_l(all)$, where the latter is the weighted average
    of the former two observables. The above mentioned inconsistency between  $A_l(lept)$
    and $A_l(had)$ is taken into account when calculating the $\overline{{\rm CL}}$ curve
    for $A_l(all)$.  Consistent results for  $\overline{{\rm CL}}$ (see Tables 9 and 10)
     are found using either
     $\chi^2_{data,WA}+\chi^2_{WA,SM}$
    directly or combining the CLs of  $\chi^2_{data,WA}$ and $\chi^2_{WA,SM}$ using Eqn(2.8).
    Only the standard interpretation of the NuTeV experiment, as a measurement of $m_W$, is
    considered in Section 4, and the $\overline{{\rm CL}}$ curves are calculated both including
    and excluding this datum. The results for  $\overline{{\rm CL}}$ are shown in Tables 9-12 
    and Figs. 2 and 3. The inconsistency of the SM interpretation is evident on inspection
    of these figures. For $m_H = 300$ GeV values of $\overline{{\rm CL}}$ differing by two orders
   of magnitude are obtained from  $A_l(lept)$ and $A_l(had)$. Also studied in Section 5 is the
   dependence of the $\overline{{\rm CL}}$ curves on the assumed values of $m_t$ and $\alp$
   (Table 13 and Figs.4 and 5). Variation of $m_t$ by plus or minus the experimental uncertainty
    changes the value of $\overline{{\rm CL}}$ by more than three orders of magnitude for 
     $m_H \simeq 300$ GeV; somewhat smaller changes are given by a similar variation of $\alp$.
    This demonstrates the importance of more precise measurements of these parameters in order
   to obtain well defined SM predictions.
   \par In Section 6 the alternative interpretation of the result of the NuTeV experiment as
   a measurement of the Z$\nu \overline{\nu}$ coupling, rather than $m_W$, is considered.
    It is pointed out that the former interpretation, mandatory in a model-independent 
    analysis where no {\it a priori} assumptions are made concerning the strengths of the
   Z$f \overline{f}$ couplings, is highly favoured by an argument based on the internal 
   consistency of LEP and NuTeV data. In this case the model-independent observable
    $\sbn$ derived from the LEP and NuTeV data differs from the SM prediction by
    3.7$\sigma$ and so is the largest single deviation observed from a SM prediction.
    The combined CL, taking into account both agreement with the SM and data consistency,
    is very similar for either interpretation of the NuTeV result. 
    \par An analysis in terms of model-independent observables similar to those previously
    published for earlier electroweak data sets~\cite{JHF1,JHF2,JHF3,JHF4} is presented in
    Section 7. Results are presented in terms of the `maximally uncorrelated' observables
   presented in Table 14. Vector and axial vector couplings of the Z to fermion pairs
    (Table 15), and the equivalent right-handed and left-handed couplings (Table 19) are also
   presented and compared with SM predictions. The history of, and the different possible 
  physical interpretations of, the $A_b$ anomaly are also discussed. Although there are no
  purely experimental reasons for doubting the correctness of the fully compatible
  LEP and SLD measurements of $A_b$  it is pointed out that the good agreement between
  the  measured values of $\sbb$ and the SM prediction (which requires the presence
  of large $m_t$ dependent quantum corrections) must be fortuitous if the
  $A_b$ anomaly is to be explained by new physics. This is an argument suggesting an
   unknown systematic bias as the cause of the effect. It is also pointed out that such
   a systematic bias would result in an anomaly predominantly in the right handed coupling
   (2.8 $\sigma$ effect observed) rather than in the left-handed one (1.7 $\sigma$ effect observed).
   This implies that only new, more precise, experiments can discriminate between new physics
   and unknown systematic bias as the cause of the observed anomaly in the right handed
   b-quark coupling. It is shown that the measured value of $\Gamma_{had}$ renders unlikely the
   possiblity that the couplings of the other d-type quarks differ from the SM prediction
   in the same way as observed for the b-quark couplings. The existing direct measurements of
   these couplings are insufficiently precise to provide any useful constraints.
    \par In Section 7 the observed and expected quantum loop corrections are also discussed
    (see Table 21). The most significant and the most precisely measured effect,
     $\delta_{Quant}^l$, defined in Eqn(7.12) is a 4$\%$ effect measured with a relative
    precision of 2.2$\%$ (measured effect 44$\sigma$ from zero). For a value of $m_H$
    of 120 GeV, the agreement of $\delta_{Quant}^l$ with the SM prediction is quite
   satisfactory (1.2$\sigma$ deviation) but, as shown in Table 20, for higher values of
   $m_H$ the level of agreement of the charged lepton couplings that determine $\delta_{Quant}^l$
   deteriorates rapidly. $\delta_{Quant}^c$ (Eqn(7.13) is also predicted to be
    $\simeq 0.04$ in good agreeement with the measurement (0.7$\sigma$ deviation) but here
     the fractional precision of the measurement is only 42$\%$.  $\delta_{Quant}^b$ 
   is predicted to be  $\simeq 0.05$ with an expected relative accuracy of 151$\%$, so that no
   significant measurement is to be expected in this case. In fact the measured value of 
   $\delta_{Quant}^b$ is much larger, 0.249(74), and so requires new physics at the tree level
   (2.7$\sigma$  deviation from the SM) if the data is correct. In contrast the 
   experimental value of  $\delta_{Quant}^{\nu}$ (Eqn(7.15) is 0.00502 with an expected 
   relative experimental uncertainty of 24$\%$ (4.2$\sigma$ deviation from zero) whereas
   the measured value is 0.0006(12) (0.5$\sigma$  deviation from zero). In this case the 
   expected quantum corrections are not observed, leading to a -3.7$\sigma$ deviation from
    SM prediction. Some theoretical interpretations of this effect have already been
    proposed~\cite{LOTW,LORTW}.

   \begin{figure}[htbp]
\begin{center}\hspace*{-0.5cm}\mbox{
\epsfysize10.0cm\epsffile{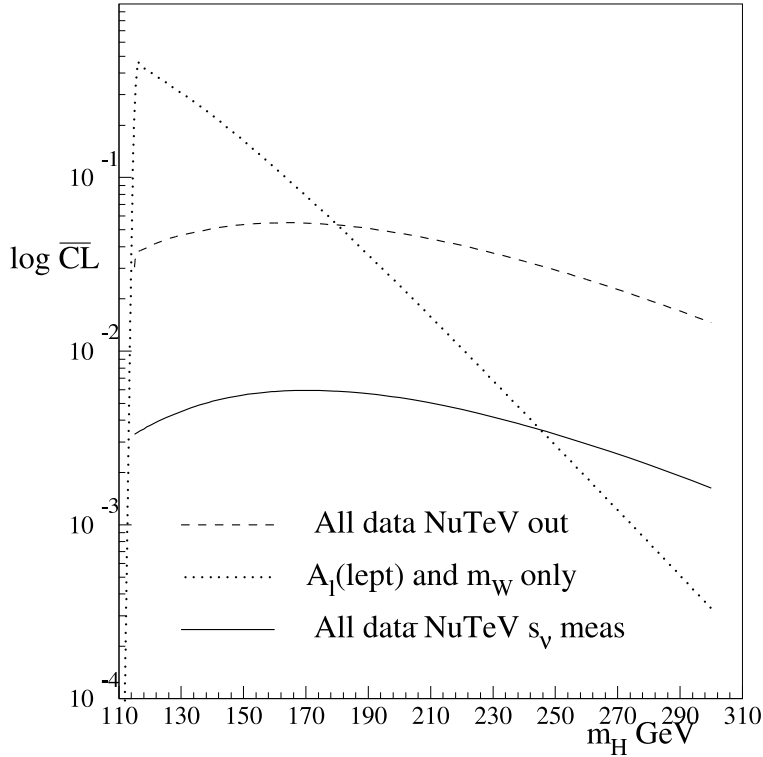}}
\caption{{ \sl Combined $m_H$ confidence levels. See Fig.6 for
  the definition of `All data'.}} 
\label{fig-fig8}
\end{center}
 \end{figure}

    \par Finally, in Section 7, combined confidence level curves $\overline{{\rm CL}}$ are derived
 including in the $\chi^2$ estimator not only the $m_H$-sensitive observables, as in Section 5,
  but all, or chosen subsets of, other model-independent observables. In order to see the impact
   of the NuTeV experiment, $\overline{{\rm CL}}$ curves are are presented excluding the
    NuTeV data or for the two alternative interpretations: $\sbn$ or $m_W$ measurements.
   The same set of  $\overline{{\rm CL}}$ curves is also obtained using only the leptonic
   observables: $A_l(lept)$, $\sbl$ and $\sbn$ in addition to $m_W$. The results are shown in
   Tables 6 and 7 and Tables 22 and 23. The curves where NuTeV is included lie about an 
   order of magnitude below those where it is excluded. The  $\overline{{\rm CL}}$ curves
   for the two different interpretations of the NuTeV experiment are similar, except that those
   corresponding to an $\sbn$ measurement lie significantly lower for large values of
    $m_H$  of $ \ge 300$ GeV.  
   \par In Section 8 a comparison is made between the CLs found previously in the
    present paper and those quoted for the latest published EWWG and EWPDG global fits.
    They are shown to be quite consistent when the various dilution effects of the
    $\chi^2$ estimators used in the global fits are taken into account. The maximum
    value of  $\overline{{\rm CL}}$ using all data and choosing the  $\sbn$ measurement
    interpretation of the NuTeV experiment of 0.0059, at $m_H = 180$ GeV, is about an order
   of magnitude lower than the CL of 4.5$\%$ quoted for the `all data' EWWG fit. The CLs of
   the global EWWG and EWPDG fits correspond to central fitted values of $m_H$ completely 
   excluded by the direct experimental lower limit of 114.4 GeV at 95$\%$ CL. Replacing the
   fitted values of $m_H$ by this lower limit, to give the largest possible CL consistent
   with all experimental data, will result in lower CLs than those quoted for the fits.
 
    \par Finally are shown, in Fig.8, the author's personal choice of the three most
    pertinent  $\overline{{\rm CL}}$ curves among the 18 different ones previously 
   presented in this paper. The `$A_l(lept)$ and $m_W$ only' curve (dotted) gives the
   most reliable estimate of $m_H$. The value of  $\overline{{\rm CL}}$ of 
   $\simeq 0.2-0.3$ for values of $m_H$ just above the direct lower limit of
    114.4 GeV is quite acceptable. However for $m_H = 300$ GeV, $\overline{{\rm CL}}$ 
   is $< 10^{-3}$ implying that, if the SM describes correctly the charged lepton
    sector and the Higgs boson exists, it must be very light indeed: $\le 180$ GeV if 
    $\overline{{\rm CL}} \ge 0.05$. Higher values of $m_H$ are favoured by the 
    `All data NuTeV out' curve (dashed). This is mainly due to the
     $A_l-A_b$ correlation following from Eqn(3,1) in the $\Afbb$ measurement,
     as discussed above. The maximum value of $\overline{{\rm CL}}$ is 
     $\simeq 0.05$ at about  $m_H = 140$ GeV. Including the NuTeV measurement
      gives the  `All data NuTeV $\sbn$ meas.' curve (solid line) with a similar shape
     but lying roughly an order of magnitude lower. The maximum value of
     $\overline{{\rm CL}}$ is $\simeq 0.006$ at   $m_H  \simeq 180$ GeV.
     This accurately reflects the best possible level
     of agreement, with the SM prediction, of the entire electroweak data set.
     It is an order of magnitude, or more, lower than the CLs quoted by the EWWG and EWPDG
     groups for their global fits to similar data sets.
     \par {\bf Acknowledgements}
     \par I thank B.Roe for teaching me the correct way to combine confidence levels and W.Metzger for
      several illuminating conversations on statistical matters. I especially thank M.Dittmar and S.Mele
     for their detailed criticisms that have enabled me to improve the clarity of the presentation
     in several places. Comments on the manuscript from G.D'Agostini, T.Aziz, M.Chanowitz, J.Erler, P.Gambino
     and P.Renton are also gratefully acknowledged.

\pagebreak

\end{document}